\documentclass[aps,prd,onecolumn,preprintnumbers,nofootinbib]{revtex4}
\pdfoutput=1

\usepackage{bm}
\usepackage{graphicx}
\usepackage{amsmath}
\usepackage{epsf}
\usepackage{amssymb} 
\usepackage{color}
\usepackage{txfonts}
\usepackage{natbib}
\usepackage{hyperref}
\usepackage{dcolumn}% Align table columns on decimal point
\usepackage{subfigure}

\newcommand{\xv}{{\bf x}}
\newcommand{\kv}{{\bf k}}

\newcommand{\qv}{{\bf q}}

\newcommand{\pin}{\phi_{\ast}}
\newcommand{\dep}{\Delta \phi}

\newcommand{\skyp}[1]{}

\newcommand{\bi}{B_{\ellt}}
\def\be{\begin{equation}}
\def\ee{\end{equation}}
\def\bea{\begin{eqnarray}}
\def\eea{\end{eqnarray}}

\def\d{\delta}
\def\Ms{\, h^{-1} \, {\rm M}_{\odot}}
\def\Mpc{\, h^{-1} \, {\rm Mpc}}

\def\cGpc{\, h^{-3} \, {\rm Gpc}^3}
\def\kMpc{\, h \, {\rm Mpc}^{-1}}
\def\icMpc{\, h^3 \, {\rm Mpc}^{-3}}
\def\nng{n_{\rm NG}}
\def\tfnl{\tilde{f}_{\rm NL}}
\def\fnl{f_{\rm NL}}

\def\fnll{f_{\rm NL}^{\rm loc.}}
\def\dfnll{\Delta f_{\rm NL}^{\rm loc.}}
\def\fnle{f_{\rm NL}^{\rm eq.}}
\def\dfnle{\Delta f_{\rm NL}^{\rm eq.}}
\def\nNG{n_{\rm NG}}

\def\kp{k_p}

\def\ellt{\ell_1\ell_2\ell_3}

\begin{document}

\title{Constraining Running Non-Gaussianity}

\author{Emiliano Sefusatti$^{1,2,3}$,  Michele Liguori$^4$, Amit P. S. Yadav$^{5}$, Mark G. Jackson$^{3,6,7}$, Enrico Pajer$^{8}$}
\email{emiliano.sefusatti@cea.fr}

\affiliation{$^1$Institut de Physique Th\'eorique, CEA, IPhT, F-91191 Gif-sur-Yvette, France}
\affiliation{$^2$Nordita, Nordic Institute for Theoretical Physics, 106 91 Stockholm, Sweden}
\affiliation{$^3$Center for Particle Astrophysics, Fermilab, Batavia, IL 60510, USA}
\affiliation{$^4$DAMTP, University of Cambridge, Wilberforce Road, Cambridge, CB3 0WA, UK}
\affiliation{$^5$Center for Astrophysics, Harvard University, Cambridge, MA 02138, USA}
\affiliation{$^6$Theory Group, Fermilab, Batavia, IL 60510, USA}
\affiliation{$^7$Instituut-Lorentz, 2333CA Leiden, The Netherlands}
\affiliation{$^8$Laboratory for Elementary Particle Physics, Cornell University, Ithaca, NY 14853, USA}

\begin{abstract}

The primordial non-Gaussian parameter $\fnl$ has been shown to be scale-dependent in several models of inflation with a variable speed of sound, such as Dirac-Born-Infeld (DBI) models. Starting from a simple ansatz for a scale-dependent amplitude of the primordial curvature bispectrum for two common phenomenological models of primordial non-Gaussianity, we perform a Fisher matrix analysis of the bispectra of the temperature and polarization of the Cosmic Microwave Background (CMB) radiation and derive the expected constraints on the parameter $\nng$ that quantifies the running of $\fnl(k)$ for current and future CMB missions such as WMAP, Planck and CMBPol. We find that CMB information alone, in the event of a significant detection of the non-Gaussian component, corresponding to $\fnl=50$ for the local model and $\fnl=100$ for the equilateral model of non-Gaussianity, is able to determine $\nng$ with a $1$-$\sigma$ uncertainty of $\Delta\nng\simeq 0.1$ and $\Delta\nng\simeq 0.3$, respectively, for the Planck mission and a factor of two better for CMBPol.
In addition, we consider a simple Fisher matrix analysis of the galaxy power spectrum to determine the expected constraints on the running parameter $\nng$ for the local model and of the galaxy bispectrum for the equilateral model from future planned and proposed photometric and spectroscopic surveys. We find that, in both cases, large-scale structure observations should achieve results comparable to or even better than those from the CMB, while showing some complementarity due to the different distribution of the non-Gaussian signal over the relevant range of scales.
Finally, we compare our findings to the predictions on the amplitude and running of non-Gaussianity of DBI inflation, showing how the constraints on a scale-dependent $\fnl(k)$ translate into constraints on the parameter space of the theory.   
\end{abstract}

\preprint{NORDITA-2009-34}
\preprint{FERMILAB-PUB-09-016-A-T}

\maketitle

%%%%%%%%%%%%%%%%%%%%%%%%%%%%%%%%%%%%%%%%%%%%%%%%%%%%%%%%%%%%%%%%%%%%%%%%%%%%%%%%
%%%%%%%%%%%%%%%%%%%%%%%%%%%%%%%%%%%%%%%%%%%%%%%%%%%%%%%%%%%%%%%%%%%%%%%%%%%%%%%%
\section{Introduction}

Present observations still have little power in discriminating between the huge number of different inflationary scenarios that have been proposed so far in the literature. Moreover, alternatives to inflation such as cyclic and ekpyrotic Universes are also compatible with the data. In order to discriminate between all these possibilities we need to move from the study of general properties shared by all the models under examination, such as flatness and scale-invariance, to more specific and strongly model-dependent predictions. For example, different inflationary models predict the primordial curvature perturbations to be {\em close to} but {\em not exactly} Gaussian. The specific departure from Gaussianity is highly model-dependent and, if detected, can be an extremely valuable tool for the purpose of discriminating between alternative scenarios.

Until fairly recently it has been generally believed that measurements of the bispectrum of CMB anisotropies, \citep{KomatsuSpergel2001,BabichZaldarriaga2004,FergussonShellard2007,YadavKomatsuWandelt2007,SmithSenatoreZaldarriaga2009}, were able to put the most stringent constraints on $\fnl$ when compared to all the other alternatives. However, the use of large-scale structure (LSS) observations as a probe of primordial non-Gaussianity has received increasing attention in the last year as it has been shown that present and future galaxy surveys allow one to obtain constraints at the same level of those from the CMB. Non-Gaussian initial conditions are expected to have a significant effect on the high-mass tail of the halo distribution, \citep{LucchinMatarrese1988,ColafrancescoLucchinMatarrese1989,ChiuOstrikerStrauss1998,RobinsonGawiserSilk2000,MatarreseVerdeJimenez2000,RobinsonBaker2000,SefusattiEtal2007,LoVerdeEtal2008}, and it has been more recently recognized in the void distribution \citep{KamionkowskiVerdeJimenez2008}. A large amplitude of the primordial bispectrum could lead to an observable initial component in the large-scale skewness of the galaxy distribution, \citep{FryScherrer1994,ChodorowskiBouchet1996,DurrerEtal2000} or on the galaxy bispectrum, \citep{Scoccimarro2000A,VerdeEtal2000,ScoccimarroSefusattiZaldarriaga2004,SefusattiKomatsu2007}. More recently, significant interest has been generated by the unexpected but large effect of local non-Gaussian initial conditions on the bias of halos and galaxies~\cite{DalalEtal2008,MatarreseVerde2008,SlosarEtal2008,McDonald2008,AfshordiTolley2008,TaruyaKoyamaMatsubara2008,Seljak2008,GrossiEtal2009}. Remarkably, this new method lead to constraints on the specific local type of non-Gaussianity from current data-sets, which are already competitive with CMB limits, \cite{SlosarEtal2008,AfshordiTolley2008}. Finally, additional confirmation of a possible detection might also come from alternative statistics such as Minkowski Functionals \citep{HikageKomatsuMatsubara2006,HikageEtal2008A}.

So far, measurements of the bispectrum of the CMB temperature fluctuations, a direct probe of the primordial bispectrum, have been found to be consistent with the Gaussian hypothesis and limits have been placed on the amplitude of a possible non-Gaussian component, \citep{KomatsuEtal2008,YadavWandelt2008,SmithSenatoreZaldarriaga2009}. Such limits are placed on the amplitude of the curvature bispectrum ---the $\fnl$ parameter--- assuming the specific dependence on the shape of the triangular configuration predicted by the inflationary model. This amplitude is usually assumed to be scale-independent. However, models such as Dirac-Born-Infeld (DBI) inflation \citep{SilversteinTong2004,AlishahihaEtal2004,Chen2005}, can be characterized by a primordial bispectrum whose amplitude varies significantly over the range of scales accessible by current cosmological probes.

Motivated by these considerations, the aim of this paper is to study what limits can be placed, in the event of a detection, on the {\it scale-dependence} of the bispectrum amplitude using present and forthcoming cosmological data-sets and how such information can be used to obtain additional constraints on inflationary models. This issue was previously addressed by \citet{LoVerdeEtal2008}, who made forecasts based on cluster number counts as a probe of non-Gaussianity on small scales combined with current CMB constraints. This work, however, simply assumed previous CMB limits on $\fnl$ as constraints on the amplitude {\it alone} at a given pivot scale. In this regard, we make a step forward by providing a complete Fisher matrix analysis of the CMB bispectrum described by two parameters: the overall amplitude, $\fnl$, and a non-Gaussian running parameter $\nng$. In addition we combine these results with simple estimates of those achievable by LSS observations such as the galaxy power spectrum and bispectrum in forthcoming surveys.

The paper is organized as follows: In Section~\ref{sec:parNG} we discuss various parameterizations of the primordial bispectrum and the phenomenological ansatz assumed in our analysis. In Section~\ref{sec:cmb} we review the CMB bispectrum, define our notation and derive the CMB Fisher matrix for the  amplitude $\fnl$ and running $\nng$, showing the results for current and future CMB missions. In Section~\ref{sec:lss} we derive the Fisher matrices for the LSS power spectrum and bispectrum and show the corresponding constraints for a sample of future photometric and spectroscopic galaxy surveys. In Section~\ref{results} we present the combined CMB plus LSS results. In Section~\ref{sec:dbi} we consider DBI as an example of inflationary model. Finally in Section~\ref{sec:conclusions} we summarize our findings and give some concluding remarks.

%%%%%%%%%%%%%%%%%%%%%%%%%%%%%%%%%%%%%%%%%%%%%%%%%%%%%%%%%%%%%%%%%%%%%%%%%%%%%%%%
%%%%%%%%%%%%%%%%%%%%%%%%%%%%%%%%%%%%%%%%%%%%%%%%%%%%%%%%%%%%%%%%%%%%%%%%%%%%%%%%
\section {Parameterizations of Primordial non-Gaussianity}
\label{sec:parNG}

The statistical properties of a Gaussian random field are encoded in its two-point correlation function or, equivalently, its power spectrum since all higher-order {\it connected} correlation functions are vanishing. On the other hand, a non-Gaussian field is given, in principle, by an infinite set of functions. Any model of inflation indeed provides predictions for all primordial correlators which are characterized not only by an overall constant amplitude but also by their peculiar dependence on the shape of the configuration of points in position space or wavenumbers in Fourier space. 

In this Section we introduce two of the most common functional forms for the curvature primordial bispectrum, the so-called {\it local} and {\it equilateral} shapes, and describe the specific ansatz for the scale-dependence of the bispectrum amplitude that we assume in our analysis.

%%%%%%%%%%%%%%%%%%%%%%%%%%%%%%%%%%%%%%%%%%%%%%%%%%%%%%%%%%%%%%%%%%%%%%%%%%%%%%%%
\subsection {Shape-dependence and amplitude of the primordial bispectrum}
\label{sec:shape}

The non-Gaussian initial conditions can be generally characterized in terms of the bispectrum $B_\Phi(k_1,k_2,k_3)$ of Bardeen's curvature perturbations $\Phi(k)$, defined as
\be
\label{eq:prim_bi}
\langle \Phi_{\kv_1}\Phi_{\kv_2}\Phi_{\kv_3}\rangle =
\delta_D^3(\kv_{123})~B_\Phi(k_1, k_2, k_3),
\ee
where we introduce the notation $\kv_{ij}\equiv\kv_i+\kv_j$. We can separate the overall amplitude $\fnl$ from the functional form $F(k_1,k_2,k_3)$ as
\be
B_\Phi(k_1,k_2,k_3)\equiv\fnl~F(k_1, k_2, k_3),
\ee
where we will define the normalization of $F$ below. Here $\fnl$ is a dimensionless parameter while the function $F(k_1,k_2,k_3)$ describes the dependence on the shape of triangular configuration defined by the three wavenumbers $k_1$, $k_2$ and $k_3$, typically having the hierarchical behavior $F(k,k,k)\sim P_\Phi^2(k)$ for equilateral configurations, with $P_\Phi(k)$ being the curvature power spectrum. In principle, the curvature trispectrum is also expected to have a significant effect on large-scale structure observables at the largest scales \citep{JeongKomatsu2009,Sefusatti2009}. However,  for simplicity we will ignore effects due to correlations functions of order higher than the bispectrum. A more detailed discussion on their role will be given in Section~\ref{sec:lss}.

Different inflationary  models predict different values for $\fnl$, starting from ${\mathcal O}(.01)$ up to very large values. CMB observations provide the upper limit $\fnl \lesssim 100$, already constraining some of the existing models. A theoretical lower bound on the detection of {\it primordial} non-Gaussianity is roughly $\fnl\gtrsim {\mathcal O}(1)$, below which second-order perturbations from post-inflationary evolution become relevant. Non-Gaussianity from canonical single-field slow-roll inflation models is predicted to be very small ${\mathcal O}(.01)$~\cite{SalopekBond1990,SalopekBond1991,FalkRangarajanSrednicki1993,GanguiEtal1994,AcquavivaEtal2003,Maldacena2003}; however,  a  large  class  of  more  general models, {\it e.g.} models with multiple scalar fields, features in inflation potential, non-adiabatic fluctuations, non-canonical   kinetic   terms,    deviations from the Bunch-Davies vacuum, among others, predict  substantially   higher level of primordial non-Gaussianity (for a review, see~\citep{BartoloEtal2004} and references therein, for recent contributions see also~\citep{ChenEastherLim2008,ByrnesChoiHall2008A,ByrnesChoiHall2008B}). For this reason alone, a detection of primordial non-Gaussianity would have the important consequence of ruling out canonical single-field slow-roll inflation as a viable inflationary scenario. 

In addition to the amplitude differences, different models of inflation also lead to different functional forms for the bispectrum, characterized by $F(k_1, k_2,k_3)$. Such functional forms can be broadly classified into three classes~\citep{BabichCreminelliZaldarriaga2004,FergussonShellard2008,KomatsuEtal2009}: a local, ``squeezed,'' shape-dependence where $F(k_1, k_2, k_3)$ is large for the configurations in which $k_1 \ll k_2$, $k_3$ (and permutations); a non-local, ``equilateral,''  where $F(k_1, k_2, k_3)$ is large for $k_1 \sim k_2 \sim k_3$ and a ``folded'' shape-dependence where $F(k_1, k_2, k_3)$ is large for flattened configurations $k_1\sim k_2\sim k_3/2$ (and permutations). Several other shapes, however, have been considered in the literature. For a more complete discussion on shapes and their correlations, see for instance \citet{FergussonShellard2008}. We limit our attention to the local and equilateral forms both because of their simplicity and because they mimic a large part of the models present in the literature.  
 
The local form can arise from a non-linear relation between the inflaton and curvature perturbations~\cite{SalopekBond1990,SalopekBond1991,GanguiEtal1994}, or alternatively in curvaton models~\cite{LythUngarelliWands2003}. Alternatives to inflation like New Ekpyrotic and cyclic models are also expected to produce a large level of non-Gaussianity of this type~\citep{KoyamaEtal2007,BuchbinderKhouryOvrut2008,LehnersSteinhardt2008A,LehnersSteinhardt2008B}. The local form of non-Gaussianity is so-called because it can be parametrized in real space by the local expression~\citep{GanguiEtal1994,VerdeEtal2000,KomatsuSpergel2001}:
\begin{equation}
\label{eqn:phiNG}
\Phi(\xv) = \Phi_G(\xv) + \fnll \left[ \Phi_G^2(\xv) - \langle \Phi_G^2(\xv) \rangle\right]\, ,
\end{equation}
where $\Phi_G$ is the Gaussian part of the perturbations. The term $\langle \Phi_G^2(\xv) \rangle$ ensures that the perturbation has zero mean. From Eq.~(\ref{eqn:phiNG}) it is easy to show that the shape dependence, $F_{\rm loc.}(k_1,k_2,k_3)$, for the local model takes the following form:
\be
\label{eq:f_local}
F_{\rm loc.}(k_1, k_2, k_3) = 2 \Delta^2_{\Phi}\left[ \frac{1}{k^{3-(n_s-1)}_1 k^{3-(n_s-1)}_2} + \frac{1}{k^{3-(n_s-1)}_1k^{3-(n_s-1)}_3}+\frac{1}{k^{3-(n_s-1)}_2 k^{3-(n_s-1)}_3}\right],
\ee  
where the normalized power spectrum $\Delta_\Phi$ is defined in terms of the tilt $n_s$ and the curvature power spectrum $P_\Phi(k)$ as $P_\Phi(k)\equiv \Delta_\Phi k^{-3+(n_s-1)}$ with $P_\Phi(k)$ defined in terms of the Gaussian component alone as $\langle\Phi_{G}({\bf k}_1) \Phi_{G}({\bf k}_2) \rangle \equiv \delta_D^{(3)}(\kv_{12}) P_\Phi(k_1)$. 

Equilateral forms of non-Gaussianity arise from models with non-canonical kinetic terms such as the DBI action~\citep{AlishahihaEtal2004}, ghost condensation~\citep{ArkaniHamedEtal2004}, or any other single-field models in which the scalar field acquires a low speed of sound~\citep{ChenEtal2007,CheungEtal2008}. Although the shapes predicted by different models are in this case not identical, it has been noted \cite{BabichCreminelliZaldarriaga2004,CreminelliEtal2006} that they are all very well-approximated by the function: 
\begin{eqnarray}
\label{eq:f_eq}
F_{eq.}(k_1, k_2, k_3) & = &  
6\Delta^2_{\Phi}\left[- \frac{1}{k^{3-(n_s-1)}_1k^{3-(n_s-1)}_2} + {\rm 2~perm.}
       - \frac{2}{(k_1k_2k_3)^{2-2(n_s-1)/3}}\right.
\nonumber\\
& &    \left. + \frac{1}{k^{1-(n_s-1)/3}_1k^{2-2(n_s-1)/3}_2k^{3-(n_s-1)}_3} + {\rm 5~perm.}\right].
\end{eqnarray}
The definition for the equilateral model follows from the local one since $\fnle$ is defined in such a way that for equilateral configurations, $F_{\rm eq.}(k,k,k)=F_{\rm loc.}(k,k,k)$ and one obtains the same value for $B_\Phi$ given $\fnle=\fnll$. 

Comparisons with observations of different forms of non-Gaussianity
lead to constraints on the different amplitude parameters. The most recent analysis of the CMB bispectrum provides for the local non-Gaussian parameter the limits $-4<\fnll<80$ at $95\%$ C.L. corresponding to a $1\sigma$ error on $\fnll$ of $\Delta \fnll\simeq 21$~\cite{SmithSenatoreZaldarriaga2009}. In the equilateral case we have $-151<\fnle<253$ with $\Delta \fnle\simeq 101$~\citep{KomatsuEtal2008}. Future missions are expected to provide uncertainties of the order of $\dfnll\sim 3$ and $\dfnle\sim 10$, \citep{KomatsuSpergel2001,BabichZaldarriaga2004,YadavKomatsuWandelt2007}. Recent constraints on local non-Gaussianity from the bias of high-redshift objects correspond to $\Delta \fnll \sim 24$ \citep{SlosarEtal2008}. Future large-scale surveys might probe $\fnll$ with $\dfnll\sim 1$, \citep{SlosarEtal2008,CarboneVerdeMatarrese2008,GrossiEtal2009}. This method, however, would lead to very mild constraints for the equilateral form of the initial bispectrum which is expected to have a negligible effect on halo bias, \citep{MatarreseVerde2008,TaruyaKoyamaMatsubara2008}. On the other hand, expected constraints on equilateral non-Gaussianity from measurements of the galaxy bispectrum in spectroscopic surveys are about $\dfnle\sim 25$~\citep{SefusattiKomatsu2007}.

Before concluding this Section, we note here that two distinct definitions of $\fnl$ are present in the literature, corresponding to a CMB convention and a LSS convention. In the CMB convention, for local non-Gaussianity, $\fnll$ is defined by Eq.~(\ref{eqn:phiNG}) with the curvature perturbations $\Phi$ evaluated at early times during matter domination, when their value its constant. In the LSS convention, one usually assumes $\Phi$ to be the value linearly extrapolated at present time, and therefore includes the late-time effect of the accelerated expansion in a $\Lambda$CDM cosmology. The two conventions are simply related by $\fnl^{\rm LSS}=[g(z=\infty)/g(0)]\fnl^{\rm CMB}$ where $g(z)$ is the suppression factor defined as $g(z)=D(z)(1+z)$ with $D(z)$ being the linear growth function of density perturbations. In our numerical analysis of the CMB bispectrum in Section~\ref{sec:cmb}, as well as for the LSS analysis of Section~\ref{sec:lss}, we assume a flat $\Lambda$CDM cosmology and adopt the WMAP 5-yr cosmological parameter values: $\Omega_b=0.044, \Omega_c=0.214, H_0=71.9, n_s=0.96, n_t=0, \tau=0.087$. For this choice one finds, for instance, $\fnl^{\rm LSS}=1.33\fnl^{\rm CMB}$. Since among our results are direct comparisons of CMB and LSS constraints, we will consistently assume the CMB convention for $\fnl$ throughout this paper.

%%%%%%%%%%%%%%%%%%%%%%%%%%%%%%%%%%%%%%%%%%%%%%%%%%%%%%%%%%%%%%%%%%%%%%%%%%%%%%%%
\subsection {Scale-dependence of the primordial bispectrum}
\label{sec:scale}

As implicitly assumed above, both for the local and equilateral type of non-Gaussianity, current observations only constrain the magnitude of $\fnl$. However if $\fnl$ is large enough, it may be also possible in the near future to constrain its possible dependence on scale. There are well-motivated models of inflation, such as single-field models with a variable speed of sound, which naturally predict non-Gaussianity of the equilateral type with a {\it scale-dependence} of $\fnl$~\cite{Chen2005,ChenEtal2007,KhouryPiazza2008}. More recently, a model with local non-Gaussianity has been shown to also present a significant running of the amplitude of the primordial bispectrum, \citep{ByrnesChoiHall2008B}. It can be well expected that for this kind of models, a measurement of, or simply a constraint on, this scale-dependence would provide additional limits on the fundamental parameters of the underlying high-energy theories.

We remark, moreover, that allowing for a running of the non-Gaussian amplitude does not imply a degradation of the error on the amplitude itself. As we will discuss at length later on, for a given observable and a given form of the initial bispectrum, it is always possible to choose a proper pivot point to define the running parameter in such a way as to minimize, or indeed remove, any degeneracy between the two parameters.

We consider a simple ansatz describing a mild scale-dependence of $\fnl$. Specifically, we parametrize the initial curvature bispectrum for both local and equilateral models as
\be
\label{eq:bisp_running_fnl}
B_\Phi(k_1,k_2,k_3)=\fnl\left( \frac{K}{\kp} \right)^{\nng}F(k_1,k_2,k_3),
\ee
corresponding to the replacement
\be\label{eq:runningfnl}
\fnl\rightarrow \fnl\left( \frac{K}{k_p} \right)^{\nng},
\ee
where we define 
\be
\label{eq:Kgeom}
K \equiv (k_1k_2k_3)^{1/3},
\ee
and $k_p$ is a pivot point. The primordial bispectrum is therefore determined in terms of two parameters: the amplitude $\fnl$ and the new parameter $\nng$ quantifying its running. 

We should point out that our definition departs from previous similar definitions introduced in the literature in two aspects. In the first, the overall scale $K$ corresponding to the triangle formed by the wavenumbers $k_1$, $k_2$ and $k_3$, is given here by the {\it geometric mean} of the triplet characterizing the bispectrum configurations. Other works, {\it e.g.}~\citep{LoVerdeEtal2008,TaruyaKoyamaMatsubara2008}, assumed the {\it arithmetic mean}, $K=(k_1+k_2+k_3)/3$, which coincide with our definition only for equilateral configurations. This second definition in terms of a sum is indeed a more accurate and correct description of the bispectrum predicted by DBI models of inflation. Our choice is dictated by the much simpler numerical implementation of the CMB estimator that takes advantage of the separability of the geometrical mean in Eq.~(\ref{eq:Kgeom})\footnote{For the Fisher matrix of large-scale structure observables, both forms can be equally considered and we will show that they both lead to similar outcomes in the equilateral case.}. While we clearly recognize the importance of a full implementation of the proper form of bispectrum, which will be addressed in future work, we argue that we do not expect significant differences in our main results on equilateral non-Gaussianity due to this choice. We notice in the first place that the two definitions differ mostly for squeezed triangular configurations, while most of the signal is concentrated in equilateral ones, for the equilateral type of non-Gaussianity. We can quantify such difference, following \citet{BabichCreminelliZaldarriaga2004}, by introducing a generalized scalar product for the bispectra $B_i$ and $B_j$ given by
\be
B_i\cdot B_j\equiv \Sigma_{k_1, k_2, k_3} \frac{B_i(k_1,k_2,k_3)B_j(k_1,k_2,k_3)}{\Delta^2B(k_1,k_2,k_3)},
\ee
where the sum runs over all triangular configurations given a specific volume and where $\Delta^2B$ represents the Gaussian variance of the bispectrum given by
\be
\Delta^2B(k_1,k_2,k_3)=\frac{1}{N_T}P(k_1)P(k_3)P(k_3),
\ee
with $N_T(k_1,k_2,k_3)$ being the number of fundamental triangular configurations in a given volume corresponding to the triplet $k_1$, $k_2$ and $k_3$. Notice that the variance $\Delta^2B$ does not depend on the form of the bispectrum. In terms of the scalar product one can define the ``cosine''
\be
\cos(B_i,B_j)\equiv\frac{B_i\cdot B_j}{\sqrt{(B_i\cdot B_i)(B_j\cdot B_j)}},
\ee 
which can be interpreted as a measure of the correlation between the two forms. 
We assume a minimum value for $k$ given by the fundamental frequency of a box the size of the Hubble volume, and a maximum value $k_{\rm max}=1\kMpc$. We compute the cosine between two bispectra both of the equilateral form, $F_{\rm eq.}$ and with a very large scale-dependence corresponding to $\nng=1$, but described respectively by  the geometric and arithmetic mean in the definition of $K$. We find that the cosine between the two is larger than $0.99$, indicating a strong correlation. In the case of the local form $F_{\rm loc.}$, the correlation we find between the two definitions of $K$ is the smaller value of about $0.91$, a result expected since most of the signal comes from squeezed triangles where the difference between geometric and arithmetic mean of the wavenumbers triplet is larger. In the local case, however, no explicit expression for a bispectrum with a strongly scale-dependent amplitude is given in the literature, so we assume our choice for $K$ to be a simple phenomenological ansatz, subject to improvements in future works.  

The second difference in our notation with respect to the literature regards the running parameter $\nng$. The notation $\nng$ was been first introduced by \citet{Chen2005} and then adopted by \citet{ChenEtal2007,LoVerdeEtal2008,KhouryPiazza2008} and also as $n_{\fnl}$ in \citet{ByrnesChoiHall2008B}, to denote a running of the parameter $\fnl$ and it has been defined in analogy to the power spectrum spectral index as $\nng-1\equiv\partial \ln|\fnl(k)|/\partial \ln k$, where the scale-dependent $\fnl(k)$ could be defined, from Eq.~(\ref{eq:bisp_running_fnl}) for equilateral configurations as $\fnl(k)\equiv B_\Phi(k,k,k)/F(k,k,k)$.  In the case of the spectral index, scale-invariance provides unity as the expected value for $n_s$, in the case of a running non-Gaussian parameter there is no theoretical argument supporting an expected value for $\nng$ of one. In our definition a constant $\fnl$ corresponds to $\nng=0$ so that we can define, for equilateral configurations,
\be
\label{eq:nNGdef}
\nng\equiv\frac{\partial \ln|\fnl(k)|}{\partial \ln k}.
\ee
We suggest this as a more natural definition to be adopted by future works on the subject. \citet{LoVerdeEtal2008} make use of the parameter $\kappa$, the dimensionless time-variation of the speed of sound of inflaton perturbations\footnote{We denote this quantity with the symbol $s$ in Section~\ref{sec:dbi}.}, related to our $\nng$ by the expression $\nng=-2\kappa$.

Finally, we choose as the value for the pivot point $k_p=0.04$ Mpc$^{-1}$, already adopted by \citep{LoVerdeEtal2008}. Such choice is arbitrary but determines, given a specific probe and for a given model, the degree of degeneracy and correlation between the two non-Gaussian parameters. This choice indeed minimizes the degeneracy for the parameters of the equilateral model as determined by the bispectrum measurement in an ideal CMB experiment. We discuss the dependence of our results on the choice of $k_p$ in detail in Appendix~\ref{app:pivot}.

%%%%%%%%%%%%%%%%%%%%%%%%%%%%%%%%%%%%%%%%%%%%%%%%%%%%%%%%%%%%%%%%%%%%%%%%%%%%%%%%
%%%%%%%%%%%%%%%%%%%%%%%%%%%%%%%%%%%%%%%%%%%%%%%%%%%%%%%%%%%%%%%%%%%%%%%%%%%%%%%%
\section {Primordial non-Gaussianity in the Cosmic Microwave Background}
\label{sec:cmb}

In this Section we describe the relation between the observed bispectrum of anisotropies in the CMB and the primordial curvature bispectrum, and we define the Fisher matrix for the two non-Gaussian parameters $\fnl$ and $\nng$. We then show the expected results corresponding to the WMAP, Planck and CMBPol experiments together with those from the ideal CMB experiment.

%%%%%%%%%%%%%%%%%%%%%%%%%%%%%%%%%%%%%%%%%%%%%%%%%%%%%%%%%%%%%%%%%%%%%%%%%%%%%%%%
\subsection {The Cosmic Microwave Background Bispectrum}

The harmonic coefficients of the CMB anisotropy $a_{lm}=T^{-1}\int d^2{\mathbf {\hat n}}\Delta T({\mathbf {\hat {\bf n}}}) Y^{*}_{\ell m}$ can
be related to the primordial fluctuation $\Phi$ as\footnote{Note that
  the formulae in this Section are written in the opposite Fourier
  Transform (FT) convention than the one used for formula \ref{eq:prim_bi},
  and that will be used in all the other Sections of this paper. The
  change of convention for this Section was adopted in order to be
  consistent with previous literature in the non-Gaussian CMB field.}
\begin{eqnarray} 
\label{phi_alm}
a_{\ell    m}^p=b_{\ell}\,4\pi (-i)^\ell\int  \frac{d^3k}{(2 \pi)^3} \Phi(\mathbf{k})   \,    g^p_{\ell}(k)   Y^*_{\ell m}(\hat \kv)+ n_{\ell m},
\end{eqnarray} 
where  $\Phi(\kv)$  is the primordial curvature perturbations, for a
comoving wavevector $\mathbf  {k}$, $g^p_{\ell}(r)$ is the radiation
transfer  function where the index $x$ refers to either temperature
($T$) or {\it E}-polarization ($E$) of the CMB. A beam function
$b_{\ell}$ and the harmonic coefficient of noise $n_{\ell  m}$ are
instrumental effects. Eq.~(\ref{phi_alm}) is written for a flat
background, but can easily be generalized. 

Any non-Gaussianity present in the primordial perturbations $\Phi(\mathbf{k})$ gets transferred to the observed CMB via Eq.~(\ref{phi_alm}). The most common way to look for non-Gaussianity in the CMB is to study the three-point function of temperature and polarization anisotropies in harmonic space. Such quantity is called the CMB angular bispectrum and is defined as 
\be
B^{pqr}_{\ell_1 \ell_2 \ell_3, m_1m_2m_3}\equiv \langle a^p_{\ell_1 m_1}a^q_{\ell_2 m_2}a^r_{\ell_3 m_3} \rangle\, ,
\ee   
and the angular-averaged bispectrum is 
\be
\bi^{pqr}=\sum_{m_1m_2m_3}  \left(\begin{array}{ccc} \ell_1 & \ell_2 & \ell_3 
\\ m_1 & m_2 & m_3 \end{array}\right)   B^{pqr}_{\ell_1 \ell_2 \ell_3, m_1m_2m_3}\, ,
\ee
 which using Eq.(\ref{phi_alm}) can be written as:
 \begin{eqnarray}
 \label{eq:cmb_bi}
 \bi^{pqr} & = & (4\pi)^3 (-i)^{\ell_1+\ell_2+\ell_3} \sum_{m_1m_2m_3} \left(\begin{array}{ccc} \ell_1 & \ell_2 & \ell_3 \\ m_1 & m_2 & m_3 \end{array} \right)  \int \frac{d^3k_1}{(2 \pi)^3}\frac{d^3k_2}{(2 \pi)^3}\frac{d^3k_3}{(2 \pi)^3} \; Y^*_{\ell_1 m_1}(\hat{\kv}_1) Y^*_{\ell_2 m_2}(\hat{\kv}_2)Y^*_{\ell_3 m_3}(\hat{\kv}_3) 
 \nonumber \\ & & 
 \times g^p_{\ell_1}(k_1)
 g^q_{\ell_2}(k_2) g^r_{\ell_3}(k_3) \; \langle\Phi(\mathbf{k}_1)\Phi(\mathbf{k}_2)\Phi(\mathbf{k}_3) \rangle \;,
 \end{eqnarray}
where $\langle\Phi(\mathbf{k}_1)\Phi(\mathbf{k}_2)\Phi(\mathbf{k}_3) \rangle$ is the primordial curvature three-point function as defined in Eq. (\ref{eq:prim_bi}). 

To forecast constraints on scale-dependent non-Gaussianity using CMB data, we will perform a Fisher matrix analysis. We will consider some non-zero $\fnl$ as our fiducial value for the Fisher matrix evaluation. Clearly, in order to be able to constrain a scale-dependence of $\fnl$, its amplitude must be large enough to produce a detection. If $\fnl$ is too small to be detected ($\fnl< 2$ is a lowest theoretical limit even for the ideal experiment), we will obviously not be able to measure any of its features, either. In the following we will then always consider a fiducial value of $\fnl$ large enough to enable a detection. Following ~\cite{KomatsuSpergel2001,BabichZaldarriaga2004,YadavEtal2008}, the Fisher matrix for the parameters $\fnl$ and $\nng$  (for generic fiducial values $\fnl\ne0$ and $\nng\ne0$) can be written as: 
\be
{\mathcal F}_{ab} = 
\sum_{\left\{i\!jk,~pqr\right\}}\sum_{\ell_1\le\ell_2\le\ell_3}
 \frac{1}{\Delta_{\ellt}} 
 \frac{\partial B^{pqr}_{\ellt}}{\partial p_a}
 \left({\bf Cov}^{-1}\right)_{i\!jk,~pqr}
 \frac{\partial B^{\,i\!jk}_{\ellt}}{\partial p_b}.
\ee
The indices $a$ and $b$ can take the values 1 and 2 corresponding to the Fisher matrix parameters $p_1\equiv\fnl$ and $p_2\equiv\nng$, respectively.  Indices $ijk$  and $pqr$ run over all the eight possible ordered combinations of temperature and polarization given by $TTT$, $TTE$, $TET$, $ETT$, $TEE$, $ETE$, $EET$ and $EEE$; the combinatorial factor $\Delta_{\ell_1  \ell_2 \ell_3}$ equals $1$ when all $\ell$'s are different, $6$ when $\ell_1 =  \ell_2 = \ell_3$, and $2$ otherwise. The covariance matrix {\bf Cov} is obtained in terms of $C_\ell^{TT}$, $C_\ell^{EE}$,  and $C_\ell^{TE}$  (see~\citep{BabichZaldarriaga2004,YadavKomatsuWandelt2007}) by applying Wick's theorem.

For non-Gaussianity of the equilateral type, for which the functional form $F(k_1,k_2,k_3)$ is given by Eq.~(\ref{eq:f_eq}), we have
\bea
\frac{\partial B^{~ijk}_{\ellt}}{\partial\fnl} & = & \sqrt{\frac{(2 \ell_1+1) (2 \ell_2+1)(2 \ell_3+1)}{4 \pi}} \left(\begin{array}{ccc} \ell_1 & \ell_2 & \ell_3 
\\ 0 & 0 & 0 \end{array}\right)  
\nonumber\\ 
& & \times ~6 \int\!\!r^2 dr \left[ -\alpha^i_{\ell_1} \beta^j_{\ell_2} \beta^k_{\ell_3} 
+ 2 \;{\rm perm.}
+ \beta^i_{\ell_1} \gamma^j_{\ell_2} \delta^k_{\ell_3} + 5 \;{\rm perm.}
- 2 \delta^i_{\ell_1} \delta^j_{\ell_2} \delta^k_{\ell_3} 
\right],  
\eea
where the functions $\alpha, \beta, \delta,$ and $\gamma$ are given by:

\begin{eqnarray}
\alpha^i_\ell(r) & \equiv & \frac2\pi \int \!\!\! dk \; k^2 \, g^i_\ell(k) ~j_\ell(k r)\left(\frac{k}{k_p}\right)^{\nng/3}, 
\\
\beta^i_\ell(r) & \equiv & \frac2\pi \int \!\!\! dk \; k^{-1} \, g^i_\ell(k) ~j_\ell(k r)\left(\frac{k}{k_p}\right)^{\nng/3}\Delta_\Phi\, k^{n_s-1}, 
\\ 
\gamma^i_\ell(r) & \equiv & \frac2\pi \int \!\!\! dk \; k \, g^i_\ell(k) ~j_\ell(k r)\left(\frac{k}{k_p}\right)^{\nng/3}\Delta_\Phi^{1/3}\,k^{(n_s-1)/3},
\\
\delta^i_\ell(r) & \equiv & \frac2\pi \int \!\!\! dk  \, g^T_\ell(k) ~j_\ell(k r)\left(\frac{k}{k_p}\right)^{\nng/3}\Delta_\Phi^{2/3}\,k^{2(n_s-1)/3}.
\end{eqnarray}

In the expression above we use the dimensionless power spectrum amplitude $\Delta_\Phi$, which is defined by $P_\Phi(k)=\Delta_\Phi k^{-3+(n_s-1)}$, 
where $n_s$ is the tilt of the primordial power spectrum.

The derivative with respect to $\nng$ produces an extra $\ln(k)$ factor in the integrals over the CMB transfer function. The explicit form is given by
\bea
\frac{\partial B^{~ijk}_{\ellt}}{\partial\nng} 
&  = & 
\sqrt{\frac{(2 \ell_1+1) (2 \ell_2+1)(2 \ell_3+1)}{4 \pi}} \left(\begin{array}{ccc} \ell_1 & \ell_2 & \ell_3 
\\ 0 & 0 & 0 \end{array}\right)   
\nonumber\\
 & & 
 \times~6\fnl^{eq.} \int_0^\infty \!\!r^2 dr \left[ -\tilde \alpha^i_{\ell_1} \beta^j_{\ell_2} \beta^k_{\ell_3} 
 -\alpha^i_{\ell_1} \tilde \beta^j_{\ell_2} \beta^k_{\ell_3} 
 -\alpha^i_{\ell_1} \beta^j_{\ell_2} \tilde \beta^k_{\ell_3} 
+ {2 \;{\rm perm.}} \right.
\nonumber \\ 
& & \left.
    +\tilde \beta^i_{\ell_1} \gamma^j_{\ell_2} \delta^k_{\ell_3} +  \beta^i_{\ell_1}
     \tilde \gamma^j_{\ell_2} \delta^k_{\ell_3} +\beta^i_{\ell_1} \gamma^j_{\ell_2} 
     \tilde \delta^k_{\ell_3} + 5 \;{\rm perm.}
         - 2 \tilde \delta^i_{\ell_1} \delta^j_{\ell_2} \delta^k_{\ell_3}
    - 2 \delta^i_{\ell_1} \tilde \delta^j_{\ell_2} \delta^k_{\ell_3}
    - 2 \delta^i_{\ell_1} \delta^j_{\ell_2} \tilde \delta^k_{\ell_3} 
\right],  
\eea
comprising a total of 30 additive contributions, where the functions $\tilde \alpha, \tilde \beta, \tilde \delta,$ and $\tilde \gamma$ are given by:

\begin{eqnarray}
\tilde \alpha^i_\ell(r) & \equiv &  \frac{2}{3\pi} \int_0^{+ \infty} \!\!\! dk \; k^2 \, g^i_\ell(k) ~j_\ell(k r) \left(\frac{k}{k_p}\right)^{\nng/3}\ln\left(\frac{k}{k_p}\right)\\
\tilde \beta^i_\ell(r) & \equiv & \frac{2}{3\pi} \int_0^{+ \infty} \!\!\! dk \; k^{-1} \, g^i_\ell(k) ~j_\ell(k r)\left(\frac{k}{k_p}\right)^{\nng/3}\ln\left(\frac{k}{k_p}\right)\Delta_\Phi\,k^{n_s-1} \\ 
\tilde \gamma^i_\ell(r) & \equiv & \frac{2}{3\pi} \int_0^{+ \infty} \!\!\! dk \; k \, g^i_\ell(k)~ j_\ell(k r)\left(\frac{k}{k_p}\right)^{\nng/3} \ln\left(\frac{k}{k_p}\right)\Delta_\Phi^{1/3}\,k^{(n_s-1)/3}\\
\tilde \delta^i_\ell(r) & \equiv & \frac{2}{3\pi} \int_0^{+ \infty} \!\!\! dk  \, g^T_\ell(k)~ j_\ell(k r)\left(\frac{k}{k_p}\right)^{\nng/3}\ln\left(\frac{k}{k_p}\right)\Delta_\Phi^{2/3}\,k^{2(n_s-1)/3}.
\end{eqnarray}

In a similar way, from Eq.~(\ref{eq:f_local}), one can derive the following expressions for the bispectrum derivatives in the local case,
\be
\frac{\partial B^{~ijk}_{\ellt}}{\partial\fnl}  =  \sqrt{\frac{(2 \ell_1+1) (2 \ell_2+1)(2 \ell_3+1)}{4 \pi}} \left(\begin{array}{ccc} \ell_1 & \ell_2 & \ell_3 
\\ 0 & 0 & 0 \end{array}\right)~ 2 \int_0^\infty \!\!r^2 dr \left[ -\alpha^i_{\ell_1} \beta^j_{\ell_2} \beta^k_{\ell_3} 
+ 2 \;{\rm perm.}\right], 
\ee
and 
\bea
\frac{\partial B^{~ijk}_{\ellt}}{\partial\nng}
& = & 
\sqrt{\frac{(2 \ell_1+1) (2 \ell_2+1)(2 \ell_3+1)}{4 \pi}} 
\left(\begin{array}{ccc} \ell_1 & \ell_2 & \ell_3 \\ 0 & 0 & 0 \end{array}\right) 
\nonumber\\
& & \times~
2 \fnll \int_0^\infty \!\!r^2 dr 
\left[ \tilde \alpha^i_{\ell_1} \beta^j_{\ell_2} \beta^k_{\ell_3} 
+ \alpha^i_{\ell_1} \tilde \beta^j_{\ell_2} \beta^k_{\ell_3} + \alpha^i_{\ell_1} \beta^j_{\ell_2} \tilde \beta^k_{\ell_3} +  {2 \;{\rm perm.}}\right].  
\eea

We will not consider a marginalization over cosmological parameters. For a discussion of the effect of uncertainties on the cosmological parameters on the determination of the non-Gaussian amplitude parameter $\fnl$ from CMB measurements, see \citep{LiguoriRiotto2008}. We modified the publicly available CMBfast code \citep{SeljakZaldarriaga1996} to compute the transfer functions $g^T_{\ell}$ and $g^E_{\ell}$.

%%%%%%%%%%%%%%%%%%%%%%%%%%%%%%%%%%%%%%%%%%%%%%%%%%%%%%%%%%%%%%%%%%%%%%%%%%%%%%%%
\subsection{Constraints from the CMB Bispectrum}

We compute here the expected uncertainties on the two non-Gaussian parameters $\fnl$ and $\nng$ from the Fisher matrix analysis of the CMB bispectrum assuming as fiducial values $\fnl=50$ and $\nng=0$ for the local model and $\fnl=100$ and $\nng=0$ for the equilateral one. We discuss in detail the dependence of these results on the choice of the fiducial values in Section~\ref{ssec:fidval}. We consider the specifications for the current Wilkinson Microwave Anisotropy Probe \citep[WMAP,][]{BennettEtal2003} and Planck \citep{Planck2006} missions, the proposed satellite mission CMBPol~\citep{BaumannEtal2008A,BaumannEtal2008B} and an ideal CMB experiment, assuming a full-sky coverage. We consider the CMB bispectrum up to $\ell_{\rm max}=1200$ for WMAP, $\ell_{\rm max}=2500$ for Planck and $\ell_{\rm max}=3000$ for CMBPol and the ideal case.

\begin{figure}[!t]
\begin{center}
{\includegraphics[width=0.48\textwidth]{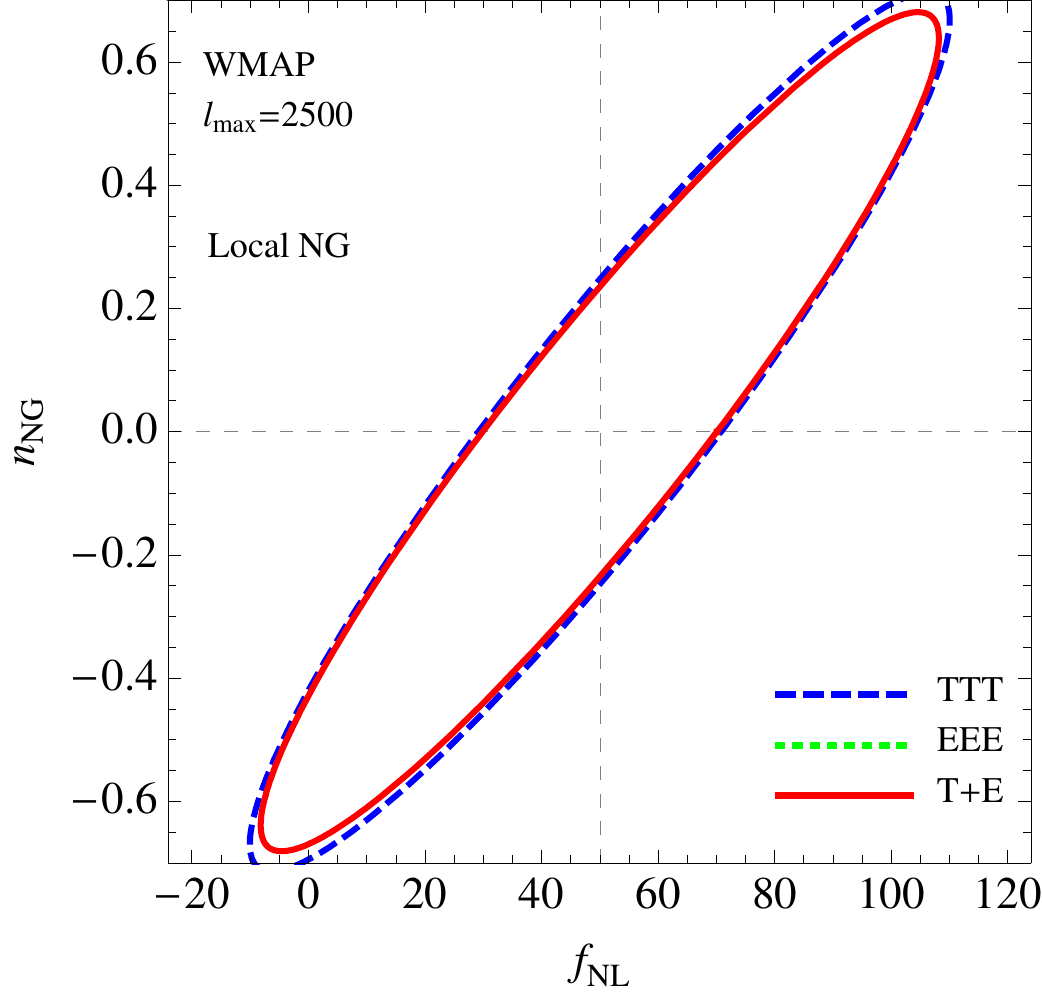}}
{\includegraphics[width=0.48\textwidth]{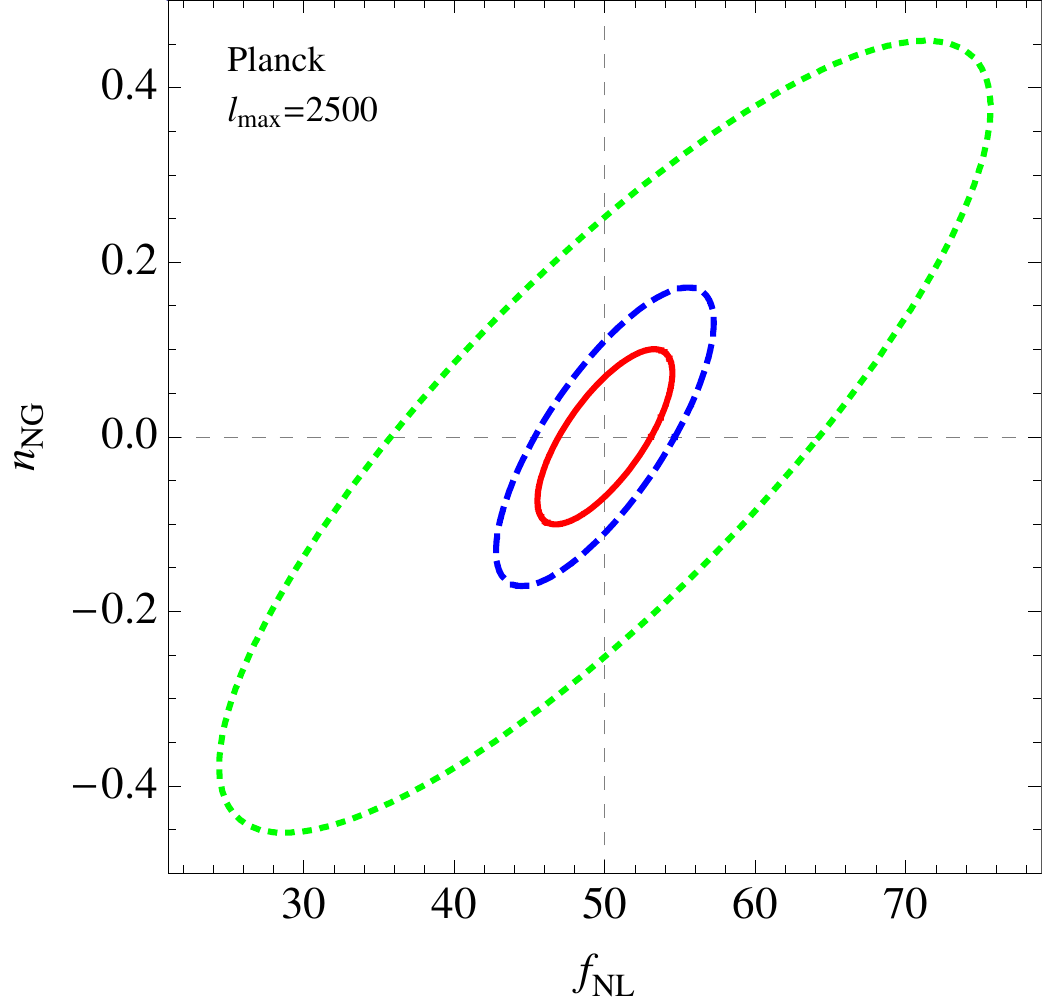}}
{\includegraphics[width=0.48\textwidth]{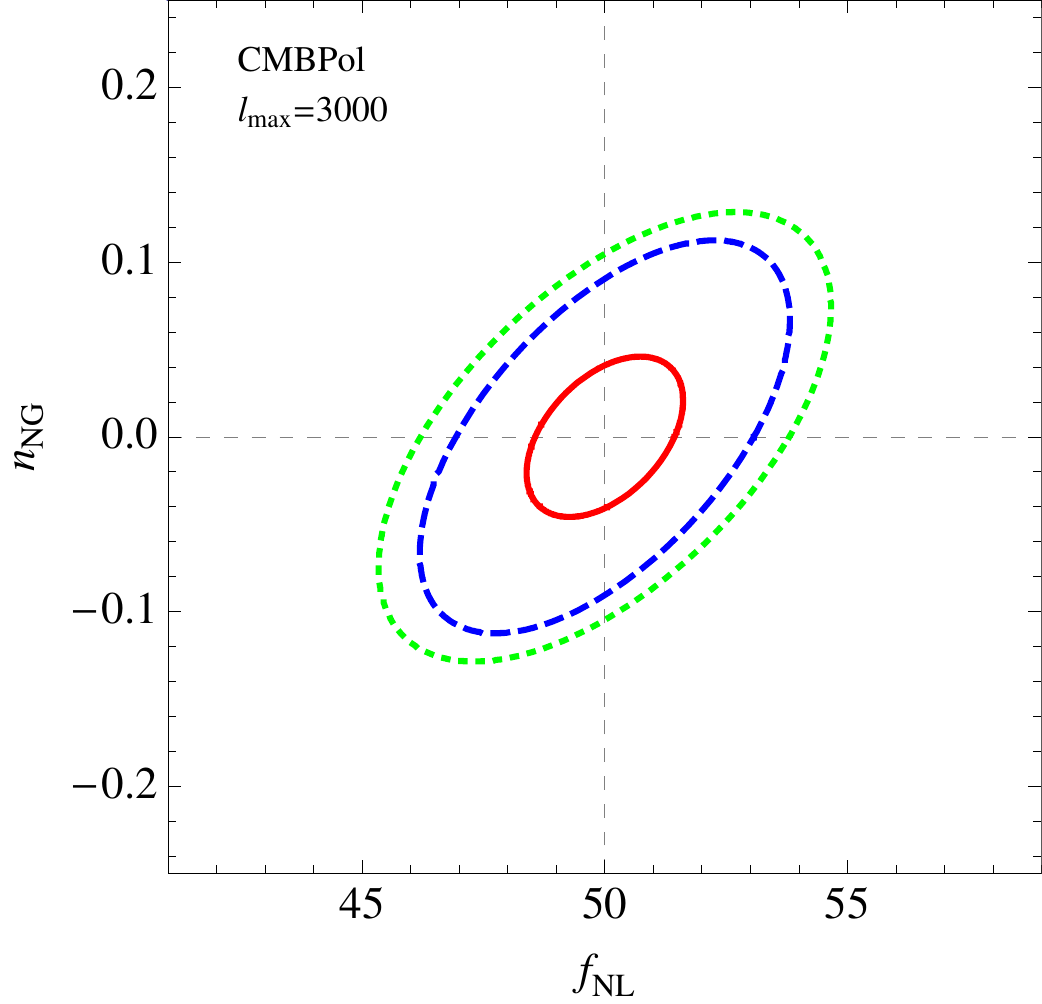}}
{\includegraphics[width=0.48\textwidth]{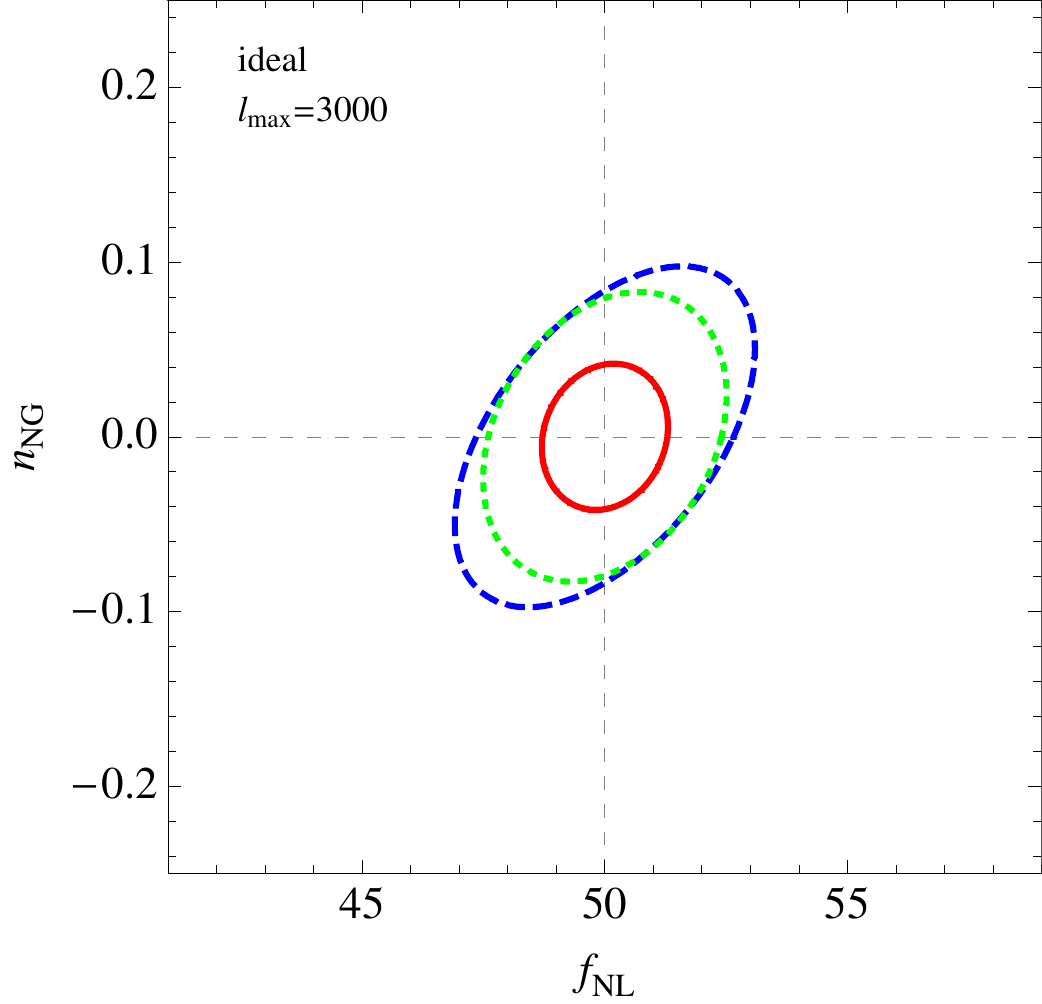}}
\caption{Local model. 1-$\sigma$ constraints on $\fnl$ and $\nng$ assuming $k_p=0.04$ Mpc$^{-1}$ and fiducial values $\fnl=50$, $\nng=0$. Dashed lines correspond to the limits from the temperature information alone, dotted lines to polarization (EEE), while the continuous lines correspond to all bispectrum combinations. We consider WMAP ({\it upper left panel}), Planck ({\it upper right}), CMBPol ({\it bottom left}) and an ideal CMB experiment ({\it bottom right}).}
\label{fig:cmbLc}
\end{center}
\end{figure}

\begin{figure}[!ht]
\begin{center}
{\includegraphics[width=0.48\textwidth]{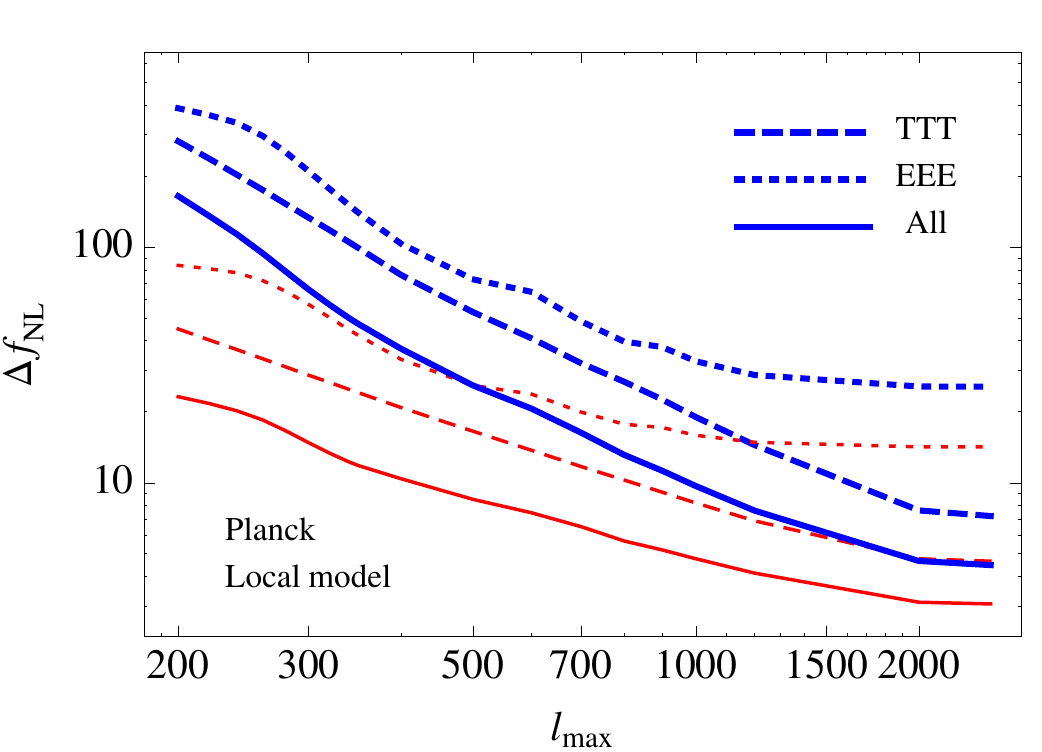}}
{\includegraphics[width=0.48\textwidth]{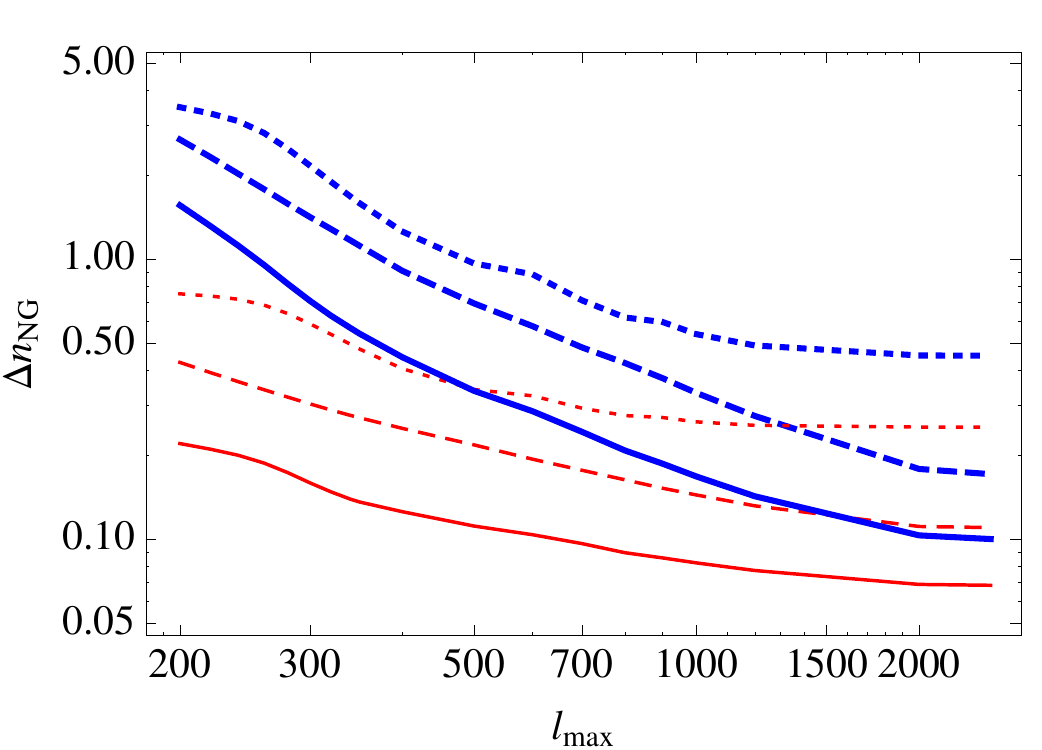}}
{\includegraphics[width=0.48\textwidth]{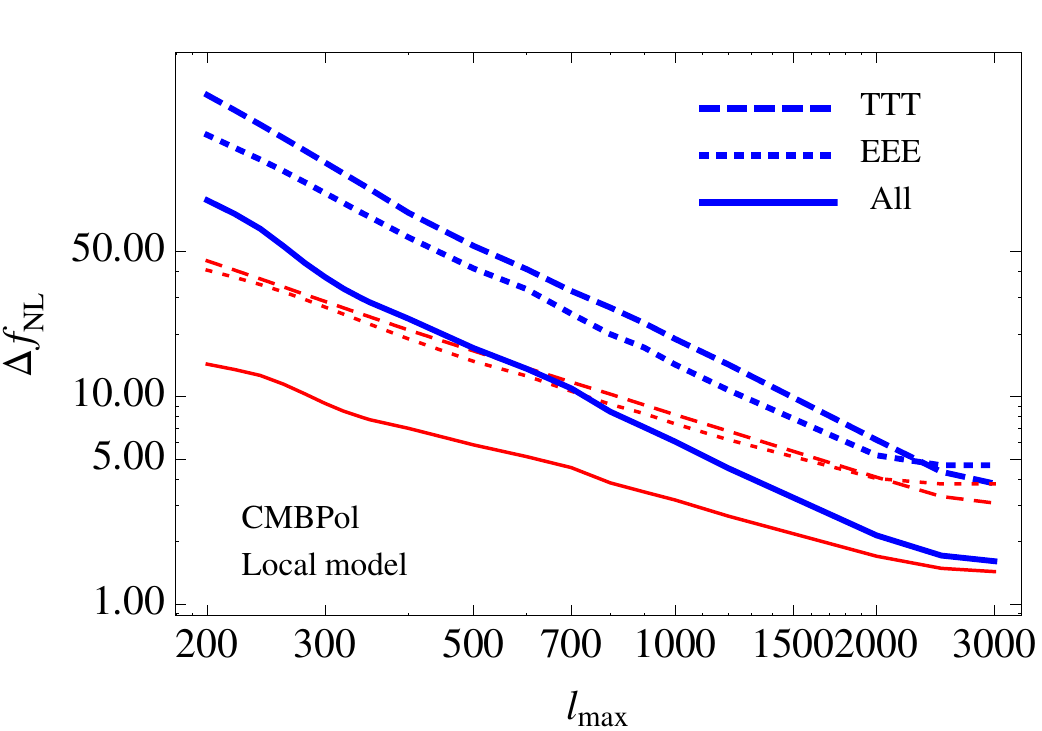}}
{\includegraphics[width=0.48\textwidth]{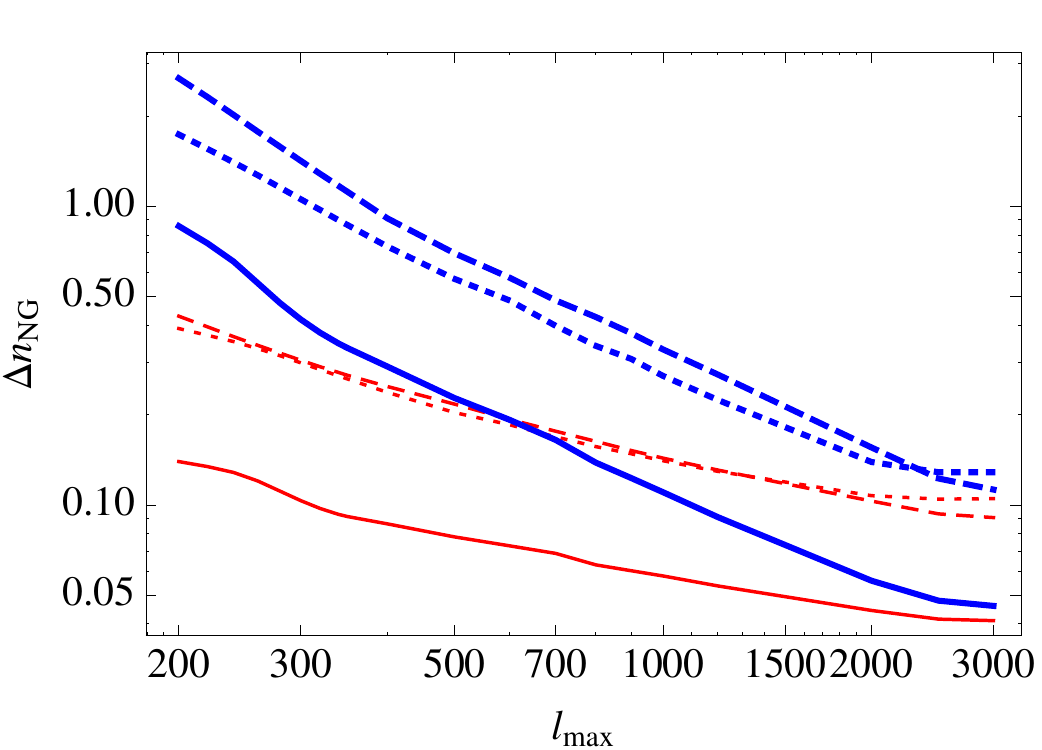}}
\caption{Local model. Expected marginalized ({\it thick, blue lines}) and unmarginalized ({\it thin, red lines}) errors for $\fnl$ ({\it left panels}) and $\nng$ ({\it right panels}) as a function of $l_{max}$ for Planck ({\it central panel}) and CMBPol ({\it lower panels}).}
\label{fig:cmbLclmax}
\end{center}
\end{figure}

In Fig.~\ref{fig:cmbLc} we compare the expected 1-$\sigma$ contours on local model parameters $\fnll$ and $\nng$ for WMAP ({\it upper left panel}), Planck ({\it upper right}), CMBPol ({\it bottom left}) and for an ideal CMB experiment ({\it bottom right}). Dashed blue lines correspond to the limits from the temperature information alone, dotted green lines to polarization (EEE), while the continuous red lines correspond to all bispectrum combinations. The figure assumes for the pivot point $k_p=0.04$ Mpc$^{-1}$, resulting in a significant
degeneracy between the two parameters. As shown in Appendix~\ref{app:pivot}, a better value for the local model CMB bispectrum would have been closer to $k_p=0.02$ Mpc$^{-1}$ for Planck and $k_p=0.03$ Mpc$^{-1}$ for CMBPol, indicating that most of the signal is coming from scales larger than the pivot point. An optimal approach would actually require one to determine the pivot scale not only depending on the shape of non-Gaussianities, but also according to the specific experiment and its multipole range and noise properties. However, it will be shown later that by combining CMB and large-scale structure observables  the degeneracy between $n_{NG}$ and $\fnl$ is largely removed. Moreover, the pivot scale $k_p = 0.04$ was the one adopted in \citep{LoVerdeEtal2008}, and it thus allows an easier comparison between the two results. For these reasons we will always assume $k_p = 0.04$ Mpc$^{-1}$ throughout the rest of this work.
In Fig.~\ref{fig:cmbLclmax} we plot the expected marginalized ({\it thick, blue lines}) and unmarginalized ({\it thin, red lines}) 1-$\sigma$ errors on the individual parameters $\fnl$ ({\it left panel}) and $\nng$ ({\it right panel}) for Planck ({\it upper panel}) and CMBPol ({\it lower panels}) as a function of $l_{max}$. 

We find that WMAP and Planck should be able to provide a 1-$\sigma$ uncertainty on the running of $\fnll$ respectively equal to
\be
\Delta\nng  \simeq  0.68~ \frac{50}{\fnll}~\frac{1}{\sqrt{f_{sky}}}\qquad [{\rm WMAP}]
\label{eq:resCMBlcWMAP},
\ee
and
\be
\Delta\nng \simeq  0.10~ \frac{50}{\fnll}~\frac{1}{\sqrt{f_{sky}}}\qquad [{\rm Planck}]\label{eq:resCMBlcPlanck},
\ee
after marginalizing over $\fnl$, while for CMBPol we find a smaller error by about a factor of two 
\be
\Delta\nng \simeq 0.05~ \frac{50}{\fnll}~\frac{1}{\sqrt{f_{sky}}}\qquad [{\rm CMBPol}]\label{eq:resCMBlcCMBPol}.
\ee
Notice that the marginalized error on $\nng$ does not depend strongly on the choice for the pivot point $\kp$. The {\it un}-marginalized 1-$\sigma$ errors, for a fiducial $\fnll=50$ and $f_{sky}=1$, are, in fact, $\Delta\nng  \simeq  0.24$, $0.07$ and $0.04$ for WMAP, Planck and CMBPol, respectively.

\begin{figure}[!t]
\begin{center}
{\includegraphics[width=0.48\textwidth]{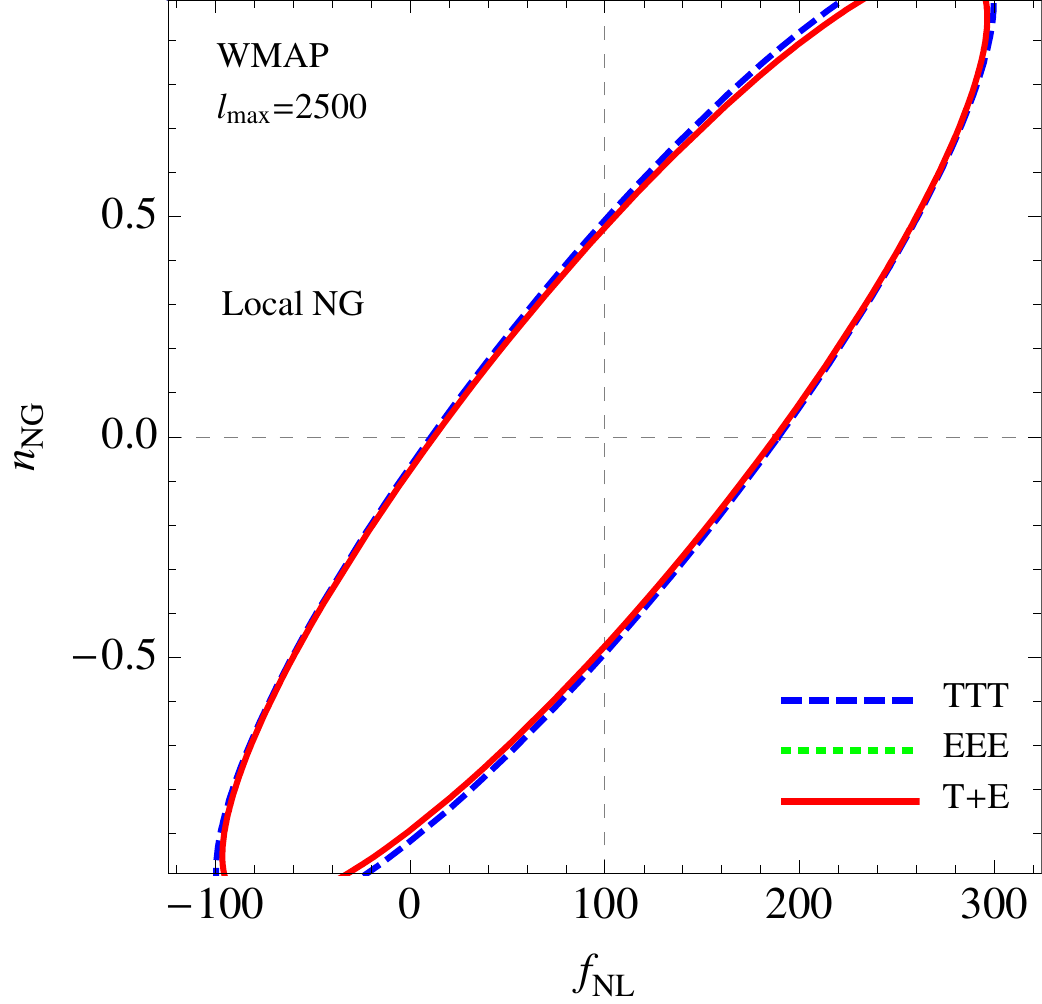}}
{\includegraphics[width=0.48\textwidth]{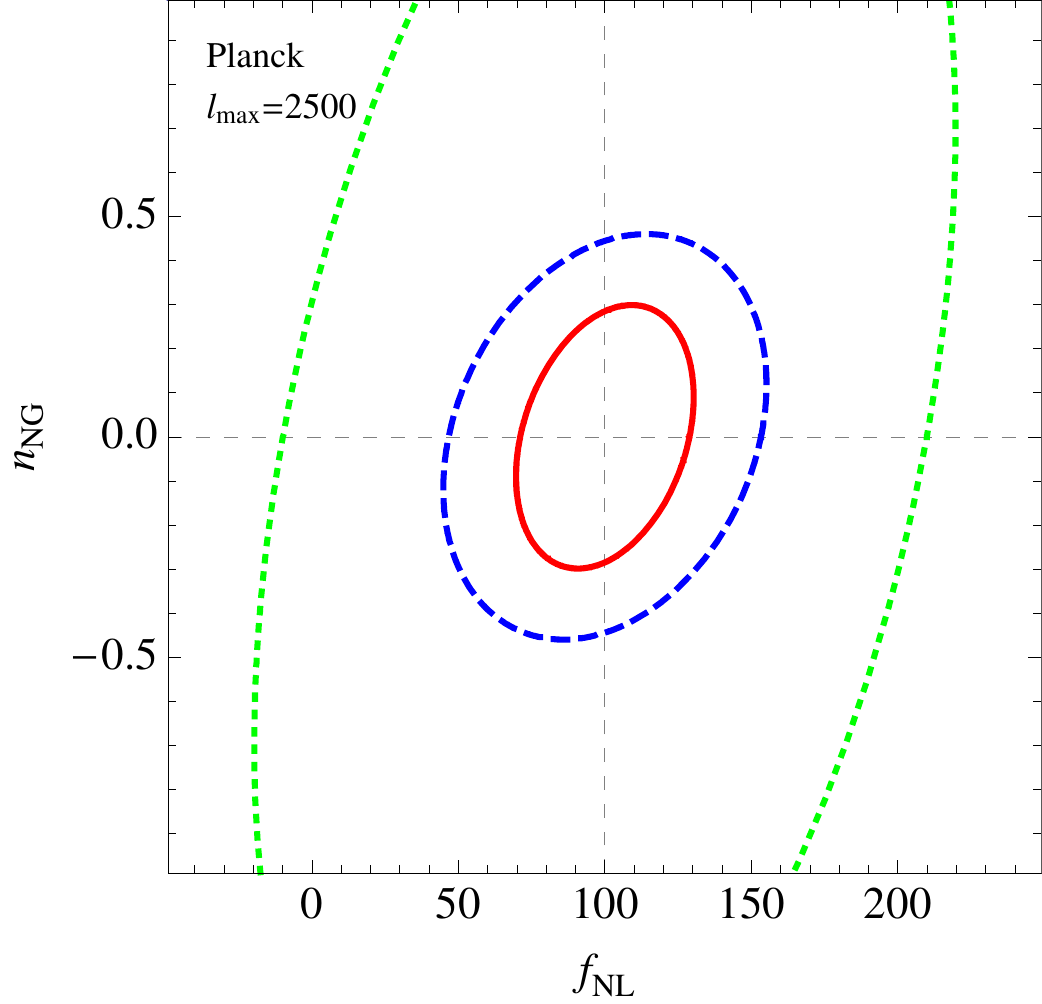}}
{\includegraphics[width=0.48\textwidth]{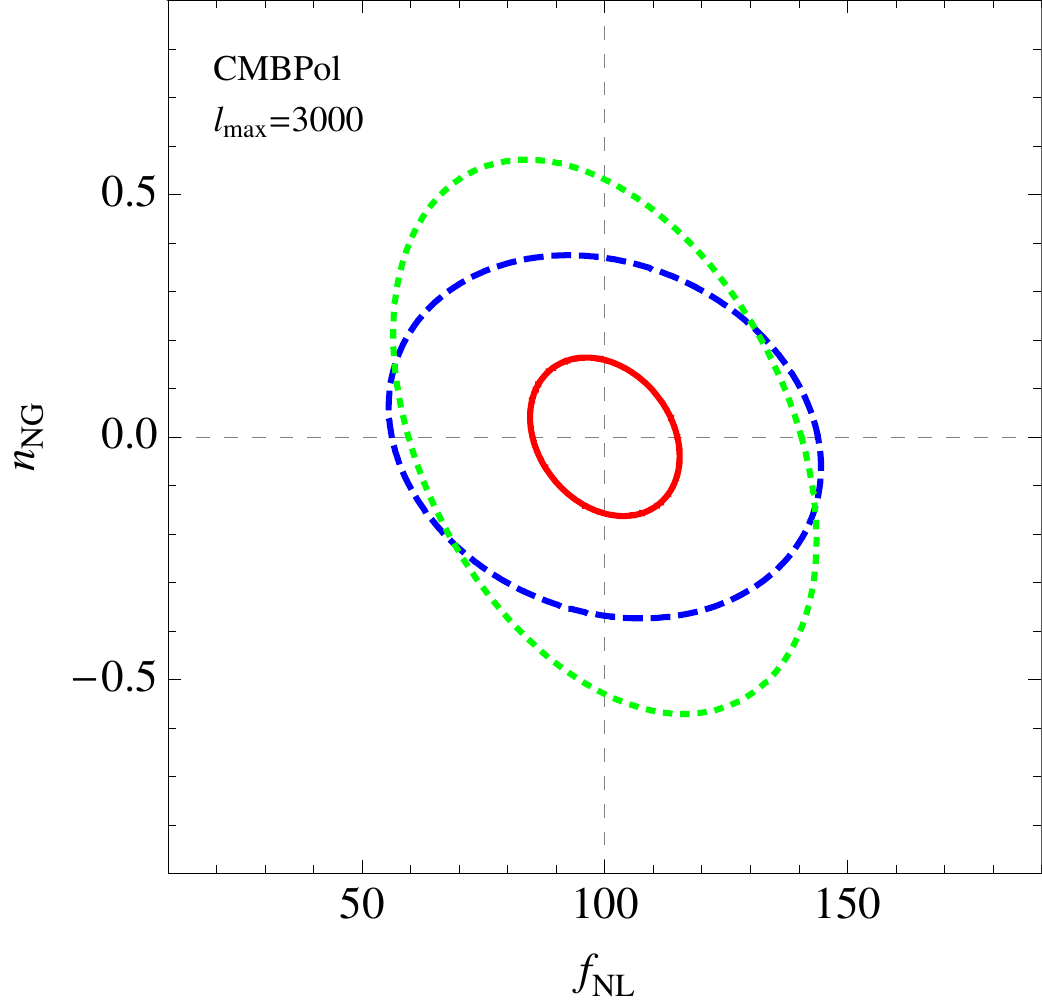}}
{\includegraphics[width=0.48\textwidth]{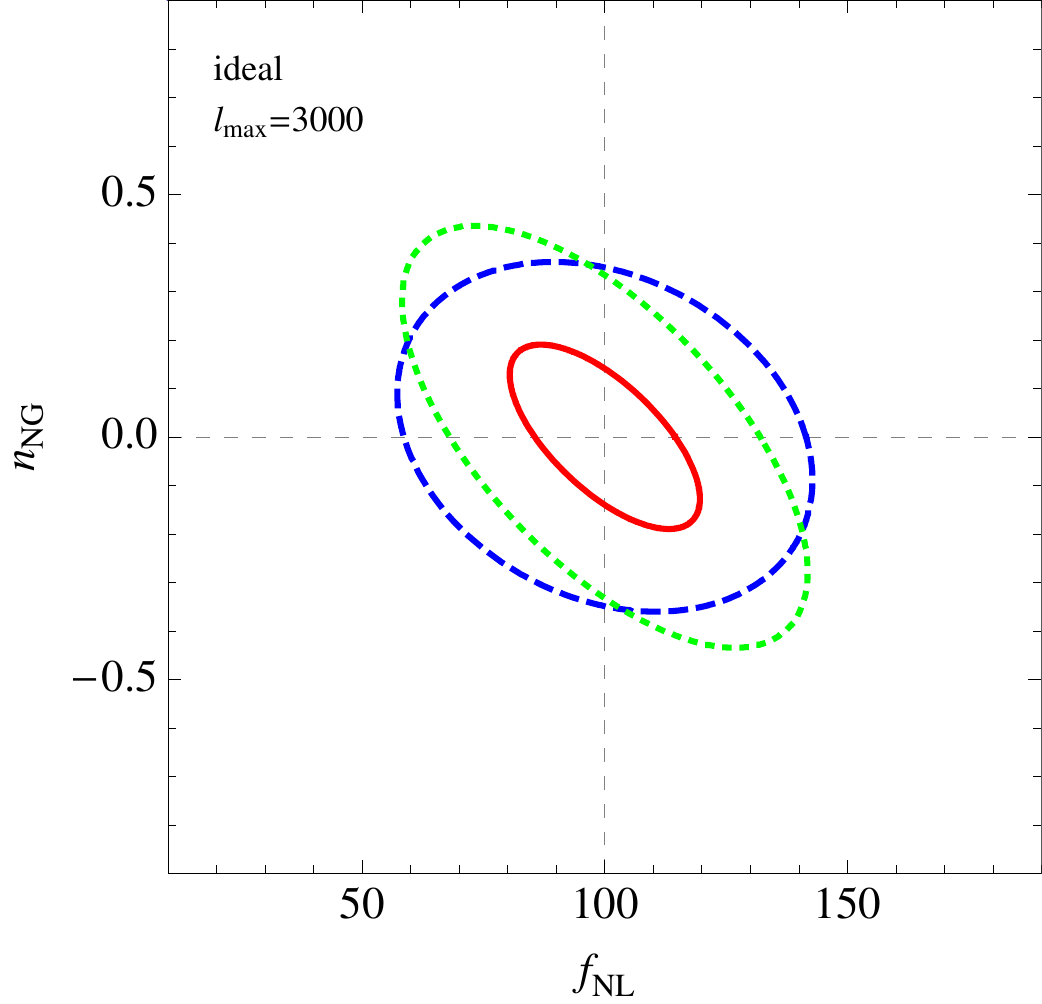}}
\caption{Equilateral model. 1-$\sigma$ constraints on $\fnl$ and $\nng$ assuming $k_p=0.04$ Mpc$^{-1}$ and fiducial values $\fnl=100$, $\nng=0$. Dashed lines correspond to the limits from the temperature information alone, dotted lines to polarization (EEE), while the continuous lines correspond to all bispectrum combinations. We consider WMAP ({\it upper left panel}), Planck ({\it upper right}), CMBPol ({\it bottom left}) and an ideal CMB experiment ({\it bottom right}).}
\label{fig:cmbEq}
\end{center}
\end{figure}
To give a sense of the magnitude of the errors provided by the Fisher matrix analysis, we can consider the change in the non-Gaussianity amplitude $\fnl(k)$ over the range of scales probed by the CMB (which roughly goes from $0.001$ to $0.1\Mpc$), and compare it to the $\fnl$ sensitivity of a given experiment. We roughly expect a departure from $n_{NG}=0$ to be undetectable if the variation it produces in $\fnl(k)$ over the available range of $k$ is below the $\fnl$ uncertainty for the experiment under examination. For example, taking a fiducial value $\fnl=50$ at $k= k_p$, a running of $\nng\simeq 0.1$,
corresponds to the two extremal values $\fnl(k=0.001\kMpc)\simeq 33$
and $\fnl(k=0.1\kMpc)\simeq 53$. At smaller scales, such as those
probed by the cluster abundance, we have $\fnl(k=0.5\kMpc)\simeq 62$. These
variations are well within current uncertainties on the amplitude
parameter. Our previous argument then suggests that $\nng = 0.1$ is too small to be detected by WMAP, in agreement with the full Fisher matrix analysis presented above. For a larger value of the running parameter as $\nng=0.4$, one obtains $\fnl(k=0.001\kMpc)\simeq 10$ and $\fnl(k=0.1\kMpc)\simeq 63$ while for the negative value $\nng=-0.5$, $\fnl(k=0.001\kMpc)\simeq 250$ and $\fnl(k=0.1\kMpc)\simeq 40$. This amount of running starts to be in the detectability range of WMAP, again in agreement with formula of Eq.~(\ref{eq:resCMBlcWMAP}). We notice that the large value obtained for the largest scale under the assumption of a running close to the expected WMAP error on $\nng$, is close to the uncertainty derived in \citep{SmithSenatoreZaldarriaga2009} for the smallest $\ell$-bin in their analysis of the WMAP bispectrum. 

\begin{figure}[!t]
\begin{center}
{\includegraphics[width=0.48\textwidth]{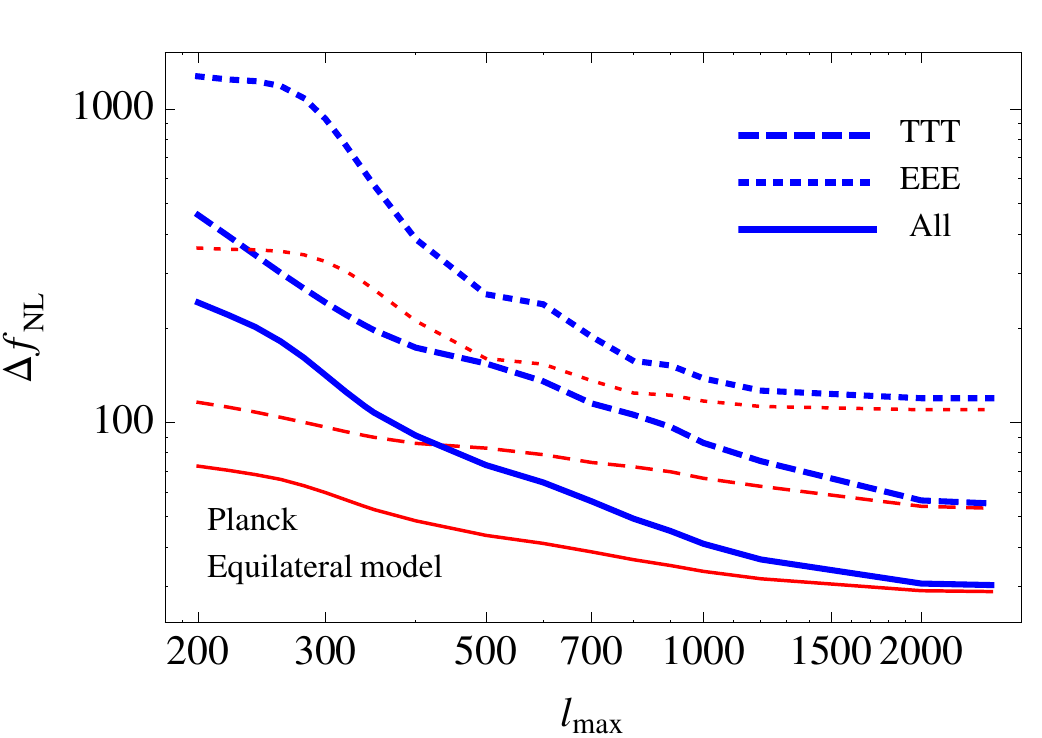}}
{\includegraphics[width=0.48\textwidth]{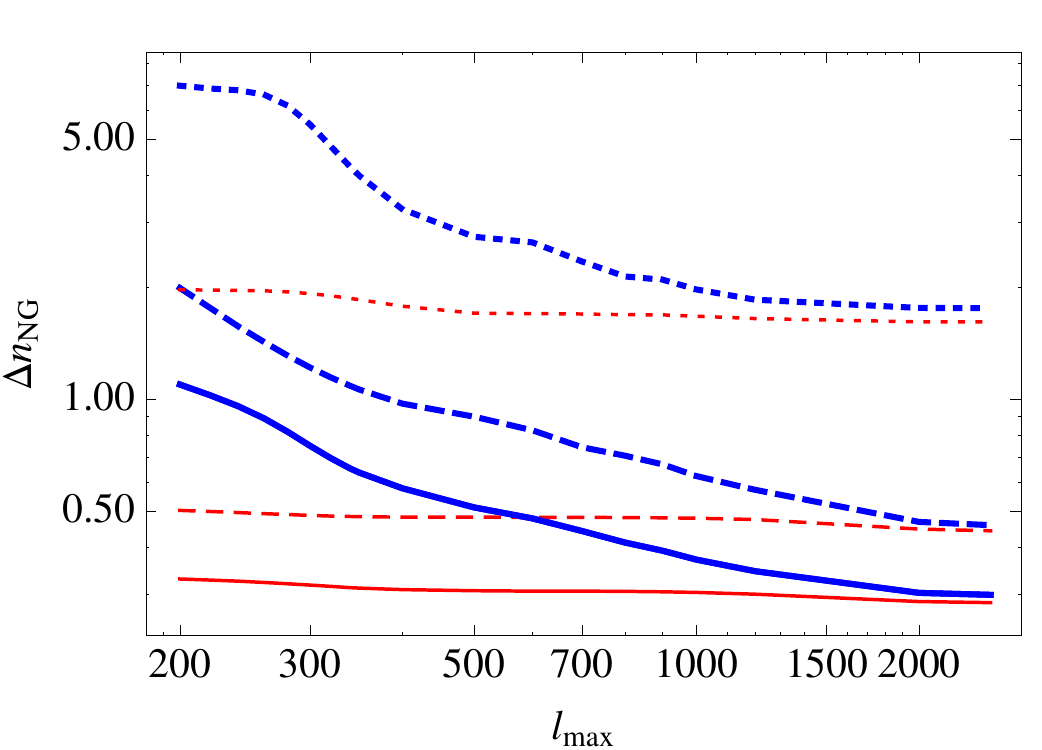}}
{\includegraphics[width=0.48\textwidth]{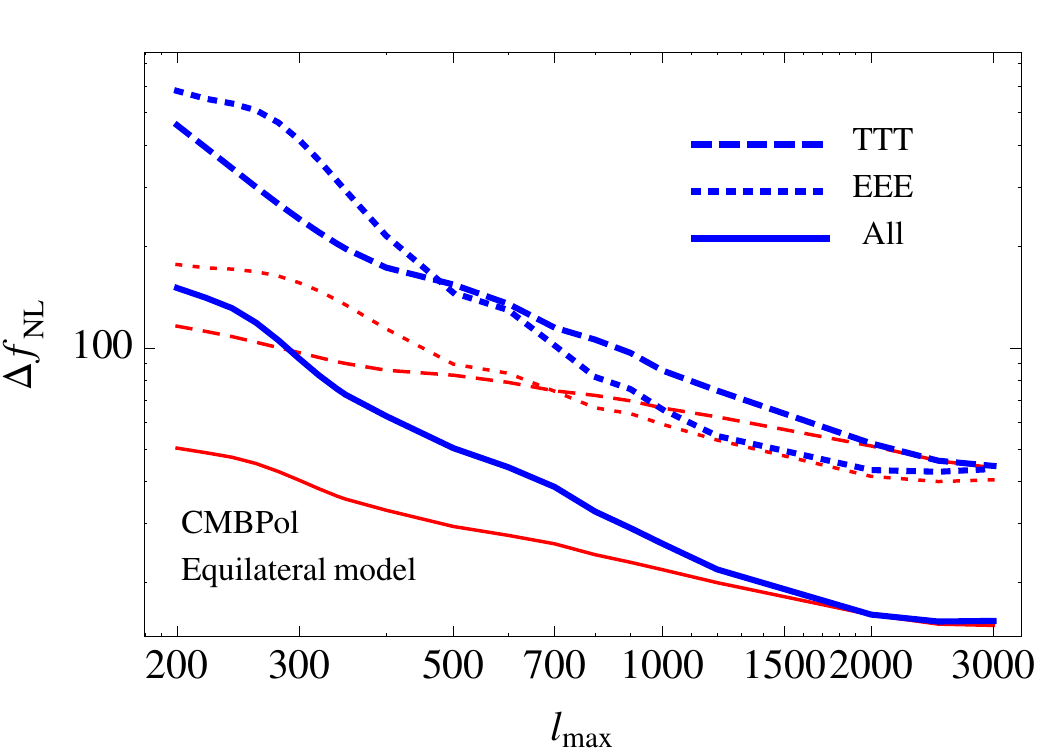}}
{\includegraphics[width=0.48\textwidth]{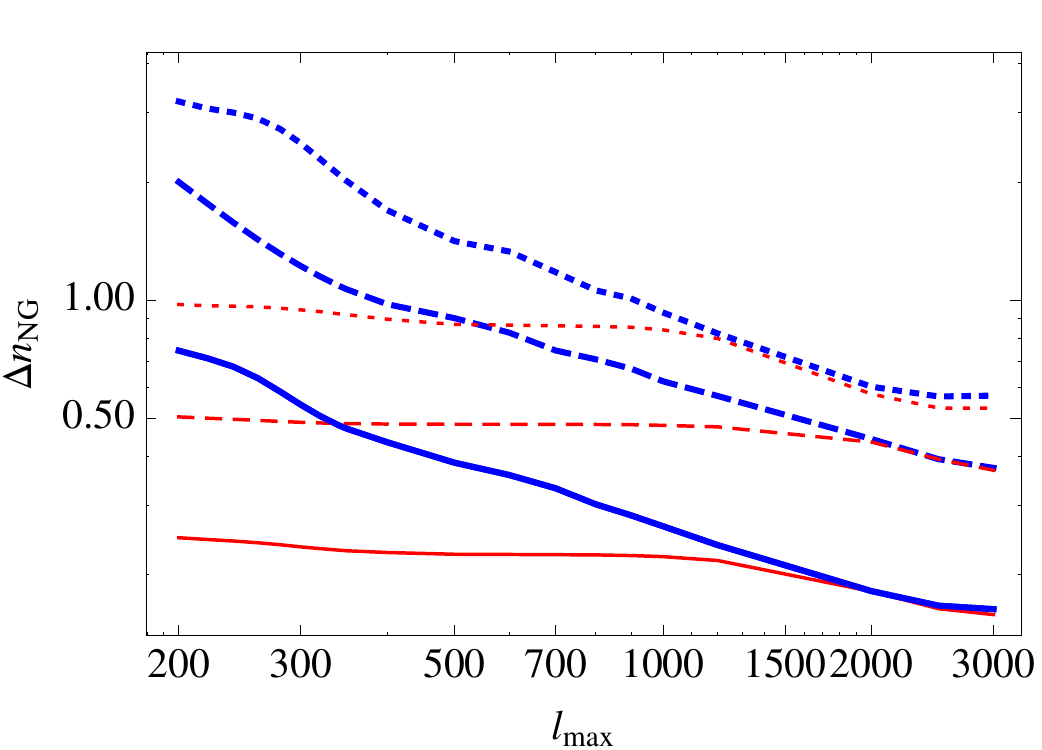}}
\caption{Equilateral model. Expected marginalized ({\it thick, blue lines}) and unmarginalized ({\it thin, red lines}) errors for $\fnl$ ({\it left panels}) and $\nng$ ({\it right panels}) as a function of $l_{max}$ for Planck ({\it central panel}) and CMBPol ({\it lower panels}).}
\label{fig:cmbEqlmax}
\end{center}
\end{figure}
Fig.~\ref{fig:cmbEq} shows the same results as Fig.~\ref{fig:cmbLc} but for the equilateral model. The reduced degeneracy with respect to the local case indicates that the chosen pivot $k_p=0.04$ Mpc$^{-1}$ is close to the optimal value for the equilateral model. This also reflects the fact that the overall signal is, for this model, relatively larger at smaller scales. This is due to the statistical weight of the number of configurations close to equilateral which increases faster with $l_{max}$ then the number of squeezed configurations correlating small scales to the few available large scales, relevant for the local bispectrum. In Fig.~\ref{fig:cmbEqlmax} we plot the expected marginalized ({\it thick lines}) and unmarginalized ({\it thin lines}) 1-$\sigma$ errors on the individual parameters $\fnl$ ({\it left panel}) and $\nng$ ({\it right panel}) for Planck ({\it upper panels}) and CMBPol ({\it bottom panels}) as a function of $l_{max}$ in the equilateral case.

The 1-$\sigma$ uncertainties on the running parameter $\nng$ expected from WMAP, Planck and CMBPol are given by
\bea
\Delta\nng  & \simeq & 1.1~ \frac{100}{\fnle}~\frac{1}{\sqrt{f_{sky}}}\qquad [{\rm WMAP}],
\label{eq:resCMBeqWMAP}\\
\Delta\nng & \simeq & 0.30~ \frac{100}{\fnle}~\frac{1}{\sqrt{f_{sky}}}\qquad [{\rm Planck}],
\label{eq:resCMBeqPlanck}\\
\Delta\nng & \simeq & 0.17~ \frac{100}{\fnle}~\frac{1}{\sqrt{f_{sky}}}\qquad [{\rm CMBPol}].
\label{eq:resCMBeqCMBPol}\eea
The corresponding unmarginalized values assuming $\fnle=100$ and $f_{sky}=1$ are $\Delta\nng \simeq 0.47$, $0.28$ and $0.16$ for WMAP, Planck and CMBPol respectively. In the equilateral case, the larger uncertainties on $\nng$ allow for a larger variation of the amplitude of non-Gaussianity over the probed
range of scale. However, we would like to remind the reader that the larger uncertainties in $\fnl$, $\nng$ for the equilateral case are only due the specific normalization choice for the equilateral bispectrum, and do not reflect a larger detection power for local non-Gaussianity with respect to equilateral non-Gaussianity \citep{FergussonShellard2008}.
Assuming $\nng\simeq 0.3$ and $\fnl(k=\kp)=100$, we have $\fnl(k=0.001\kMpc)\simeq 30$ and $\fnl(k=0.1\kMpc)\simeq 120$ while for $\nng=-0.3$, $\fnl(k=0.001\kMpc)\simeq 330$ and $\fnl(k=0.1\kMpc)\simeq 84$. These are, again, relatively small variations, stressing the constraining power of the CMB bispectrum on the running parameter.

%%%%%%%%%%%%%%%%%%%%%%%%%%%%%%%%%%%%%%%%%%%%%%%%%%%%%%%%%%%%%%%%%%%%%%%%%%%%%%%%
\subsection{Dependence on the fiducial values}
\label{ssec:fidval}

\begin{figure}[t]
\begin{center}
{\includegraphics[width=0.48\textwidth]{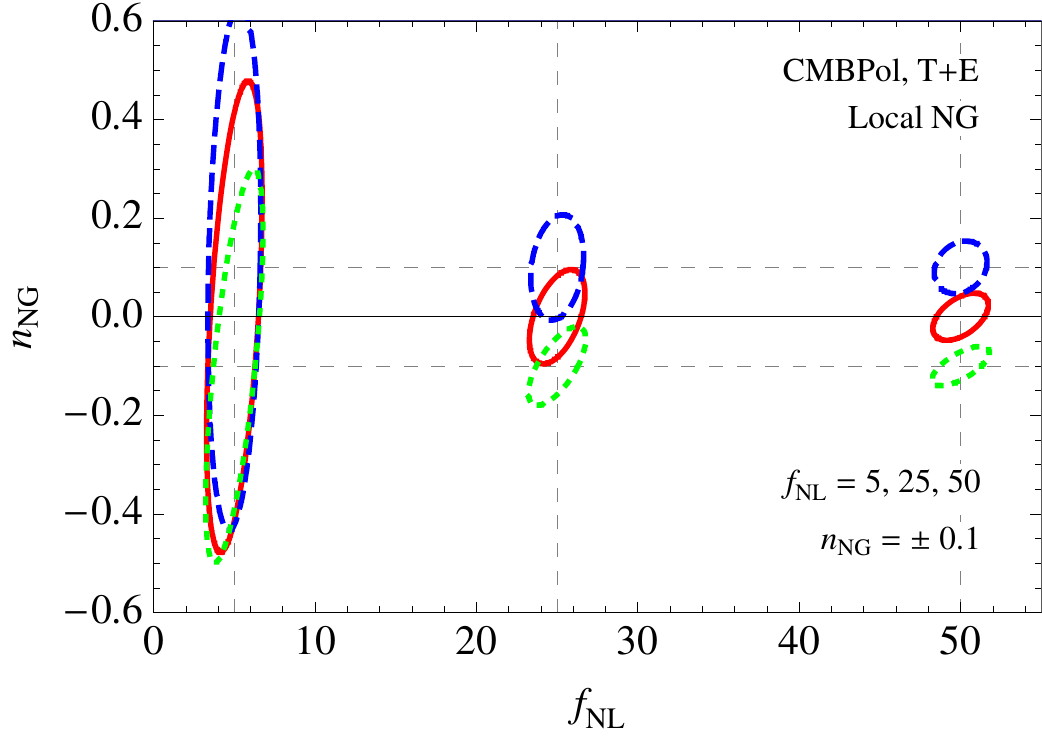}}
{\includegraphics[width=0.48\textwidth]{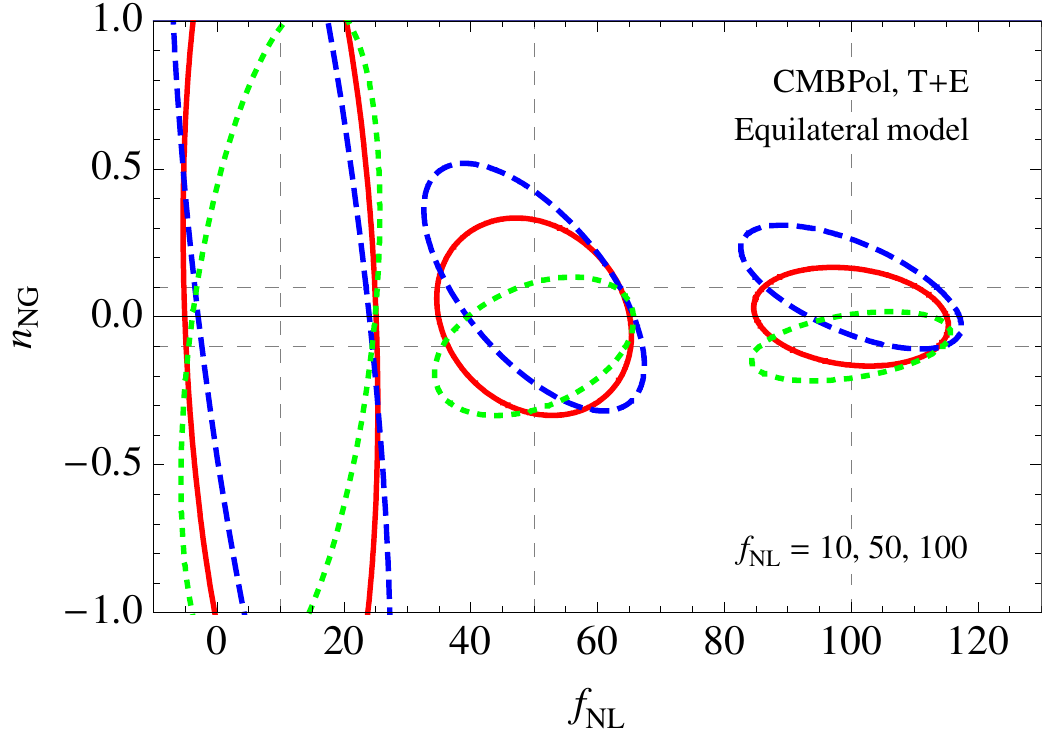}}
\caption{$1-\sigma$ contours for CMBPol assuming different fiducial values for the parameters. {\it Left panel}: local model for $\fnl=5, 25, 50$. {\it Right panel}: equilateral model for $\fnl=10, 50, 100$. In both cases we also consider $\nng=0$ ({\it continuous, red lines}), $0.1$ ({\it dashed, blue lines}) and $-0.1$ ({\it dotted, green lines}).}
\label{fig:cmbVarFid}
\end{center}
\end{figure}
The dependence of the previous results, both for the local and equilateral models, on the fiducial value $\fnl^*$ of $\fnl$ is trivial since for the Fisher matrix, $F_{ij}$, with $i,j=1$ corresponding to $\fnl$ and $i,j=2$ to $\nng$, we have the scaling $F_{12}\sim \fnl$ and $F_{22}\sim \fnl^2$. While the errors on $\fnl$ clearly do not depend, in our approximation, on the assumed value for the same parameter, for the errors, both marginalized and not, on $\nng$ we have in general
\be
\Delta \nng \sim 1/\fnl^*,
\ee
as emphasized by the expressions in Eq.s~(\ref{eq:resCMBlcWMAP}) to 
%(\ref{eq:resCMBlcCMBPol}) and (\ref{eq:resCMBeqWMAP}) to
(\ref{eq:resCMBeqCMBPol}).
The correlation coefficient between the two parameters, defined as $c_{12}\equiv (F^{-1})_{12}/\sqrt{(F^{-1})_{11}(F^{-1})_{22}}$ is independent of $\fnl^*$ so that the value of the latter does not affect the level of degeneracy. On the other hand, the fiducial value $\nng^*$ of $\nng$ clearly has a large impact on the distribution of the signal over the relevant range of scales. We can therefore expect it to determine in part the level and direction of the degeneracy between $\fnl$ and $\nng$.

These considerations are illustrated in Fig.~\ref{fig:cmbVarFid}, where we compare the contours for temperature and polarization shown in Fig.~\ref{fig:cmbLc} and Fig.~\ref{fig:cmbEq} to the same contours derived assuming different fiducial values for $\fnl$ and $\nng$. On the left panel we show the local model results for $\fnl=5$, $25$ and $50$ and for $\nng=0$ and $\pm 0.1$. On the right panel we show instead the
equilateral model for $\fnl=10$, $50$ and $100$ with $\nng=0$ and $\pm 0.1$. In this last case, the effect on the parameter degeneracy is particularly evident, although we would like to stress again that, once a given experimental configuration has been chosen, the degeneracy can always be removed through an appropriate choice of pivot and it is not going to be an issue in the experimental determination of $\fnl$ and $\nng$.

%%%%%%%%%%%%%%%%%%%%%%%%%%%%%%%%%%%%%%%%%%%%%%%%%%%%%%%%%%%%%%%%%%%%%%%%%%%%%%%%
%%%%%%%%%%%%%%%%%%%%%%%%%%%%%%%%%%%%%%%%%%%%%%%%%%%%%%%%%%%%%%%%%%%%%%%%%%%%%%%%
\section{Primordial non-Gaussianity in the Large Scale Structure}
\label{sec:lss}

The impact of non-Gaussian initial conditions on the evolution of the matter and galaxy distributions and their effect on large-scale structure observables has been the subject of several works over the last two decades focusing, for the most part, on the galaxy clusters abundance \citep{LucchinMatarrese1988,ColafrancescoLucchinMatarrese1989,ChiuOstrikerStrauss1998,RobinsonGawiserSilk2000,MatarreseVerdeJimenez2000,KoyamaSodaTaruya1999}, on higher-order moments of the density probability distribution \citep{FryScherrer1994,ChodorowskiBouchet1996,DurrerEtal2000}, on higher-order galaxy correlation functions, \citep{Scoccimarro2000A,VerdeEtal2000,ScoccimarroSefusattiZaldarriaga2004,SefusattiKomatsu2007} and the Minkowski Functionals of the density field \citep{HikageKomatsuMatsubara2006,HikageEtal2008A}. More recently, a strong effect due to non-Gaussian initial conditions of the local type on the bias of dark matter halos has been measured in $N$-body simulations by \citet{DalalEtal2008}. This result, somehow anticipated by earlier works on high-peak clustering \citep{GrinsteinWise1986,MatarreseLucchinBonometto1986}, spurred a series of works \citep{MatarreseVerde2008,SlosarEtal2008,Slosar2008,Seljak2008,CarboneVerdeMatarrese2008,McDonald2008,AfshordiTolley2008,TaruyaKoyamaMatsubara2008,DesjacquesSeljakIliev2008,PillepichPorcianiHahn2008,WandsSlosar2009} indicating that significant constraints on the amplitude of local non-Gaussianity, comparable to those obtained from the CMB bispectrum, can be derived from measurements of the power spectrum of galaxies and quasars in current large-scale surveys \citep{SlosarEtal2008,AfshordiTolley2008}. At the same time, further studies of the effect of non-Gaussianities on the halo mass function are being proposed \citep{MaggioreRiotto2009C,Oguri2009,LamSheth2009,LamShethDesjacques2009}.

We will not attempt here to perform a complete analysis of LSS probes including all relevant observables (and their covariance) as this would be beyond the scope of this work. We will derive, instead, simple estimates of the expected errors on the amplitude and running parameters $\fnl$ and $\nng$, choosing specifically a representative observable for each model: namely the bias of galaxies as determined from {\it power spectrum} measurements in galaxy surveys for the {\it local} non-Gaussian model, and the galaxy {\it bispectrum} for the {\it equilateral} model. As we will argue in the next Section, we expect the two different models to affect in quite a different way the large-scale clustering of biased objects. We therefore make, in choosing these specific and different measurements, what we consider a reasonable assumption on the impact of different non-Gaussian models, warning that further theoretical work, possibly supported by numerical simulations, will be needed to strengthen our understanding of this phenomenon.

Regarding the possibility of a scale-dependent non-Gaussianity, \citet{LoVerdeEtal2008} provided the first analysis of the possibility of constraining a running $\fnl$ parameter by combining current limits from the CMB with future measurements of cluster abundance. Focusing in particular on the equilateral model for the curvature bispectrum, this work assumes the amplitude of $\fnl(k)=\fnl(k/\kp)^{-2s}$ to be constrained by the CMB bispectrum at the pivot point $\kp=0.04$ Mpc$^{-1}$ and derives the expected constraints on the running parameter\footnote{The dimensionless variation of the speed of sound $s$ was denoted $\kappa$ in \citep{LoVerdeEtal2008}.} $s$ by considering the effective amplitude of $\fnl(k)$ at the smaller scales ($k\sim 0.3-0.6\kMpc$) probed by cluster surveys (see Fig.~1 in \citep{LoVerdeEtal2008}). For an all-sky cluster survey up to redshift $z_{max}=1.3$ with single mass bin defined by the threshold $M>1.45\times 10^{14}\Ms$ they find, in terms of our running parameter $\nng=-2s$, the 1-$\sigma$ constraints, marginalized over $\Omega_m$, $\sigma_8$ and $h$, assuming the fiducial values $\fnl=38$ and $\nng=0$, $\Delta\nng \simeq 3.38$ with a WMAP prior $\Delta \fnl(k=\kp)=150$, and $\Delta\nng \simeq 2$ with a Planck prior $\Delta \fnl(k=\kp)=40$. Note that their analysis does not include the {\it simultaneous} limits that measurement of the CMB bispectrum {\it alone} is expected to provide on both the amplitude $\fnl$ and running $\nng$ that have been discussed in the previous Section. In terms of these fiducial values, Eqs.~(\ref{eq:resCMBeqWMAP}) and (\ref{eq:resCMBeqPlanck}) give, for instance, $\Delta\nng\simeq 2.9$ for WMAP and $\Delta\nng\simeq 0.8$ for Planck, a factor of $1.2$ and $2$, respectively, better than the results of \citep{LoVerdeEtal2008}. It is to be remarked, however, that in the event of a detection, an independent confirmation from an observable probing quite different scales such as the cluster abundance would be of great importance.

The matter overdensity in Fourier space $\d_\kv$ is related to the curvature perturbations $\Phi_\kv$ at early times during matter domination by the Poisson equation as
\be
\d_\kv(z)=M(k,z)~\Phi_\kv,
\ee
where we introduced the function
\be
M(k,z)=\frac{2}{3}\frac{k^2T(k)D(z)}{\Omega_mH_0^2},
\ee
with $T(k)$ being the matter transfer function and $D(z)$ the growth function. The linear matter power spectrum is therefore given by 
\be
P_0(k)=M^2(k,z)P_\Phi(k).
\ee
A departure from Gaussianity in the initial conditions results in non-vanishing higher-order, connected, correlation functions for the matter field proportional to the corresponding correlation functions for the curvature fluctuations. In particular, for the initial matter bispectrum $B_0$ we have
\be
B_0(k_1,k_2,k_3)=M(k_1,z)M(k_2,z)M(k_3,z)B_\Phi(k_1,k_2,k_3).
\ee
For simplicity, we will ignore the non-Gaussianity represented by a non-vanishing initial trispectrum---and by higher-order correlation functions---although its effect on the galaxy bispectrum can be significant, see \citep{JeongKomatsu2009,Sefusatti2009}. On the other hand, their expressions involve extra parameters, other than $\fnl$, and their scale-dependence deserves a specific study.

In the rest of this Section we will summarize and discuss in some details the recent literature regarding the effect of non-Gaussian initial conditions on the bias of galaxies and clusters and we motive our choice for the observables considered for our simple analysis. We will then derive the expected uncertainties on the two parameters for the two models of the curvature bispectrum introduced in Section~\ref{sec:parNG} considering measurements of the galaxy power spectrum for the local model and the galaxy bispectrum for the equilateral model.

%%%%%%%%%%%%%%%%%%%%%%%%%%%%%%%%%%%%%%%%%%%%%%%%%%%%%%%%%%%%%%%%%%%%%%%%%%%%%%%%
\subsection{Halo bias and primordial non-Gaussianity}
\label{ssec:HBPNG}

As first shown by \citet{DalalEtal2008} and later confirmed by \citet{DesjacquesSeljakIliev2008,PillepichPorcianiHahn2008,GrossiEtal2009} in numerical simulations with non-Gaussian initial conditions of the local kind, the large-scale bias of halos can be greatly affected by relatively small values of $\fnl$. \citet{DalalEtal2008,SlosarEtal2008,AfshordiTolley2008} provided expressions for the halo bias in the presence of local non-Gaussianity based on the peak-background split formalism \citep{ColeKaiser1989}. They find the correction $\Delta b_{h,1}(k,M,z,\fnl)$ to the {\it Gaussian} linear halo bias $b_{h,1}(M)$, defined as $b_{h,1}^{NG}=b_{h,1}+\Delta b_{h,1}$, to be given by 
\be\label{eq:dblocal}
\Delta b_{h,1}(k,M,z,\fnl) = 2 \tfnl[b_{h,1}(M,z)-1]\delta_c\frac{1}{M(k,z)}
\ee
where $\d_c\simeq 1.68$ is the critical overdensity for spherical collapse and $\tfnl\simeq 1.33\fnl$ is the non-Gaussian parameter in the large-scale structure convention. The strong scale- and redshift-dependence of such correction allowed \citet{SlosarEtal2008} and \citet{AfshordiTolley2008} to place significant constraints on $\fnll$ already from present large-scale structure data. 

A different approach has been followed instead by \citet{MatarreseVerde2008}  and by \citet{TaruyaKoyamaMatsubara2008} to derive a similar expression for the correction to the power spectrum of biased tracers of the dark matter distribution. In particular, \citet{MatarreseVerde2008} considered a peak number density defined by the expression $\rho_{p,M}(\xv)=\theta[\delta_R(\xv)-\delta_c]$, where $\theta$ is a step function. Based on earlier works by \citet{GrinsteinWise1986} and \citet{MatarreseLucchinBonometto1986}, they derive the correction to the peak correlation function in terms of the three-point correlation function. Their result can be rewritten in the form of a linear {\it peak} bias correction given by
\be\label{eq:dbMV}
\Delta b_{p,1}(k,M,z,\fnl) = \frac{\nu^2}{\sigma_R^2}\frac{1}{2P_{0,R}(k)}\int d^3q B_{0,R}(k,q,|\kv-\qv|)
\ee
where $\nu\equiv\delta_c/\sigma_R$ represents the peak height with $\sigma_R$ being the r.m.s. of matter fluctuations filtered on the scale $R$ and where $P_{0,R}=M_R^2(k,z)P_\Phi(k)$ and $B_{0,R}(k_1,k_2,k_3)=M_R(k_1,z)M_R(k_2,z)M_R(k_3,z)B_\Phi(k_1,k_2,k_3)$ are the initial, filtered matter power spectrum and bispectrum and where we introduce, as short-hand notation, $M_R(k,z)=W_R(k)M(k,z)$ with $W_R(k)$ being the smoothing function. In the large-scale limit ($\kv\rightarrow 0$) and assuming the local form for the curvature bispectrum $B_\Phi$ {\it with no running}, one can use the approximation
\be
\frac{1}{P_{0,R}(k)}\int d^3q B_{0,R}(k,q,|\kv-\qv|)\simeq 4\tfnl\frac{\sigma_R^2}{M_R(k,z)},
\ee
which leads to the correction to the linear {\it peak} bias
\be
\Delta b_{p,1}(k,M,z,\tfnl) \simeq 2\tfnl\frac{\nu^2}{\sigma_R^2}\frac{\sigma_R^2}{M_R(k,z)}=2\tfnl(b_{p,1}-1)\delta_c\frac{1}{M_R(k,z)},
\ee
equivalent to Eq.~(\ref{eq:dblocal}) after identifying $b_{p,1}=1+\nu/\sigma_R$ as the Eulerian linear peak bias.

\citet{TaruyaKoyamaMatsubara2008} computed the 1-loop corrections to the galaxy power spectrum due to a non-vanishing primordial bispectrum, assuming a {\it local} bias relation between galaxies and the smoothed matter density $\delta_R(\xv)$,
\be
\label{eq:localbias}
\delta_g(\xv)=b_1\delta_R(\xv)+\frac{1}{2}b_2\delta_R^2(\xv)+...~,
\ee
finding a correction to the linear bias parameter that can be written as 
\be\label{eq:dbTKM}
\Delta b_{1}(k,M,z,\tfnl) = b_2\frac{1}{2P_{0,R}(k)}\int d^3q B_{0,R}(k,q,|\kv-\qv|)
\ee
which, in the large-scale limit becomes
\be\label{eq:dbTKMapprox}
\Delta b_{1}(k,M,z,\tfnl) \simeq 2\tfnl b_2\frac{\sigma_R^2}{M_R(k,z)},
\ee
and which notably involves the quadratic bias parameter $b_2$ from Eq.~(\ref{eq:localbias}).

The three expressions of Eq.~(\ref{eq:dblocal}), Eq.~(\ref{eq:dbMV}) and Eq.~(\ref{eq:dbTKM}) all show the same scale- and redshift-dependence but with a different constant of proportionality which depends on the specific assumptions. In other words, references \citep{MatarreseVerde2008} and \citep{TaruyaKoyamaMatsubara2008} derive an effect on the halo/galaxy bias that, for the local model, can be shown to be equivalent, at large scales, to the one described by Eq.~(\ref{eq:dblocal}), and they actually coincide if one considers as well the high-peak limit $\nu/\sigma_R\gg 1$ since, in this case
\be
[b_{h,1}-1]\delta_c\simeq b_2\sigma_R^2\simeq\nu^2=\frac{\delta_c^2}{\sigma_R^2},
\ee 
once one assumes an expression such as the one derived by \citet{MoWhite1996} for the galaxy quadratic bias parameter. On the other hand, both \citep{MatarreseVerde2008} and \citep{TaruyaKoyamaMatsubara2008} provide an expression that can be applied to any model of primordial non-Gaussianity as they depend generically on the primordial curvature bispectrum. For instance, as pointed out in \citep{TaruyaKoyamaMatsubara2008}, while for a local non-Gaussianity both Eq.~(\ref{eq:dbMV}) and Eq.~(\ref{eq:dbTKM}) predict a large effect as the one measured in $N$-body simulations, in the {\it equilateral} case the effect is largely suppressed. As shown in Fig.~6 of \citep{TaruyaKoyamaMatsubara2008}, the relative correction to the Gaussian bias parameter at redshift $z=1$, with $b_1=2$ and $b_2=0.5$ stays below $1\%$, even when a running of $\fnl$ corresponding to $\nng=0.6$ is considered. 

It should be noted that while the parameterization of a scale-dependent $\fnl(k_1,k_2,k_3)$ described in Section~\ref{sec:parNG}, which is to be understood as a parameterization of the initial bispectrum, can be easily and unambiguously implemented in Eq.~(\ref{eq:dbMV}) and Eq.~(\ref{eq:dbTKM}), the same is not true for Eq.~(\ref{eq:dblocal}). In fact while Eq.~(\ref{eq:dblocal}) suggests one replace a constant $\fnl$ as indicated by Eq.~(\ref{eq:runningfnl}) with $k$ being the unequivocal scale of the problem, in the other case, assuming $k=(k_1k_2k_3)^{1/3}$ in the expression for the curvature bispectrum, the scale dependence is partially integrated over, leading, as we will show in detail in the next Section, to a quite different behavior. Based on these considerations we assume for our analysis an expression for the correction to the linear bias parameter proportional to the integral of Eq.~(\ref{eq:dbTKM}) and Eq.~(\ref{eq:dbMV}): this provides an  expression for the correction to the galaxy power spectrum in the case of local non-Gaussianity and, at the same time, a plausible argument for neglecting altogether any effect on the galaxy power spectrum in the equilateral case.

For equilateral non-Gaussianity we consider instead its effect on the galaxy {\it bispectrum}, and, in particular, we assume such effect to be given simply by the {\it primordial component}, neglecting corrections due to non-linear galaxy bias. This is a quite conservative choice. In fact, it has been recently shown by \citet{JeongKomatsu2009} and by \citet{Sefusatti2009} that significant corrections at large scales, similar to those affecting the galaxy power spectrum, are expected as well for the galaxy bispectrum. Since these early results still require a comparison with numerical simulation to properly test their theoretical predictions, we will consider for equilateral non-Gaussianity the sole initial component present as well in the matter bispectrum.
 
 A complete analysis of large-scale structure data should naturally include two-point statistics as well as higher-order correlators. In this work, however, we neglect for the time being the extra information that could be extracted, for instance, from the galaxy bispectrum on local non-Gaussianity, leaving this subject for a future work. As for the equilateral model, we consider as a large-scale structure probe the galaxy bispectrum, extending the analysis of \citet{SefusattiKomatsu2007} to the case of a running $\fnle$, keeping in mind that this analysis needs to be improved by taking into account the large-scale corrections due to non-local bias studied in \citep{JeongKomatsu2009,Sefusatti2009} that will probably lead to smaller uncertainties on the non-Gaussian parameter $\fnl$ both for the local and equilateral models.

%%%%%%%%%%%%%%%%%%%%%%%%%%%%%%%%%%%%%%%%%%%%%%%%%%%%%%%%%%%%%%%%%%%%%%%%%%%%%%%%
\subsection{Constraints on local non-Gaussianity from the galaxy power spectrum}
\label{ssec:localHB}

\begin{figure}[t]
\begin{center}
{\includegraphics[width=0.48\textwidth]{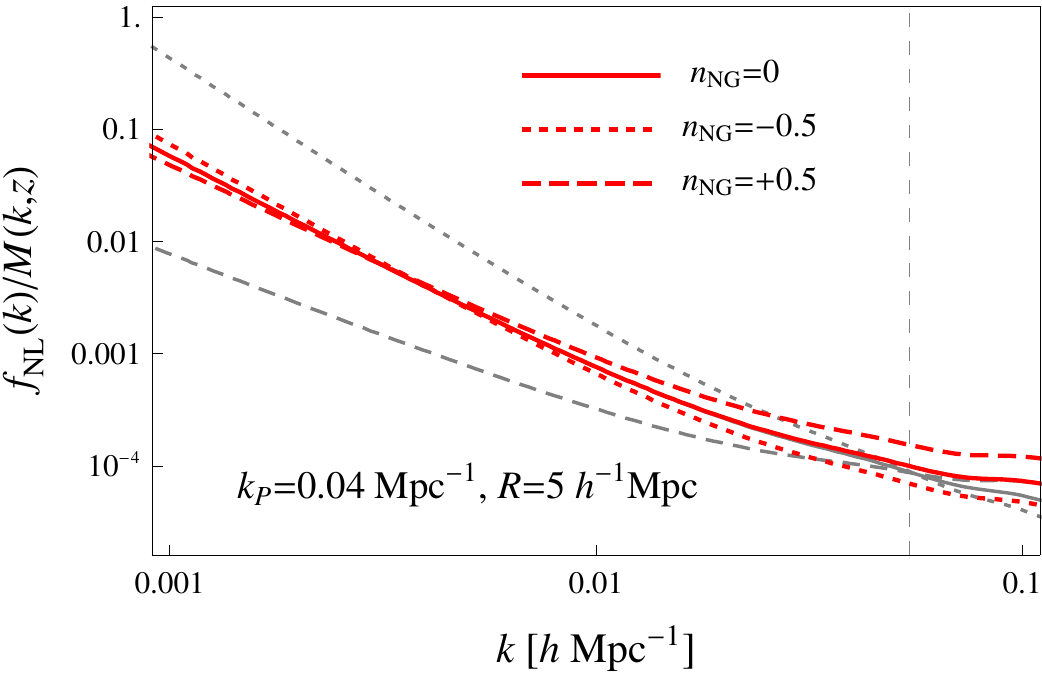}}
{\includegraphics[width=0.48\textwidth]{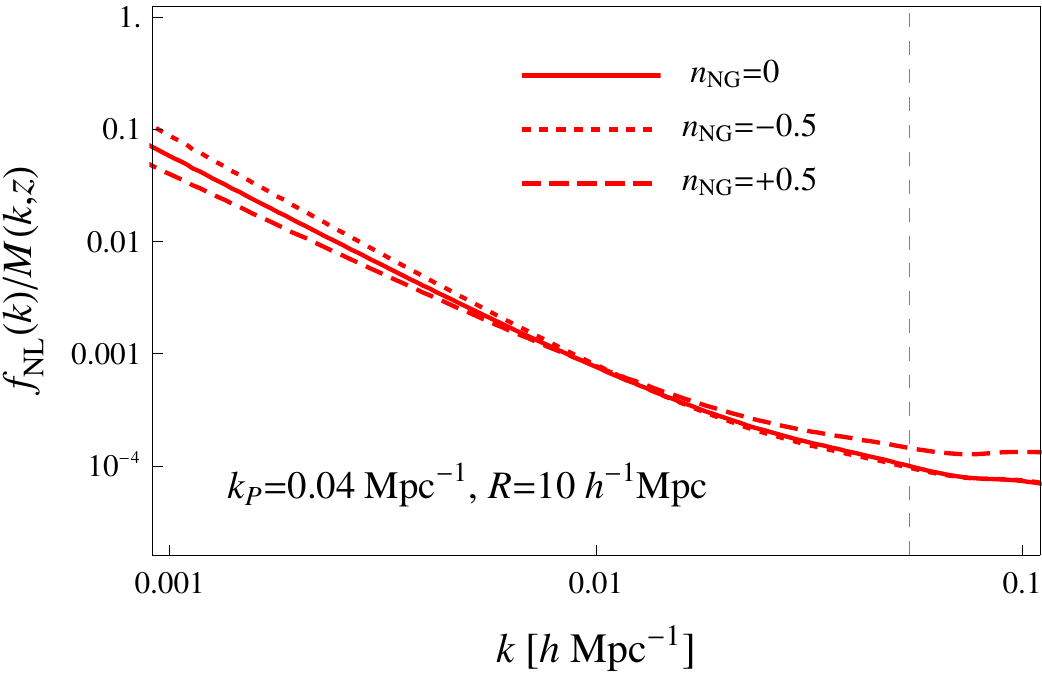}}
\caption{Scale-dependent factor, $\fnl(k)/M(k,z)$,  of the bias correction for $\fnl=1$ and $\nng=0$ (continuous red line), $\nng=0.5$ (dashed red line) and $\nng=-0.5$ (dotted red line). The vertical dashed line indicates the pivot scale assumed, $\kp=0.04$ Mpc$^{-1}$. {\it Left}: smoothing scale of Eq.~(\ref{eq:fnlII}), $R=5\Mpc$. {\it Right}: $R=10\Mpc$. The thin gray lines on the left panel show the behaviour correspondent to a simple substitution of $\fnl$ with $\fnl(k)=(k/\kp)^{\nng}$ in Eq.~(\ref{eq:dblocal}).}
\label{fig:dblocal}
\end{center}
\end{figure}

In this Section we perform a Fisher matrix analysis to obtain an estimate of the constraints achievable from power spectrum measurements in future galaxy surveys on $\fnll$ and $\nng$. We assume the correction to the galaxy linear bias parameter $b_1$ to be given by
\be\label{eq:dblocal2}
\Delta b_{1}(k,z,\fnl) = 2~q~\fnl(k)~(b_1-1)~\delta_c\frac{1}{M_R(k,z)}\frac{1}{1+(a k)^n}.
\ee
This expression assumes the normalization of the bias correction to be given by comparison to $N$-body simulation and encodes the actual dependence on the {\it scale-dependent} curvature bispectrum in the function $\fnl(k)$. Specifically, we include, to the theoretical expression of Eq.~(\ref{eq:dblocal}), the correcting factor $q=0.75$ introduced, and motivated in terms of the ellipsoidal collapse model, by \citet{GrossiEtal2009} and the small-scale suppression factor $1/[1+(a k)^n]$ derived by \citet{DesjacquesSeljakIliev2008} from a fit to numerical simulations at smaller scales, with $a=23.5$ and $n=1.69$ which we will assume valid for all redshift and densities. Such damping has been observed as well by \citet{PillepichPorcianiHahn2008}, which also provided a fitting formula consistent with the one we assume on relevant scales. In other terms, we are assuming the {\it amplitude} of the bias correction to be proportional to $[b_{1}(M,z)-1]/[1+(a k)^n]$ since this factor has been compared to simulations in the case of a constant $\fnl$. However, we assume at the same time a scale-dependence consistent with Eq.~(\ref{eq:dbTKM}), so that
\bea\label{eq:fnlII}
\fnl(k)& \equiv&\frac{M_R(k,z)}{4\sigma_R^2P_{0,R}(k)}\int d^3qB_{0,R}(k,q,|\kv-\qv|)
\nonumber\\
&= & \frac{1}{4\sigma_R^2}\int d^3qM_R(q)M_R(|\kv-\qv|)\frac{B_\Phi(k,q,|\kv-\qv|)}{P_\Phi(k)}
\nonumber\\
&= & \frac{1}{2\sigma_R^2}\int d^3qM_R(q)M_R(|\kv-\qv|)
\nonumber\\
& & \times\fnl\left[\frac{(kq|\kv-\qv|)^{1/3}}{k_p}\right]^{\nng}\left[P_\Phi(q)+P_\Phi(|\kv-\qv|)+\frac{P_\Phi(q)P_\Phi(|\kv-\qv|)}{P_\Phi(k)}\right],
\eea
where the last equality assumes the curvature bispectrum of the local model. The dependence on the smoothing scale $R$ is very weak at large scales if $\nNG=0$ (see also Fig.~1 in \citep{MatarreseVerde2008}), since in this case, for the local model, we have
\bea
\fnl(k)
& = &
\frac{\fnl}{2\sigma_R^2}\int d^3qM_R(q)M_R(|\kv-\qv|)
\left[P_\Phi(q)+P_\Phi(|\kv-\qv|)+\frac{P_\Phi(q)P_\Phi(|\kv-\qv|)}{P_\Phi(k)}\right]
\nonumber\\
& \stackrel{k\rightarrow 0}{\simeq} &
\frac{\fnl}{2\sigma_R^2}\int d^3qM_R^2(q)
2P_\Phi(q)
 =  \fnl,
\eea
but it is significant, {\it even at large scales}, when the running parameter is different from zero. In Fig.~\ref{fig:dblocal} we plot the scale-dependent part of the bias correction, $\fnl(k)/M(k,z)$ according to Eq.~(\ref{eq:fnlII}) assuming the amplitude $\fnll=1$, the pivot $\kp=0.04$ Mpc$^{-1}$ and a smoothing scale $R=5\Mpc$ (left panel) and $R=10\Mpc$ (right panel). Thick lines show the behaviour for the different values $\nng=0$ (continuous line), $\nng=+ 0.5$ (dashed) and $\nng=-0.5$ (dotted). The thin, dashed vertical line indicates the position of the pivot point, $\kp$ in $\kMpc$.

In the left panel of Fig.~\ref{fig:dblocal}, the thin (gray) lines show the behaviour corresponding to a simple substitution of $\fnl$ with $\fnl(k)=\fnl(k/\kp)^{\nng}$ in Eq.~(\ref{eq:dblocal}). One can clearly see that such na\"\i ve substitution would lead to a very different effect on halo bias, particularly on the relevant range of scales. The integration in Eq.~(\ref{eq:fnlII}) in general tends to dampen the effect of a running $\fnl$ since most of the scale-dependence is integrated over. Moreover, it induces an ``effective" pivot scale that depends on the value of $R$ and it is greater, in momentum space, for larger values of $R$. This can be understood in terms of the expected relative value of the integral in Eq.~(\ref{eq:fnlII}), at the pivot point $\kp$ with respect to the value of $\sigma_R$. Since both the chosen values $R=5\Mpc$ and $R=10\Mpc$ correspond to momenta much larger than $\kp=0.04$ Mpc$^{-1}$, for the smaller value of $R=5\Mpc$ we expect the integral to provide a larger value for positive $\nng$ relative to $\sigma_R$ than the case of $R=10\Mpc$. The dependence on the assumed value of $\kp$ is instead more trivial, as a shift in $\kp$ corresponds to a shift in the effective pivot point of the bias correction. It is interesting to note that this peculiar behaviour might be relevant to correctly identify a theoretical description of this effect, once the scale-dependence of the curvature bispectrum is properly implemented in numerical simulations.  

In our analysis we will assume for simplicity a single representative value of $R=5\Mpc$, although the smoothing scale should in principle be related to the halo population of interest. We notice, however, that while the degeneracy between the two non-Gaussian parameters is affected to some extent by this choice, the overall value of their marginalized uncertainties does not change significantly for our choice of the pivot scale. 

For a given survey, defined uniquely by its redshift range, sky coverage and expected galaxy number density $n_g$, we consider shells in redshift of maximum size $\Delta z = 0.5$ and evaluate all quantities at the mean redshift $\bar{z}_j$. This implies that, even in the ideal case of an all-sky survey, the volume of each shell does not exceed $100\cGpc$ for our cosmology, so that the largest scale probed corresponds to a wavenumber $k_{min}\sim 0.0013\kMpc$. We conservatively ignore large-scale correlations among different redshift bins. 

The Fisher matrix for the non-Gaussian parameters $\fnll$ and $\nng$ is therefore given by
\be
{\mathcal F}_{ab}\equiv\sum_{j=1}^{N_z}\sum_{i=1}^{N_k}
\frac{\partial P_g(k_i,\bar{z}_j)}{\partial p_a}
\frac{\partial P_g(k_i,\bar{z}_j)}{\partial p_b}
\frac{1}{\Delta P_g^2(k_i,\bar{z}_j)}
\ee
where $N_z$ is the number of redshift bins and $N_k$ is the total number of wavevectors $k_i$ from $k_f=2\pi/V^{1/3}$ to a maximum $k_{max}=0.03\kMpc$ in steps of $k_f$ in the given redshift shell.

We will consider, again conservatively, only linear corrections in $\fnl$ to the galaxy Gaussian linear bias $b_1$ according to Eq.~(\ref{eq:dblocal2}), so that
\be
P_g(k)\simeq b_1^2\left[1+2\Delta b_1(b_1,k,z,\fnl)\right]P(k)
\ee
with $P(k)$ being the linear matter power spectrum. The power spectrum variance $\Delta P_g^2(k)$ is approximated by its Gaussian component
\be\label{eq:PVar}
\Delta P_g^2(k)\simeq\frac{k_f^2}{2\pi k^2}P_{tot}^2(k)\left[1+\frac{4\Delta b_1(k,\fnl)P(k)}{P_{tot}(k)}\right]
\ee
where $P_{tot}$ is the Gaussian galaxy power spectrum including shot-noise
\be
P_{tot}(k)=P_g(k)+\frac{1}{(2\pi)^3 n_g}
\ee
with $n_g$ representing the galaxy density of the survey. The last factor in Eq.~(\ref{eq:PVar}) corresponds to the linear correction to the Gaussian variance due to effect on bias of a non-vanishing $\fnl$, which we will assume as fiducial value. We ignore for simplicity corrections due to higher-order correlations. The galaxy bias parameter is assumed to be known and it is computed by means of the halo model prescription and halo occupation distribution described in \citep{SefusattiKomatsu2007}, to which we refer the reader, which in turn is based on \citep{MoJingWhite1997,ShethTormen1999,ScoccimarroEtal2001A,TinkerEtal2005,ConroyWechslerKravtsov2006}.

\begin{table*}[b]
\caption{\label{tab:surveys} 
Characteristics of the galaxy surveys considered: sky coverage ($\Delta\Omega$), redshift range ($z_{min}<z<z_{max}$), volume ($V$) and mean galaxy density ($n_g$) together with the expected $1$-$\sigma$ uncertainties on the amplitude and running parameters $\fnl$ and $\nng$ from the Fisher matrix analysis of the {\it galaxy power spectrum for the local model} with fiducial values $\fnll=50$ and $\nng=0$ and from the Fisher matrix analysis of the {\it galaxy bispectrum} (assuming the {\it only} the primordial non-Gaussian component, see text) {\it for the equilateral model} with fiducial values $\fnle=100$ and $\nng=0$. Uncertainties are marginalized over the other non-Gaussian parameter, while cosmology and bias factors are assumed as known.}
\begin{ruledtabular}
\begin{tabular}{lcccc|cc|cc}
 Survey & $\Delta\Omega$ [deg$^2$] & $z_{min}<z<z_{max}$ & $V$ [$\cGpc$] & $n_g$ [$\icMpc$] & $\Delta\fnll$ & $\Delta\nng$\footnote{Assumes a fiducial $\fnll=50$.} & $\Delta\fnle$ & $\Delta\nng$\footnote{Assumes a fiducial $\fnle=100$.}\\
\hline\hline
\multicolumn{5}{l}{\it Spectroscopic} \\
\hline
BOSS        & $10,000$ & $0<z<0.7$   & $5.8$  & $2.7\times 10^{-4}$ & $52$ & $2.6$ & $75$  & $1.5$\\
ADEPT       & $28,000$ & $1<z<2$     & $114$  & $9.4\times 10^{-4}$ & $9.7$ & $0.49$ & $22$  & $0.25$\\
Euclid	    & $20,000$ & $0<z<2$     & $108$  & $1.5\times 10^{-3}$ & $8.4$ & $0.42$ & $17$  & $0.16$\\
HETDEX      & $200$    & $2<z<4$     & $2.8$  & $5.0\times 10^{-4}$  & $96$ & $3.8$ & $86$  & $0.60$\\
CIP         & $300$    & $3.5<z<6.5$ & $3.6$  & $5.0\times 10^{-3}$ & $61$ & $2.3$ & $35$ & $0.16$ \\
\hline
\multicolumn{5}{l}{\it Photometric} \\
\hline
%DES         & $5,000$  & $0.2<z<1.3$ & $11.9$ & $1.8\times 10^{-4}$ & $34$ & $1.7$ & - & - \\
%PAU         & $8,000$  & $0.1<z<0.9$ & $8.5$  & $10^{-3}$  & $34$ & $1.7$ & - & - \\
LSST        & $30,000$ & $0.3<z<3.6$ & $390$  & $2.8\times 10^{-3}$& $3.5$ & $0.17$ & - & -\\
PanSTARRS   & $30,000$ & $0<z<1.2$   & $60.8$ & $1.7\times 10^{-3}$ & $11$ & $0.59$ & - & - \\
\end{tabular}
\end{ruledtabular}
\end{table*}
In Table~\ref{tab:surveys} we show the specifications of LSS surveys that we will assume for the analysis in the following Sections. For the power spectrum analysis we consider both spectroscopic and photometric surveys, under the assumption that photometric uncertainties have a negligible impact on the determination of galaxies clustering properties on the interesting range of scales. Among the surveys, proposed or under construction, we choose those able to provide constraints on the non-Gaussian parameters, comparable to those provided by future CMB observations. 
%PAU, \citep{BenitezEtal2008}
%LSST, \citep{IvezicEtal2009}

\begin{figure}[t]
\begin{center}
{\includegraphics[width=0.48\textwidth]{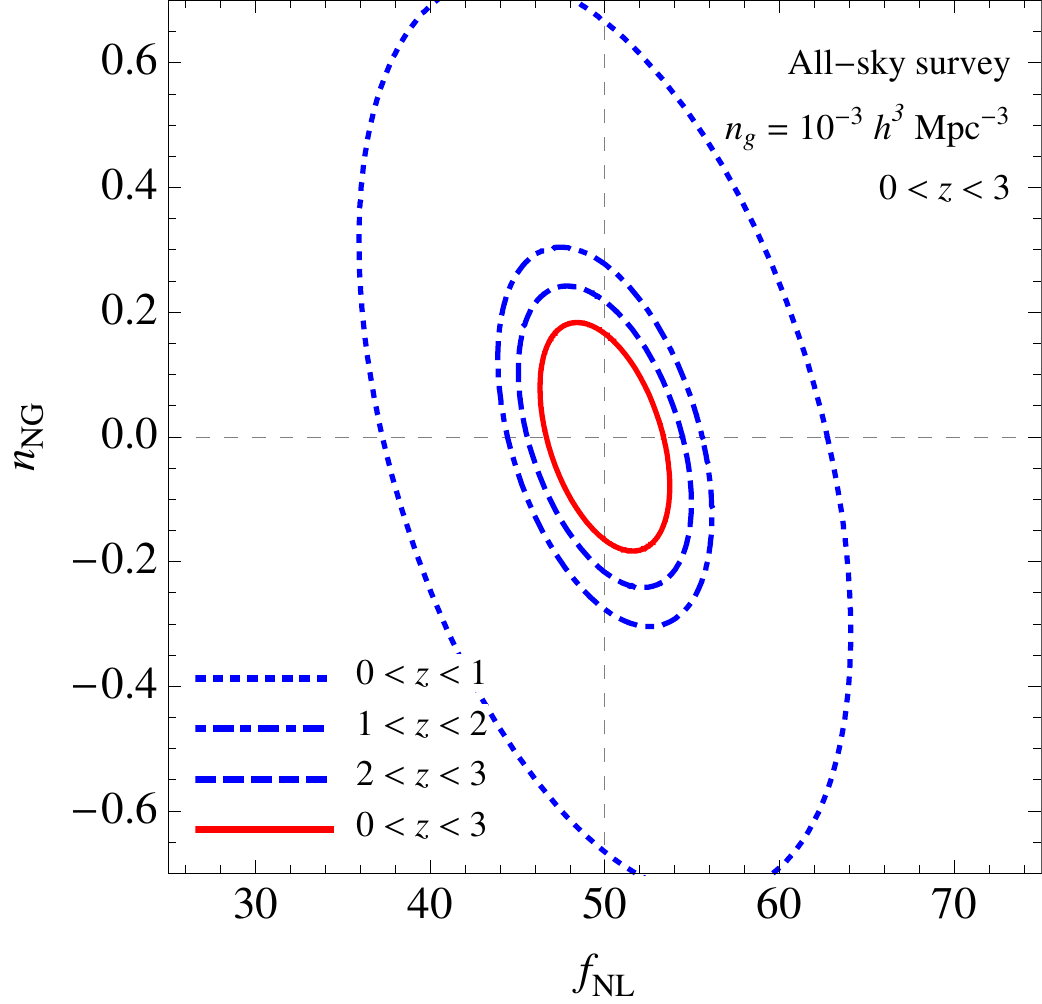}}
{\includegraphics[width=0.48\textwidth]{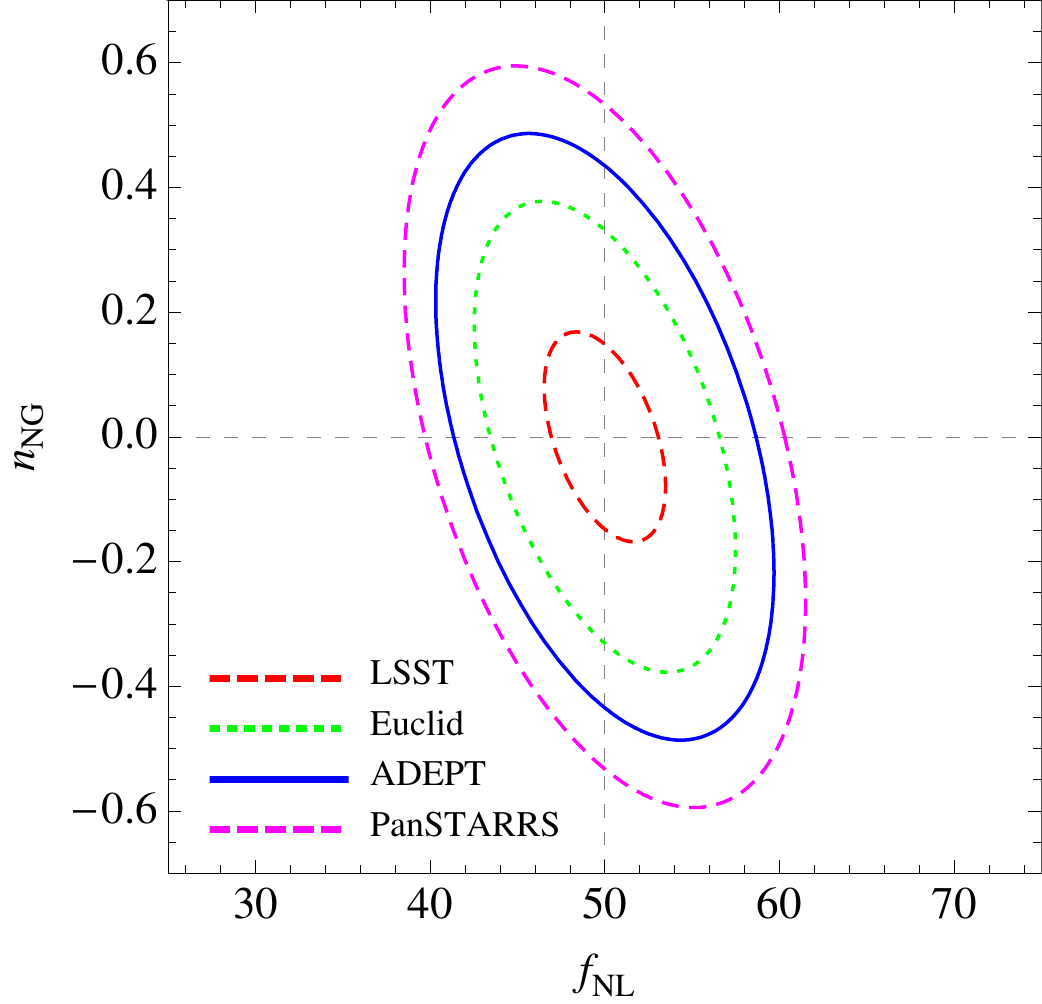}}
\caption{1-$\sigma$ contour plots for $\fnl$ and $\nng$ for the surveys considered for the local model from the power spectrum analysis. The left panel assumes an ideal all-sky survey from redshift $z=0$ to $z=3$ and galaxy number density $n_g=10^{-3}\icMpc$. The right panel refers to the survey characteristics specified in Table~\ref{tab:surveys}. Notice that the choice of redshift bins of size $\Delta z=0.5$ and fixed minimal scale corresponding to $k_{max}= 0.03\kMpc$ implies that the same range of scales is probed by all surveys, resulting in the same degree of degeneracy between the two non-Gaussian parameters $\fnl$ and $\nng$. }
\label{fig:LSSlocPS}
\end{center}
\end{figure}
In Fig.~\ref{fig:LSSlocPS} we show the 1-$\sigma$ contours of the uncertainty on the non-Gaussian parameters $\fnll$ and $\nng$ determined from the Fisher matrix analysis of the galaxy power spectrum assuming fiducial values $\fnll=50$ and $\nng=0$. On the left panel we consider an ideal all-sky survey with a galaxy number density of $10^{-3}\icMpc$ up to redshift $3$. We plot the contours corresponding separately to the redshift intervals $0<z<1$ ({\it dotted line}), $1<z<2$ ({\it dot-dashed}), $2<z<3$ ({\it dashed}) and  $0<z<3$ ({\it continuous}). On the right panel, we show the contours for a few proposed large-scale surveys. The unmarginalized errors on the amplitude parameter $\fnl$ are larger by a factor of a few with respect to the results of the similar analysis of \citep{CarboneVerdeMatarrese2008}, depending on the survey characteristic and specifically on the value of the linear Gaussian bias $b_1$. The reason for this is maily the inclusion in the variance for the power spectrum of the correction to the bias, since we are assuming a fiducial value for $\fnl$ different from zero. This reduces the signal-to-noise at the largest scale. We notice that, for a fiducial $\fnl=0$, we are able to recover the results of \citep{CarboneVerdeMatarrese2008} only replacing the factor $(b_1-1)$ in Eq.~(\ref{eq:dblocal2}) with $b_1$, as it seems to be assumed by Eq.~(9) defining the Fisher matrix in \citep{CarboneVerdeMatarrese2008}.

The direction of the degeneracy displayed by all contours reflects the choice for the pivot point $\kp$ and indicates that most of the signal is coming from scales larger than the pivot scale. In fact, we remind the reader that the Fisher matrix analysis is limited to $0.001\Mpc < k < 0.03\Mpc$ by choice and this results in the quite uniform results for the different surveys in the left panel of Fig.~\ref{fig:LSSlocPS}. In other words we are conservatively assuming no redshift evolution for the maximum value of $k$ in the Fisher matrix analysis. This appears to be confirmed by $N$-body simulations on the range $0<z<2$ in \citep{DesjacquesSeljakIliev2008,PillepichPorcianiHahn2008}.

It is not straightforward to derive a simple prescription for the expected uncertainties as a function of redshift and sky-coverage. Our simple analysis shows that high-redshift surveys ($z>1$) covering a large fraction of the sky corresponding to a volume of about $100\cGpc$ might provide a 1-$\sigma$ error on the running parameter of the order of $\Delta\nng\simeq 0.4(50/\fnll)$.

%%%%%%%%%%%%%%%%%%%%%%%%%%%%%%%%%%%%%%%%%%%%%%%%%%%%%%%%%%%%%%%%%%%%%%%%%%%%%%%%
\subsection{Constraints on equilateral non-Gaussianity from the galaxy bispectrum}
\label{ssec:eqBisp}

As shown by \citep{TaruyaKoyamaMatsubara2008}, the large correction to the galaxy power spectrum given by Eq.~(\ref{eq:dbTKM}) for the local model is essentially negligible for the equilateral one. On the other hand, higher-order correlation functions of the galaxy distribution are expected to present an initial component when primordial perturbations are non-Gaussian. Previous analyses indicated that while current galaxy surveys are not yet competitive with CMB observations, in the future they could in principle provide an important confirmation of a possible CMB detection, \citep{VerdeEtal2000,ScoccimarroSefusattiZaldarriaga2004,SefusattiKomatsu2007}. These results, however, were based on the tree-level expression for the galaxy bispectrum in perturbation theory, which assumes the non-Gaussian contribution to the galaxy bispectrum to be given just by the primordial component to the matter bispectrum. In a couple of recent works, \citet{JeongKomatsu2009} and \citet{Sefusatti2009} have shown that non-linearities in the galaxy bias relation can induce significant 1-loop corrections, corresponding to the leading contributions at large scales. In particular, such corrections depend on the effects of non-Gaussianity on the matter trispectrum. The effect on the galaxy, or halo, bispectrum is therefore similar to that on the power spectrum, where large-scale non-Gaussian corrections are due to the matter bispectrum. 

These preliminary results still require proper comparisons with $N$-body simulations before being able to provide accurate predictions. For this reason, we make the very conservative choice to assume that the galaxy bispectrum is described by its tree-level expression in perturbation theory, where the only non-Gaussian effect is given by the initial matter bispectrum. In fact, it is reasonable to expect that the corrections studied in \citep{JeongKomatsu2009,Sefusatti2009}--which we neglect---would significantly improve the constraints that we can place on primordial non-Gaussianity once reliable predictions for the bias parameters, derived for instance from numerical simulations, are given.
 
 \begin{figure}[t]
\begin{center}
{\includegraphics[width=0.48\textwidth]{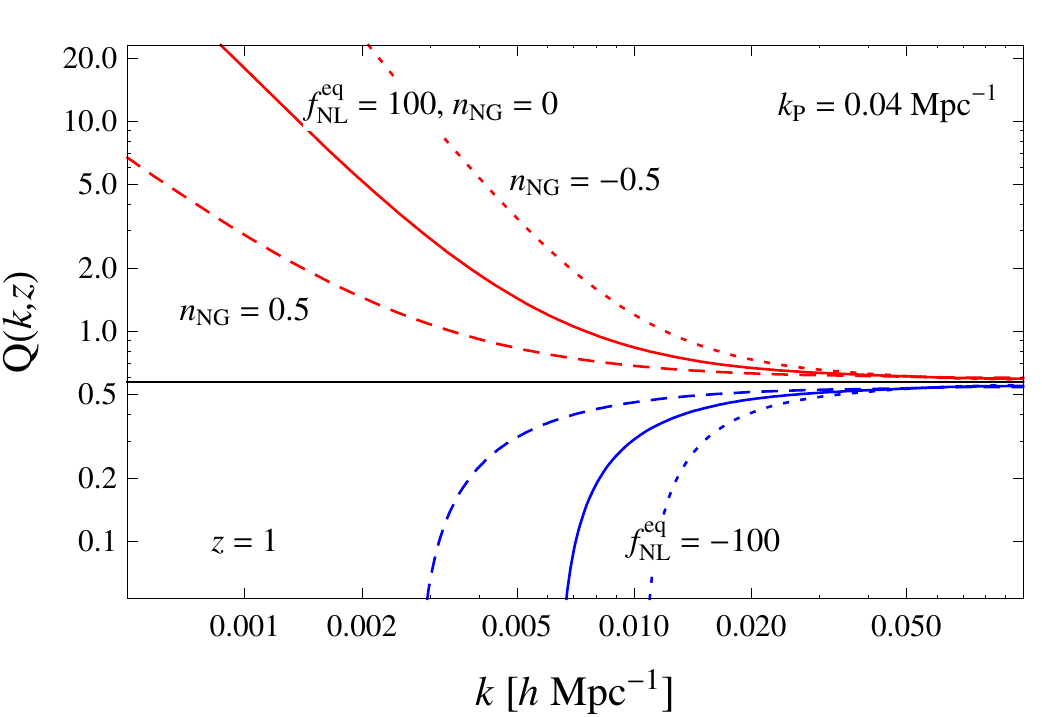}}
{\includegraphics[width=0.47\textwidth]{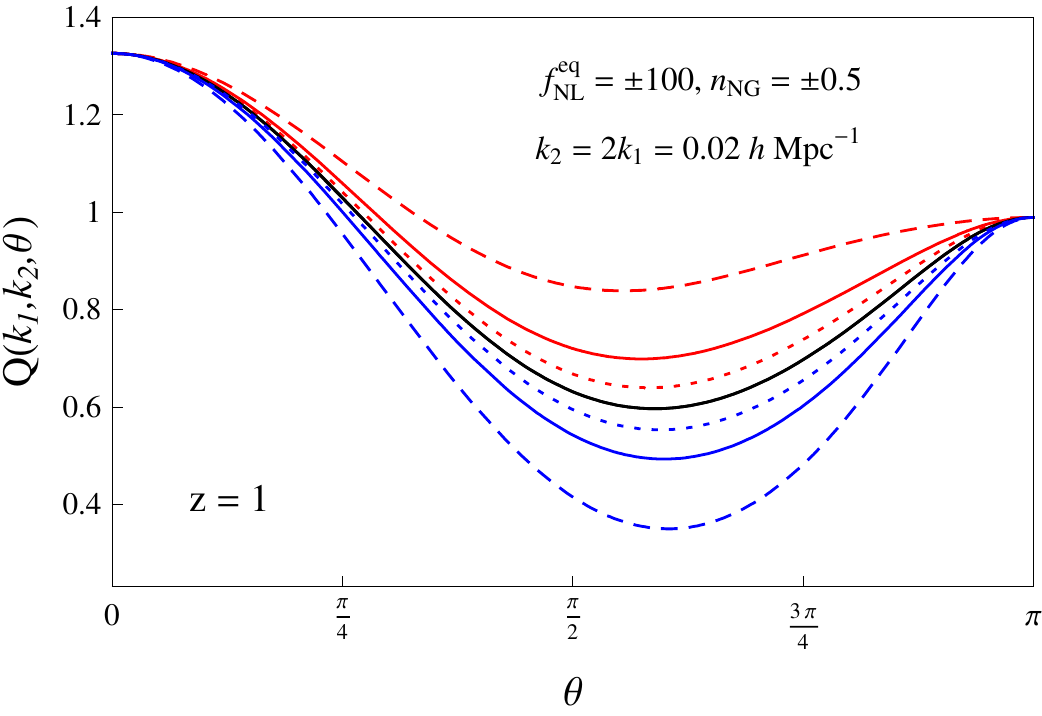}}
\caption{Effect of a running non-Gaussian component on the matter bispectrum. {\it Left panel:} reduced matter bispectrum for equilateral configurations, $Q(k)$, as a function of the wavenumber $k$. {\it Right panel:} reduced matter bispectrum as a function of the angle $\theta$ between two fixed wavenumbers $k_1=0.01\kMpc$ and $k_2=2 k_1$. In both cases, the black continuous line indicates the Gaussian component while the red and blue continuous lines correspond to the full bispectrum including the non-Gaussian component for $\fnl=+100$ and $\fnl-100$, respectively, at redshift $z=1$. In addition, the dashed lines correspond to $\nng=+0.5$, dotted lines to $\nng=-0.5$ for both $\fnl=\pm 100$.}
\label{fig:bisp}
\end{center}
\end{figure}
We will therefore consider, for our Fisher matrix analysis, the reduced galaxy bispectrum $Q_g$ given by
\bea
Q_g(k_1,k_2,k_3)&\equiv&\frac{B_g(k_1,k_2,k_3)}{P_g(k_1)P_g(k_2)+{\rm cyc.}}
\nonumber\\
&=&\frac{1}{b_1}\left[Q_G(k_1,k_2,k_3)+Q_I(k_1,k_2,k_3)\right]+\frac{b_2}{b_1^2},
\eea
where $Q_G$ represents the reduced matter bispectrum induced by gravitational non-linearities, evaluated at second order in perturbation theory, $Q_I$ represents the reduced matter bispectrum due to non-Gaussian initial conditions and where the only effect of non-linear, quadratic bias is given by the constant term $b_2/b_1$.  

In Figure~\ref{fig:bisp} we show the effect on the tree-level reduced matter bispectrum of equilateral non-Gaussian initial conditions. On the left panel the reduced bispectrum for equilateral configurations is plotted as a function of  $k=k_1=k_2=k_3$. The upper continuous curve corresponds to $\fnle=100$ with $\nng=0$, while the dashed and dotted curves corresponds to $\nng=0.5$ and $\nng=-0.5$ respectively. The horizontal continuous curve represents the tree-level prediction for Gaussian initial conditions. The lower curves are for $\fnle=-100$ with $\nng=0$ and $\pm 0.5$. On the left panel we show the reduced bispectrum for configurations with sides $k_1=0.01\kMpc$ and $k_2=0.02\kMpc$ as a function of the angle between the vectors $\kv_1$ and $\kv_2$. One can notice how the largest non-Gaussian corrections correspond to triangular configurations close to equilateral, while they vanish for squeezed configurations with $\theta=0$ and $\theta=\pi$ as expected for the equilateral model.

For the details of the Fisher matrix analysis we refer the reader to Section III of \citep{SefusattiKomatsu2007}, since our calculation simply adds the extra running parameter and assumes a slightly different cosmology. We remind the reader, however, that unlike the case of the power spectrum, we consider here, for a given survey, all measurable triangular configurations with sides in the range defined by the fundamental frequency $k_f=2\pi/V^{1/3}$ and a maximum wavenumber $k_{max}$ identified by the non-linear scale which therefore increases with the median redshift of the survey or the specific survey redshift bin. This implies that different geometries and redshifts will probe different ranges in scales resulting in different degeneracies between the two non-Gaussian parameters $\fnle$ and $\nng$ for our choice of the pivot point $\kp=0.04$ Mpc$^{-1}$. 

\begin{figure}[t]
\begin{center}
{\includegraphics[width=0.48\textwidth]{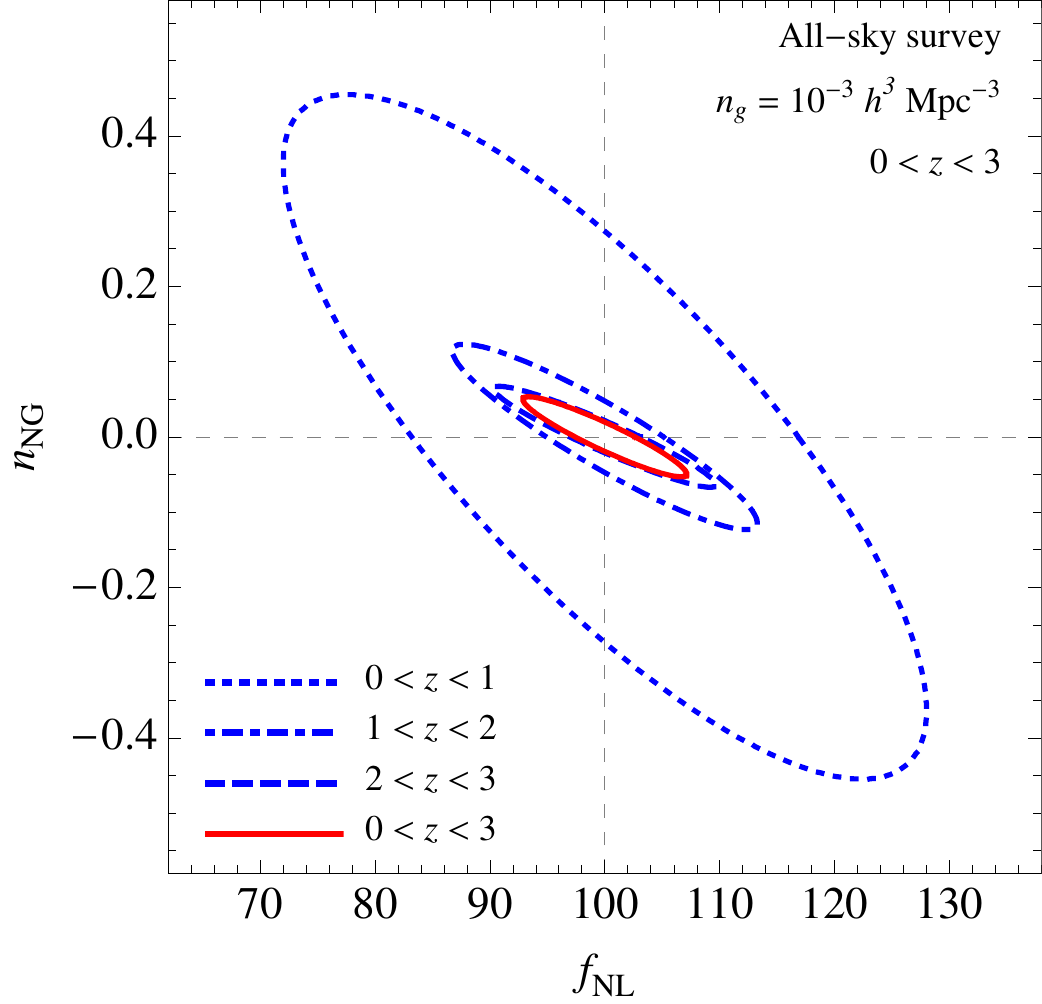}}
{\includegraphics[width=0.48\textwidth]{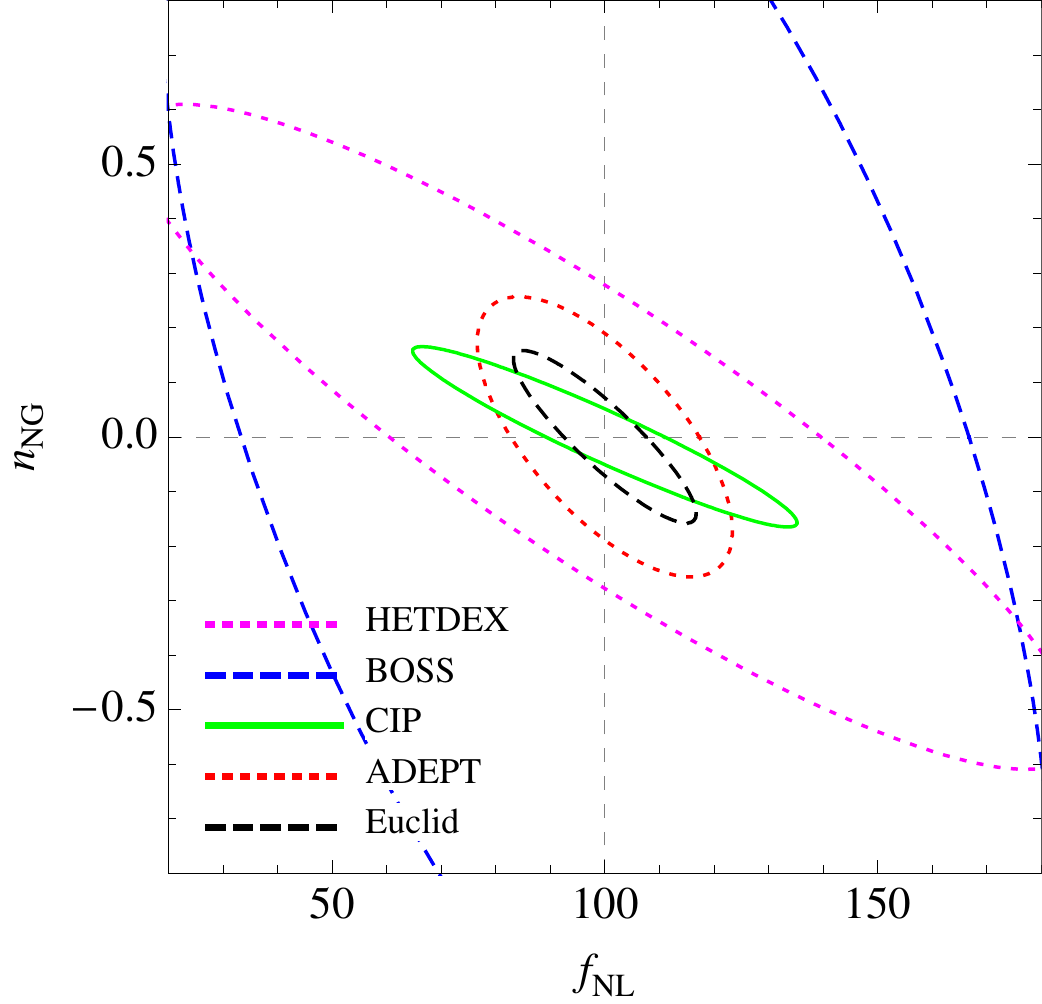}}
\caption{1-$\sigma$ contour plots for $\fnle$ and $\nng$ for the equilateral model from the galaxy bispectrum analysis. {\it Left panel}: ideal all-sky survey up to redshift $z=3$ in redshift intervals of size $\Delta z=1$ and galaxy number density $n_g=10^{-3}\icMpc$. {\it Right panel}: expectations for the surveys BOSS ({\it outer dashed line}), HETDEX ({\it outer dotted line}), CIP ({\it continuous line}), ADEPT ({\it inner dotted line}) and Euclid ({\it inner dashed line}). Note that the degeneracy between the two parameters depends on the choice of $k_{max}$ for each survey, as explained in the text.}
\label{fig:LSSeqBS}
\end{center}
\end{figure}
As for the power spectrum analysis of Section~\ref{ssec:localHB}, we assume the bias parameters to be known and given by the halo model prediction plus a prescription for the halo occupation distribution (HOD) depending on the expected galaxy number density. While this might seem to be a strong assumption, we remark that an equally strong correlation between the linear and quadratic bias parameters $b_1$ and $b_2$ is expected in the halo model framework and observed in $N$-body simulations and observations, see \citep{NishimichiEtal2007} and references therein. As shown in \citet{SefusattiKomatsu2007}, if the halo model prediction for the halo bias functions, which can be measured in numerical simulations is assumed then the uncertainty on the non-Gaussian parameter $\fnle$ marginalized over the HOD parameters is very close to the unmarginalized ones. A marginalization over the galaxy bias parameters $b_1$ and $b_2$, in other words, would be an unduly conservative approach as it would introduce essentially unphysical degeneracies.

In Fig.~\ref{fig:LSSeqBS} we show the results of the Fisher matrix analysis of the galaxy bispectrum in terms of the non-Gaussian parameters $\fnle$ and $\nng$. The left panel shows the 1-$\sigma$ contours for an ideal all-sky survey up to redshift $z=3$ in redshift intervals of size $\Delta z=1$ and galaxy number density $n_g=10^{-3}\icMpc$. Each interval assumes redshift bins of size $\Delta z=0.5$ with all quantities evaluated at the mean redshift of the bin and we ignore cross-correlations between different bins. In the right panels we instead show the results corresponding to the spectroscopic surveys of Table~\ref{tab:surveys}, where the marginalized constraints on $\fnle$ and $\nng$ are given. 

It is interesting to notice how future redshift surveys such as Euclid will be able to place constraints on the running parameter of the order of $\Delta \nng\simeq 0.3$ at $95\%$ C.L., if a significant non-Gaussian component of $\fnle\simeq 100$ is detected. On the other hand, high-redshift surveys will be able to probe smaller scales and provide complementary information as shown by the different degeneracy in the $\fnl$-$\nng$ plane for experiments such as HETDEX and CIP. We remind the reader that these simple results do not take into account the large-scale corrections to the galaxy bispectrum that come from the non-Gaussian effects on the matter trispectrum described in \citep{JeongKomatsu2009,Sefusatti2009} and that we expect to improve the constraints on both non-Gaussian parameters $\fnl$ and $\nng$. A proper and {\it complete} assessment of the potential of LSS observations, including a joint analysis of power spectrum and bispectrum, requires a deeper and more quantitative understanding of effects which have just began to be studied and must necessarily wait for future work.

%%%%%%%%%%%%%%%%%%%%%%%%%%%%%%%%%%%%%%%%%%%%%%%%%%%%%%%%%%%%%%%%%%%%%%%%%%%%%%%%
%%%%%%%%%%%%%%%%%%%%%%%%%%%%%%%%%%%%%%%%%%%%%%%%%%%%%%%%%%%%%%%%%%%%%%%%%%%%%%%%
\section{Constraints from joint CMB and LSS observations}
\label{results}

In this Section we consider the combination of the results from the CMB analysis of Section \ref{sec:cmb} with the results from the analysis of large-scale structure observables of the previous Section, both for the local and the equilateral model. In Table~\ref{tab:CMBLSS} we show the expected $1$-$\sigma$ constraints on both non-Gaussian parameters $\fnl$ and $\nng$ for the local and equilateral model for all CMB and LSS experiments previously considered. We also compute the joint constraints between CMB and LSS for experiments providing results of the same order of magnitude. All errors are marginalized over the other non-Gaussian parameters while we assume cosmological and bias parameters as known. All errors also assume the pivot point $k_p=0.04$ Mpc$^{-1}$. In order to give a quantitative idea of how much the degeneracy between $\fnl$ and $\nng$---which strongly depends on the choice of $k_p$---is affecting the results, we include in parenthesis the corresponding unmarginalized errors.
\begin{table*}[b]
\caption{\label{tab:CMBLSS} 
Expected $1$-$\sigma$ uncertainties on the amplitude and running parameters $\fnl$ and $\nng$ from the Fisher matrix analysis of the CMB bispectrum, the {\it galaxy power spectrum for the local model} with fiducial values $\fnll=50$ and $\nng=0$ and from the Fisher matrix analysis of the {\it galaxy bispectrum} (assuming {\it only} the primordial non-Gaussian component, see text) {\it for the equilateral model} with fiducial values $\fnle=100$ and $\nng=0$, together with the corresponding joint constraints from CMB and large-scale structure observations. Uncertainties are marginalized over the other non-Gaussian parameter assuming the pivot point $k_p=0.04$ Mpc$^{-1}$, while cosmology and bias factors are assumed as known. In parentheses we also provide the corresponding {\it unmarginalized} values. Joint constraints are shown only for the combinations which provide a sensible improvement over the CMB results.}
\begin{ruledtabular}
\begin{tabular}{l||cccc|cccc||cccc|cccc}
   & \multicolumn{4}{c}{$\Delta\fnll$} & \multicolumn{4}{c}{$\Delta\nng$\footnote{Assumes a fiducial $\fnll=50$.}} & \multicolumn{4}{c}{$\Delta\fnle$} & \multicolumn{4}{c}{$\Delta\nng$\footnote{Assumes a fiducial $\fnle=100$.}}\\
\hline\hline
\multicolumn{17}{l}{CMB} \\
\hline
WMAP        & & $58.0$ & ($20.0$) & & & $0.68$ & ($0.24$) & & & $196$ & ($88$) & & & $1.1$ & ($0.47$) &  \\
Planck      & & $4.7$ & ($3.1$) & & & $0.10$ & ($0.07$) & & & $30$ & ($29$) & & & $0.30$ & ($0.29$) &  \\
CMBPol      & & $1.6$ & ($1.4$) & & & $0.05$ & ($0.04$) & & & $15$ & ($15$) & & & $0.17$ & ($0.16$) &\\
\hline\hline
\multicolumn{17}{l}{LSS, {\it spectroscopic}} \\
\hline
BOSS      & & $52$ & ($40$) & & & $2.6$ & ($2.0$) & & & $75$ & ($57$) & & & $1.5$ & ($1.1$) &\\
ADEPT     & & $9.7$ & ($8.7$) & & & $0.49$ & ($0.43$) & & & $22$ & ($12$) & & & $0.25$ & ($0.14$) &\\
Euclid    & & $8.4$ & ($7.3$) & & & $0.42$ & ($0.37$) & & & $16.7$ & ($6.7$) & & & $0.16$ & ($0.06$) &\\
HETDEX    & & $96$ & ($48$) & & & $3.8$ & ($1.9$) & & & $86$ & ($32$) & & & $0.60$ & ($0.22$) &\\
CIP       & & $61$ & ($30$) & & & $2.3$ & ($1.1$) & & & $35.2$ & ($8.9$) & & & $0.16$ & ($0.04$) &\\
\hline
\multicolumn{17}{l}{LSS, {\it photometric}} \\
\hline
%PAU       & & $34$  & ($26$) & & & $1.7$ & ($1.3$) & & & - & - & & & - & - &\\
%DES       & & $34$  & ($26$) & & & $1.7$ & ($1.3$) & & & - & - & & & - & - &\\
LSST      & & $3.5$ & ($3.1$) & & & $0.17$ & ($0.15$) & & & - & - & & & - & - &\\
PanSTARRS & & $11$  & ($10$) & & & $0.59$ & ($0.53$) & & & - & - & & & - & - &\\
\hline\hline
\multicolumn{17}{l}{CMB+LSS }\\
\hline
Planck + HETDEX  & & - & - & & & - & - & & & $24$ & ($23$) & & & $0.20$ & ($0.20$) &\\ 
 Planck + ADEPT   & & - & - & & & - & - & & & $16$ & ($15$) & & & $0.17$ & ($0.16$) &\\  
Planck + Euclid  & & $3.6$ & ($2.8$) & & & $0.09$ & ($0.07$) & & & $12.7$ & ($7.3$) & & & $0.12$ & ($0.07$)  &\\
Planck + LSST    & & $2.3$ & ($2.2$) & & & $0.07$ & ($0.06$) & & & - & - & & & - & - &\\ 
\hline
CMBPol + ADEPT   & & - & - & & & - & - & & & $12$ & ($11$) & & & $0.13$ & ($0.12$) &\\
CMBPol + Euclid  & & - & - & & & - & - & & & $10.6$ & ($6.7$) & & & $0.10$ & ($0.07$) &\\
CMBPol + Euclid + CIP & & - & - & & & - & - & & & $9.5$ & ($5.7$) & & & $0.07$ & ($0.04$) &\\
CMBPol + LSST    & & $1.4$ & ($1.3$) & & & $0.04$ & ($0.04$) & & & - & - & & & - & - &
\end{tabular}
\end{ruledtabular}
\end{table*}

As a general remark, we expect large-volume and high-redshift galaxy surveys such as ADEPT, Euclid and LSST to provide constraints on the amplitude, as well as running of primordial non-Gaussianity comparable or even better than the constraints we expect from a CMB mission like Planck. In fact, from our simple analysis, we can expect a sensible improvement even over the results of the proposed CMBPol mission. 

\begin{figure}[t]
\begin{center}
%{\includegraphics[width=0.48\textwidth]{fig_CMB+LSS_LcPS_Allsky_z0to5_kP0p04}}
{\includegraphics[width=0.48\textwidth]{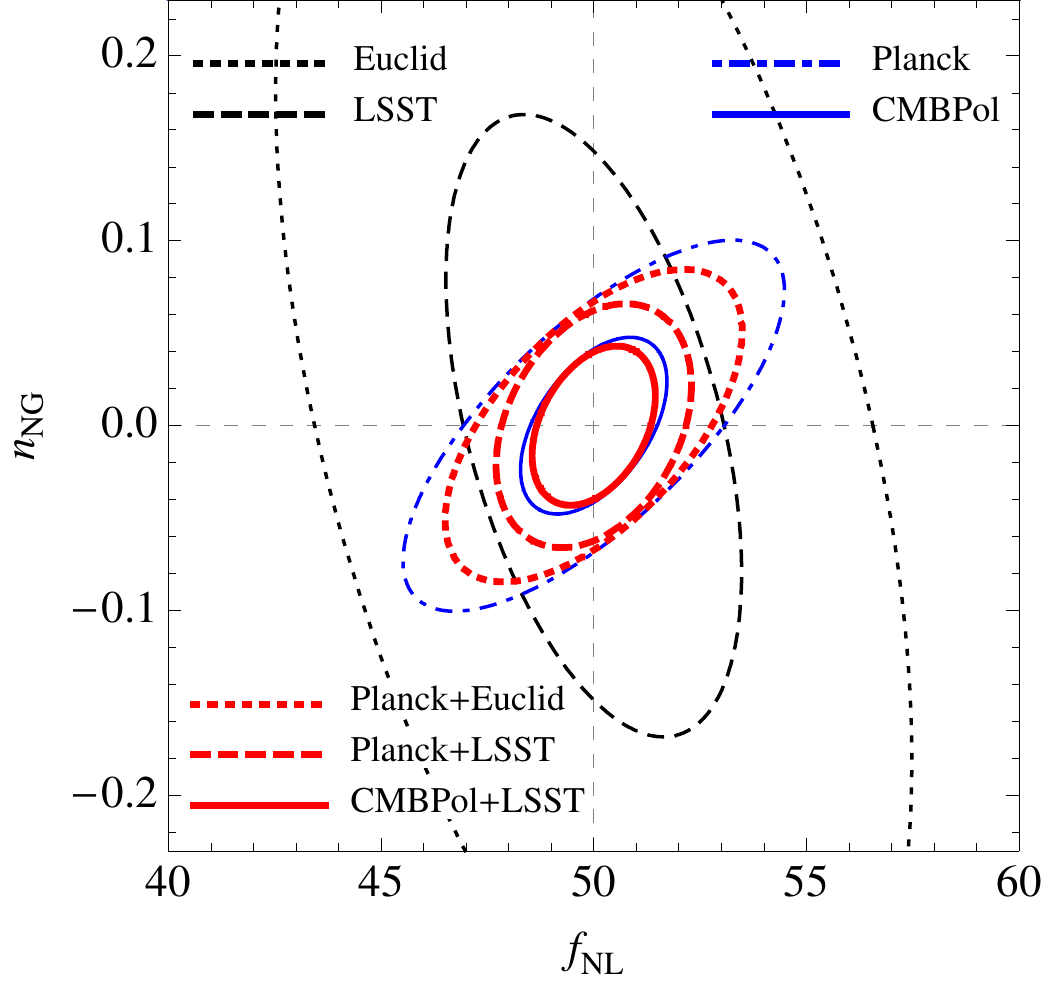}}
\caption{1-$\sigma$ contour plots for $\fnl$ and $\nng$ for the local model for the Planck ({\it dot-dashed blue}) and CMBPol ({\it continuous, blue}) experiments compared to the LSS surveys Euclid ({\it dotted, black}) and LSST ({\it dashed, black}) together with the combined contours given by Planck and Euclid ({\it thick, dotted, red}), Planck and LSST ({\it thick, dashed, red}) and by CMBPol and LSST ({\it thick, continuous, red}).}
\label{fig:CMBLSSLc}
\end{center}
\end{figure}

In Fig.~\ref{fig:CMBLSSLc} we show the 1-$\sigma$ uncertainties contours from the CMB bispectrum and LSS power spectrum analysis together with the combined constraints for the local model assuming the fiducial values $\fnll=50$ and $\nng=0$. In particular, we compare the CMB experiments Planck and CMBPol with the LSS surveys Euclid and LSST together with the specific combinations of Planck with Euclid and LSST and CMBPol with LSST. The limits provided by Planck are comparable and complementary to those of LSST, the best large-scale structure probe in our choice. CMBPol would instead provide constraints quite close to the ideal case, only slightly improved by power spectrum measurements. We notice that although the choice of the pivot is not optimal for neither the CMB or the LSS observations as both individually show a significant degeneracy between the non-Gaussian parameters, in the combined constraint, such degeneracy is quite reduced.  In particular, this shows the complementarity of the different probes when a running of $\fnll$ is considered. Notice that in the case of Planck, and even more so for CMBPol, such complementarity is a consequence of the different distribution of the non-Gaussian signal, for the specific local model, over the relevant range of scales, since the two observables are essentially probing the same range. Clearly the galaxy power spectrum, under the simple assumptions of our analysis provides a marginal improvement to the constraints from the ideal CMB results. It is reasonable to expect, however, that comparable or better constraints on local non-Gaussianity can be derived from an analysis of the galaxy bispectrum in light of the recent results of \citep{JeongKomatsu2009,Sefusatti2009}, which we did not consider here but that will be studied in future works.

\begin{figure}[t]
\begin{center}
{\includegraphics[width=0.48\textwidth]{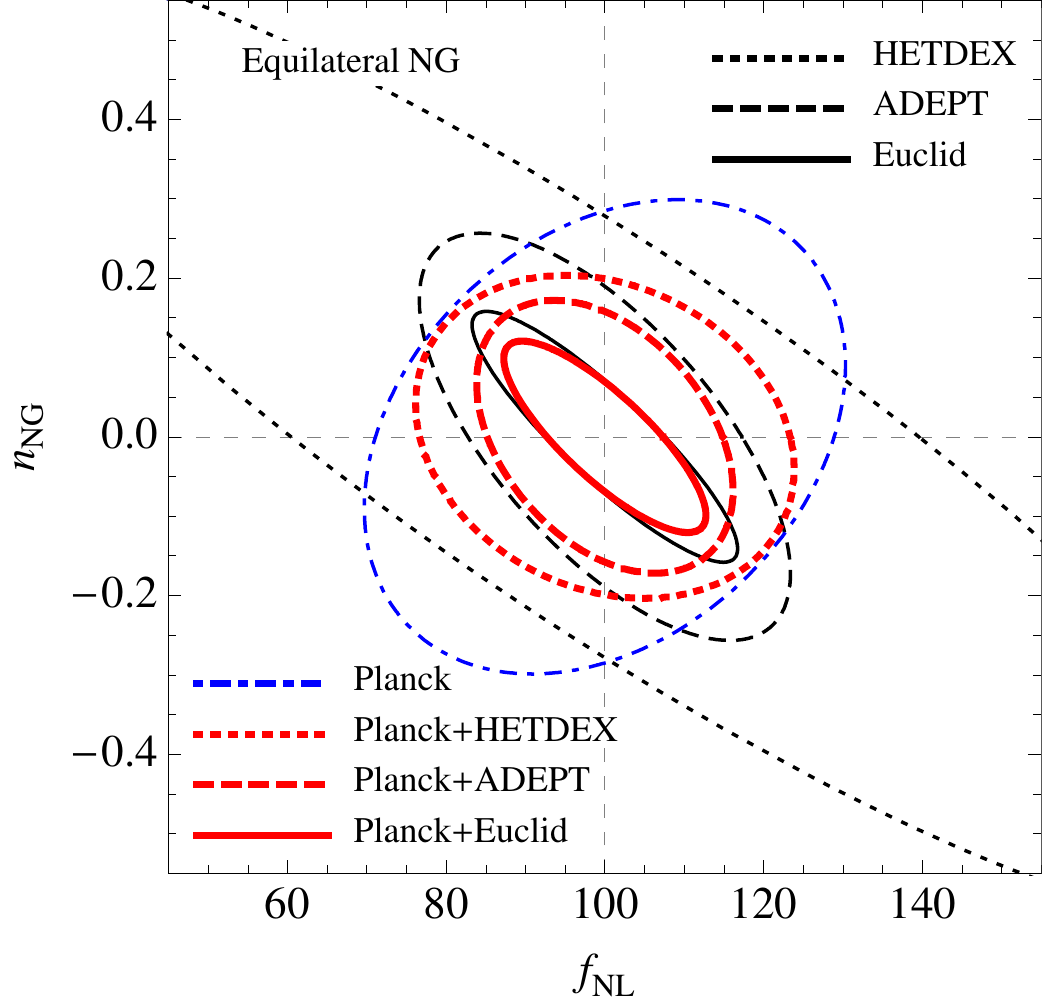}}
{\includegraphics[width=0.48\textwidth]{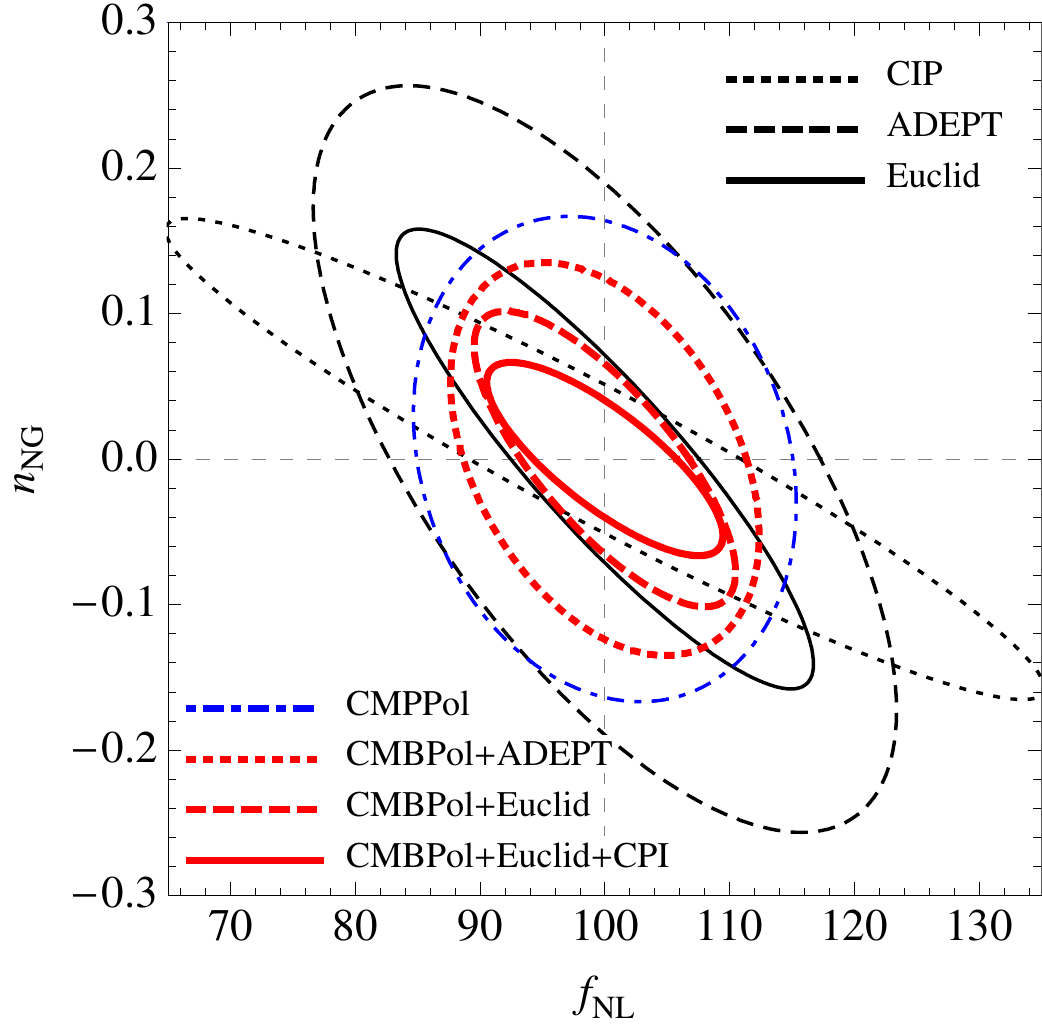}}
\caption{1-$\sigma$ contour plots for $\fnl$ and $\nng$ for the equilateral model from CMB and LSS measurements. {\it Left panel}: Planck ({\it dot-dashed, blue}) experiment compared to the LSS surveys HETDEX ({\it dotted, black}), ADEPT ({\it dashed, black}) and Euclid ({\it continuous, black}) together with the combined contours given by Planck with HETDEX ({\it thick, dotted, red}), with ADEPT ({\it thick, dashed, red}) and with Euclid ({\it thick, continuous, red}). {\it Right panel}: CMBPol ({\it dot-dashed, blue}) experiment compared to the LSS surveys CIP ({\it dotted, black}), ADEPT ({\it dashed, black}) and Euclid ({\it continuous, black}) together with the combined contours given by CMBPol with ADEPT ({\it thick, dotted, red}), with Euclid ({\it thick, dashed, red}) and with Euclid {\it and} CIP ({\it thick, continuous, red}).}
\label{fig:CMBLSSEq}
\end{center}
\end{figure}

In Fig.~\ref{fig:CMBLSSEq} we show the 1-$\sigma$ uncertainties contours from the CMB and LSS bispectrum analysis together with the combined constraints for the equilateral model assuming the fiducial values $\fnle=100$ and $\nng=0$. The left panel shows the Planck experiments compared to the redshift surveys HETDEX, ADEPT and Euclid, together with their combinations. The right panel shows instead the CMBPol against the surveys ADEPT, CIP and Euclid, together with their combined contours. In the case of the equilateral non-Gaussianity, future and proposed CMB and LSS experiments are expected to provide comparable constraints. In addition, they also tend to provide similar degeneracies between $\fnl$ and $\nng$. 

This analysis essentially extends to the running parameter $\nng$ the analysis of \citet{SefusattiKomatsu2007} on the constraints on the amplitude of non-Gaussianities from the galaxy bispectrum in high-redshift surveys. Again, we should keep in mind that such analysis is quite conservative since we are neglecting large-scale non-Gaussian corrections due to non-linear bias \citep{JeongKomatsu2009,Sefusatti2009}.

%%%%%%%%%%%%%%%%%%%%%%%%%%%%%%%%%%%%%%%%%%%%%%%%%%%%%%%%%%%%%%%%%%%%%%%%%%%%%%%%
%%%%%%%%%%%%%%%%%%%%%%%%%%%%%%%%%%%%%%%%%%%%%%%%%%%%%%%%%%%%%%%%%%%%%%%%%%%%%%%%
\section{DBI Inflation: a case study}
\label{sec:dbi}

In this Section we present an example of how our analysis can translate into constraints of a specific model of inflation. As a case study we choose models in which the kinetic term is of the Dirac-Born-Infeld (DBI) type instead of the canonical one. One of the virtues of such models is that they arise naturally in string theory \cite{SilversteinTong2004} when inflation is driven by the motion of a D-brane \cite{DvaliTye1999}. Models of DBI inflation are particularly interesting in the present context because they typically lead to large scale-dependent non-Gaussianity in the primordial spectrum of perturbations. We will first review the basics of DBI inflation and then possible concrete realizations.

The dynamics of a D-brane in string theory is governed by the DBI action \cite{Polchinski1995}. Consider a compactification from ten to four dimensions with a generic warp factor. For a spacetime-filling D3-brane the four-dimensional low-energy effective action is given by\footnote{Our convention for the metric signature is $(-,+,+,+)$.}
\be 
\label{dbi}
S_{DBI} = - \int d^4 x \sqrt{-g} \left[ f(\phi)^{-1} \sqrt{1 + f(\phi) \partial_{\mu} \phi\partial^{\mu} \phi } - f(\phi)^{-1} + V(\phi) \right]\,,
\ee
where $f(\phi)$ is proportional to the warp factor at the position of the D3-brane in the compact space. The position is given by six scalar fields; in general this would lead to a multi-field inflationary model.  In the explicit realizations we will consider, we assume that the brane moves along a single direction parameterized by the single scalar field $\phi$.

Homogeneous solutions for the above action coupled to gravity can be found \cite{AlishahihaEtal2004,SilversteinTong2004} using the Hamilton-Jacobi formalism. One finds an upper bound for the velocity of the inflaton, $\dot\phi\leq f(\phi)^{-1/2}$. In the presence of strong warping, $f(\phi)$ can become very large\footnote{$f(\phi)$ has the dimension of an inverse energy density. Here ``large'' means $f m_s^4\gg1$, where $m_s$ is the string scale.} leading to a slow evolution that can support a prolonged stage of inflation in the presence of a slow-roll steep potential $V(\phi)$. In analogy with special relativity, it is convenient to introduce the function
\be 
\gamma\equiv \left[1-f(\phi)\dot\phi^2\right]^{-1/2} \,.
\ee
For flat potentials and small warping, $\gamma\simeq 1$, the dynamics of the system is the same as in slow-roll inflation. On the other hand, for steep potentials, {\it i.e.}~a large slow-roll parameter $\eta \gtrsim 1$, and strong warping we get $\gamma\gg 1$. In this ``relativistic'' regime all higher-derivative terms in the expansion of the square root in Eq.~\ref{dbi} become important. The speed of sound of $\phi$-perturbations becomes very small, $c_s=\gamma^{-1}$. As first pointed out in \cite{AlishahihaEtal2004} and calculated in detail in \cite{Chen2005,ChenEtal2007}, this can induce large non-Gaussianity in the primordial spectrum of density perturbations. The largest contribution comes from the equilateral configurations and can be estimated by\footnote{In \cite{AlishahihaEtal2004,Chen2005,ChenEtal2007}, the sign convention of Maldacena \cite{Maldacena2003} is used which is opposite to the WMAP convention. We adopt the latter, which explains the minus sign in Eq.~\ref{e:NG}.}
\be\label{e:NG}
f_{NL}\simeq -0.32 \,\gamma^2=-0.32\,c_s^{-2}\,.
\ee
The running is straightforwardly calculated and is given by
\be\label{e:run}
n_{NG}\equiv\frac{\partial \ln |f_{NL}|}{\partial \ln k}\simeq -2 \frac{\dot c_s}{c_s H}\equiv -2 s\,,
\ee
where we defined $s$ as the dimensionless time variation of the speed of sound\footnote{This is in analogy with the slow-roll parameter $\eta\equiv \dot \epsilon/(\epsilon H)$}. 

The actual value of $f_{NL}$ and $n_{NG}$ depends on the model and in particular on the functions $V(\phi)$ and $f(\phi)$ in Eq.~\ref{dbi}. These in turn depend on how the DBI inflation idea is specifically realized: in particular one has to specify how all the moduli are stabilized and the compact geometry in which the D3-brane moves. At present, no explicit realization of DBI inflation in string theory has been found that is consistent with the current CMB data. Rather than consider this to be a failure of the DBI inflation idea, we feel this reflects our limited knowledge of both moduli stabilization and the geometry of (compact) Calabi-Yau manifolds. 

In light of the above consideration, we deem it most sensible to consider effective models which are inspired by DBI inflation with phenomenological but realistic choices of $f(\phi)$ and $V(\phi)$. The hope is that once a feasible embedding of DBI inflation in string theory is found, it might be reasonably approximated by the effective models that we analyze here. The string theory realizations proposed until now that have most closely reproduced existent data can be divided in three broad classes: UV models with $\dot f[\phi(t)]>0$, IR models with $\dot f[\phi(t)]<0$ and angular models with $\dot f[\phi(t)]=0$. In the following, we describe each class and provide an archetypal choice of $f(\phi)$ and $V(\phi)$. The resulting non-Gaussian parameters are plotted in Fig \ref{fig:dbi} together with the constraints from the data.

%%%%%%%%%%%%%%%%%%%%%%%%%%%%%%%%%%%%%%%%%%%%%%%%%%%%%%%%%%%%%%%%%%%%%%%%%%%%%%%%
\subsection{The UV model}

In a broad class of models (initiated in \cite{KachruEtal2003}), the inflating D3-brane starts somewhere in the bulk of the compact dimensions and falls inside a region with increasingly stronger warping, referred to as the throat.  As mentioned previously, we assume that the brane moves exclusively along the \textit{radial} direction of the throat that we parameterize with $\phi$. We choose as zero for $\phi$ the tip of the throat where the warping is maximal. Inflation starts in the UV region (small warping) of the throat and ends somewhere close to the tip. Models in this class share the feature that $f(\phi)$ is monotonically increasing with time, as $\phi$ rolls towards zero. 

In the best-studied examples \cite{KlebanovStrassler2000,GiddingsKachruPolchinski2002}, the throat is a conifold \cite{CandelasDelaOssa1990}. The resulting warp factor depends on how and where the singularity is cut off. A toy model that captures the features of a realistic solution ({\it e.g.}~\cite{KlebanovStrassler2000}) is given by $f(\phi)\simeq\lambda/(\phi^2+\mu^2)^2$, where $\lambda$ is a dimensionless parameter proportional to the D3-brane tension and $\mu$ determines the IR cut-off. Notice that the warp factor depends only on the radial and not on the angular position. Hence, a model of so-called angular inflation then takes place at constant $f$. We analyze this case in Section \ref{angular}.
 
As regards the potential, a series of detailed analyses \cite{KachruEtal2003,BurgessEtal2007,KrausePajer2008,BaumannEtal2008,BaumannEtal2009} have confirmed that generically a mass term $m^2 \phi^2$ is present, with $m$ of the order of the Hubble scale. This fact constitutes a serious problem for slow-roll inflation ($\eta\sim \mathcal{O}(1)$), but not for DBI inflation that proceeds with arbitrarily steep potentials. Other $\phi$-dependent terms can of course be present, but they will be subleading in the small-$\phi$ limit.

In \cite{BeanEtal2007} the simplest realistic UV model of DBI inflation was compared with the data.  It was found that no model exists that obeys all the constraints imposed by the internal consistency of string theory \textit{and} that is compatible with the data. In \cite{PeirisEtal2007}, the analysis was generalized to include other plausible $f(\phi)$ and $V(\phi)$, but resulted in the same conclusion.  As pointed out in \cite{BaumannMcAllister2007}, the reason for these negative results can be drawn back to the existence of a geometrical limit for the allowed range of $\phi$. The upper bound is given by $(\Delta \phi/M_{pl})^2<4/N$, where $N$ is a quantum number that characterizes the strength of the warping and for consistency $N\gg1$. 

If one could relax this bound the cosmological data could be easily reproduced \cite{BeanEtal2007,PeirisEtal2007,LorenzMartinRingeval2008}. Several ideas and speculations about how this could be achieved in string theory have been proposed. Nevertheless, until now no explicit model has emerged that is fully under control and where the geometric bound is violated.

\begin{figure}[t]
\begin{center}
{\includegraphics[width=0.48\textwidth]{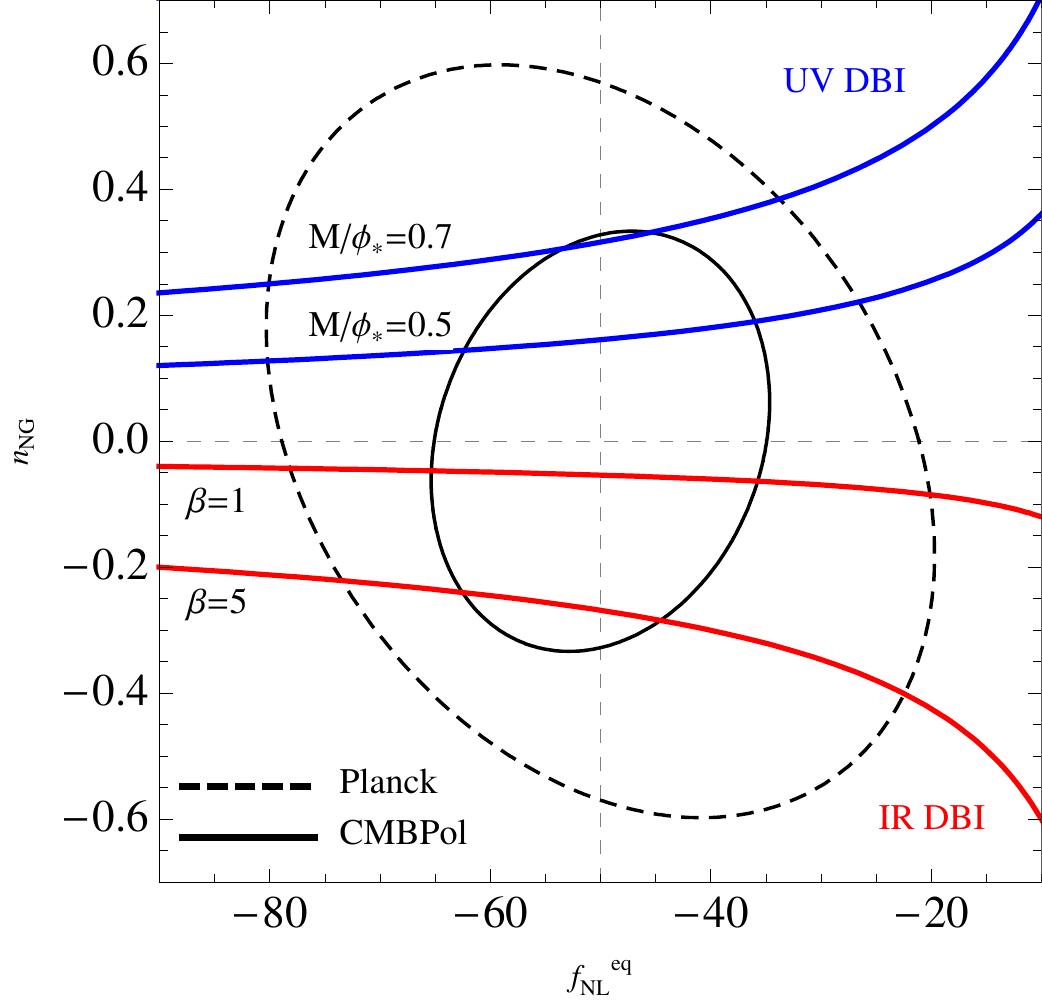}}
{\includegraphics[width=0.48\textwidth]{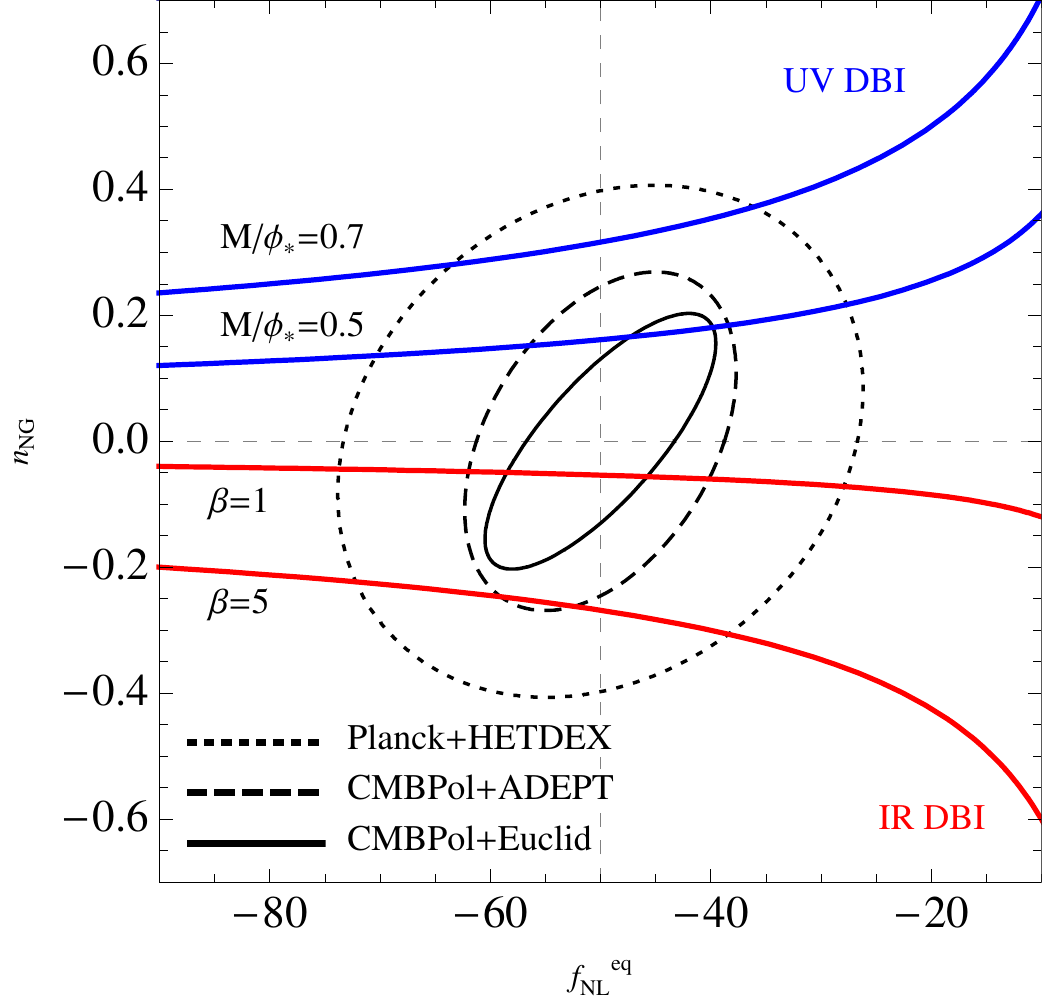}}
\caption{1-$\sigma$ contour plots for CMB alone ({\it left}) and combined CMB and LSS measurements ({\it right}) {\it vs.} DBI predictions. Assumes fiducial values $\fnle=-50$ and $\nng=0$, the first corresponding to the lower 1-$\sigma$ limit from the WMAP bispectrum analysis \citep{KomatsuEtal2008}.}
\label{fig:dbi}
\end{center}
\end{figure}

Given all the above considerations, we focus on a string theory-inspired model of DBI inflation, for which we neglect the geometrical bound and select the simplest realistic functions $f(\phi)$ and $V(\phi)$. Our archetypal UV DBI model is chosen to be defined by
\be
\quad f(\phi)=\frac{\lambda}{(\phi^2+\mu^2)^2}\,, \quad V(\phi)=\frac12 m^2\phi^2\,,
\ee
which is known to induce a power-law period of inflation \cite{SilversteinTong2004}. When $\phi\gg\mu$ the warp factor approximates that of AdS, $f(\phi)\sim \lambda \phi^{-4}$. Defining $p^2\simeq m^2 \lambda/(6M_{pl}^2)$ and assuming $p\gg1$, we find that the scale factor evolves as $a(t)\propto t^p$. In \cite{Chen2005}, the $f_{NL}$ parameter produced with the above choices was estimated for the equilateral configurations. The result in the WMAP notation is
\be
f_{NL}\simeq -1.3\,\frac{p^2M_{pl}^4}{\phi_{\ast}^4}\,,
\ee
where $\phi_{\ast}$ indicates the value of $\phi$ around which the CMB perturbations exit the horizon. The running of the non-Gaussianity $n_{NG}$ is straightforwardly calculated as\footnote{We use a different convention (see Eq.~\ref{eq:nNGdef}) with respect to \cite{Chen2005,ChenEtal2007} where a scale invariant $\fnl$ would correspond to $\nng=1$.}
\be 
\nng\equiv  \frac{\partial\ln|\fnl|}{\partial \ln k } \simeq \frac 4 p\,.
\ee
A few comments are in order. First, $n_{NG}>0$, which implies that the speed of sound decreases as time evolves. This is because in the UV models the brane advances towards regions with larger warping, {\it i.e.}~$\dot f >0$. Second, we notice that for simplicity we have chosen an archetypal model with three parameters\footnote{From an effective field theory point of view, there would be another additional parameter, {\it i.e.}~the reheating temperature $T_R$, that determines the number of e-foldings between the end of inflation and the time at which the CMB perturbations have exited the horizon. This parameter would be degenerate with say $\mu$ in the present model. On the other hand, once an explicit embedding of both inflation and the standard model in a string theory construction is specified, $T_R$ should be determined.} $\lambda$, $\mu$ and $m$. In the present context, one parameter is fixed by the COBE normalization while the other two still allow for enough freedom to cover most of the interesting region in the sector $f_{NL}<0\,\cap\, n_{NG}>0$. For illustrative purposes, we plot in Fig. \ref{fig:dbi} the prediction of the model in the $\fnl$-$\nng$ plane for different (constant) values of $\phi_{\ast}/M_{pl}$ together with the expected $1$-$\sigma$ constraints from CMB experiments ({\it left panel}) and combined CMB and large-scale structure constraints ({\it right panel}). This clearly shows that the geometric bound on the range of $\phi$ \cite{BaumannMcAllister2007} must be evaded in order to reconcile UV DBI models with experiments.

%%%%%%%%%%%%%%%%%%%%%%%%%%%%%%%%%%%%%%%%%%%%%%%%%%%%%%%%%%%%%%%%%%%%%%%%%%%%%%%%
\subsection{The IR model}

In a second class of models (initiated in \cite{Chen2005A}), the inflating D3-brane starts at the tip of a throat (the IR region), where the warp factor is maximal, and evolves towards the bulk of the compact space. This motion still takes place along the radial direction, but opposite to the UV models since $f(\phi)$ is now monotonically decreasing with time. 

In \cite{BeanEtal2008} an extensive comparison of this class of models with the data has been performed. Similar to the UV models, no IR DBI model was found which is compatible with the data and which obeys all the consistency conditions imposed by the embedding into string theory. The main obstacle is that the largest allowed number of e-foldings during the DBI regime is smaller than order ten. Even if inflation can be prolonged by a phase of fast or slow roll, the perturbations produced during the relativistic DBI regime are too red-tilted. References \cite{Chen2005B,Chen2005,BeanEtal2008} concluded that the model might be saved by assuming that the CMB perturbations are produced in a phase preceding the DBI ones. This would be a Hagedorn phase ({\it i.e.}~the Hubble temperature is larger than the warped string scale) in which the string partition function formally diverges and all string states are infinitely populated. This scenario is too poorly understood to allow for a rigorous treatment of the production of perturbations, and so we prefer to neglect this possibility until a better description is available.

As we did for the UV models, we consider an archetypal effective model of IR DBI inflation. To get a model compatible with current data, we neglect some of the constraints present in the known string theory constructions. In particular we relax the bound on the inflaton field range \cite{BaumannMcAllister2007} and allow for an arbitrary number of e-foldings in the ultra-relativistic DBI regime. We choose
\be 
\quad f(\phi)=\frac{\lambda}{(\phi^2+\mu^2)^2}\,, \quad V(\phi)=V_0-\frac12 m^2 \phi^2\,,
\ee
and we will often use the reparameterization $m^2\equiv \beta H^2$. As we mentioned above, considering various corrections \cite{KachruEtal2003,BurgessEtal2007,KrausePajer2008,BaumannEtal2008,BaumannEtal2009} one finds generically $\beta \sim\mathcal{O}(1)$. In \cite{Chen2005}, the $f_{NL}$ parameter produced by the above model in the equilateral configurations was estimated to be
\be
f_{NL}\simeq -0.036 \beta^2 N_{\ast}^2\,,
\ee
where $N_{\ast}$ is the number of e-foldings prior the end of inflation around which the CMB perturbations are produced. The running of the non-Gaussianity $n_{NG}$ was estimated to be
\be 
n_{NG}\simeq -\frac{2}{N_{\ast}}\,.
\ee
In contrast to the UV model, $n_{NG}<0$, a consequence of the fact that $\dot f <0$. Notice that for a large class of potentials, the sign of $n_{NG}$ does not depend on $V(\phi)$; as we will soon see, this is not the case in the angular model. In addition, our archetypal effective model has enough parameters to cover most of the interesting region in the sector $f_{NL}<0\,\cap\, n_{NG}<0$. We show in Fig. \ref{fig:dbi} some $n_{NG}(f_{NL})$ curves for various values of $\beta$. The curves are similar to those in the UV model but with opposite sign.

%%%%%%%%%%%%%%%%%%%%%%%%%%%%%%%%%%%%%%%%%%%%%%%%%%%%%%%%%%%%%%%%%%%%%%%%%%%%%%%%
\subsection{The angular model}\label{angular}

We now consider a third model where $f(\phi)$ is constant, which in a sense lies in between the UV and IR models. In a string theory construction, this can be achieved if the inflationary D3-brane moves along the angular directions of a warped conifold, and so maintains a constant radius. An explicit realization of this was constructed in \cite{Pajer2008} (see also \cite{DeWolfeEtal2007} for an earlier attempt). The D3-brane moves along the tip (which is an $S^3$) of a warped deformed conifold, hence the name \textit{inflation at the tip}. The motion is driven by the F-term potential generated by the presence of a stack of D7-branes with gaugino condensation. In this Section the field $\phi$ refers to the angular position of the D3-brane, while the radial position is constant.

The analysis of \cite{Pajer2008} shows that, even for a mildly warped throat, a phase of ultra-relativistic DBI inflation generally occurs. On the other hand, once the bounds induced by the consistency of the string theory construction are imposed, not enough e-foldings of DBI inflation can be generated\footnote{On the contrary, if one allows for fine tuning, models of slow-roll inflation can be found that are compatible with all present data.}. Although some ideas have been proposed to overcome these difficulties \cite{Pajer2008}, no successful concrete model of angular DBI inflation has been found.

Analogously to the IR and UV cases, in the lack of a genuine string theory construction we consider an archetypal effective model. For simplicity, we propose the most minimal model with just two parameters,\footnote{In this model $N_{\ast}\sim \gamma (\pin/2)^2$ which shows that a superplanckian field range is required to get enough e-foldings. Hence, a string theory construction needs to go beyond this minimal model. A problematic field range bound is therefore common to all three models of DBI inflation.}
\be
\quad f(\phi)=\text{const}\,, \quad V(\phi)=\frac12 m^2\phi^2\,.
\ee
The resulting non-Gaussianity can be estimated using Eq.~\ref{e:NG} and the approximated solution obtained in Appendix \ref{a:mia}
\be
\fnl\simeq -0.3 \gamma ^2 \sim -\frac15 m^2 f\,.
\ee
The running of $\fnl$ has been calculated in Appendix \ref{a:mia}. The result is
\be
\nng\simeq -\frac{8}{f m^2 \pin^4}\,,
\ee
where again $\phi=\pin$ corresponds to the time around which CMB perturbations exit the horizon. If we use COBE normalization to fix one of the two parameters $f$ or $m$ and express the other in terms of $\fnl$, we find that $\nng(\fnl)\sim 10^{-4}$, {\it i.e.}~it is constant at leading order. A small $\nng$ could have been anticipated: $\nng$ takes the opposite sign in the UV and IR models, respectively, and the angular model lies in the middle, {\it i.e.}~$\dot f_{UV}>\dot f_{ang}=0 > \dot f_{IR}$, close to where $n_{NG}$ passes through zero. As can be seen from Fig. \ref{fig:dbi}, such a small $n_{NG}$ is much below the experimental sensitivity. Contrary to the UV and IR models, in the angular model the sign of $n_{NG}$ is not fixed by $f(\phi)$ but it depends on the potential. In particular, a convex $V(\phi)$ gives a negative $\nng$ and vice versa.

%%%%%%%%%%%%%%%%%%%%%%%%%%%%%%%%%%%%%%%%%%%%%%%%%%%%%%%%%%%%%%%%%%%%%%%%%%%%%%%%
\section{Conclusions}
\label{sec:conclusions}

We studied the possibility of constraining a scale-dependence of the amplitude of a non-Gaussian component in the primordial perturbations in future observations. Focusing on two models---{\it local} and {\it equilateral}---for the bispectrum of the curvature fluctuations at early times, we performed a simple Fisher matrix analysis of the CMB bispectrum in terms of the non-Gaussian {\it amplitude} parameter $\fnl$ {\it and} a {\it running} parameter $\nng$.
In addition, we considered a simple Fisher matrix analysis of the large-scale structure power spectrum for the local model and of the galaxy bispectrum for the equilateral model of non-Gaussianity. We compared and combined these results to those expected from CMB observation to explore possible complementarities between such different probes of the early Universe.

We find that, in the event of a detection of a relative large non-Gaussian component, upcoming and future CMB missions such as Planck and CMBPol are capable of providing significant constraints on a possible scale-dependence such as predicted in certain inflationary models like DBI inflation. Assuming the current central values of the limits from WMAP observations for the local and equilateral models, that is $\fnll=38$ \citep{SmithSenatoreZaldarriaga2009} and $\fnle=51$ \citep{KomatsuEtal2008}, our Fisher matrix analysis indicates that we could expect 1-$\sigma$ uncertainties on the running parameters of the order of $\Delta\nng\simeq 0.15$ and $\Delta\nng\simeq 0.6$ from the Planck mission. The proposed CMBPol mission should reach uncertainties of the order of $\Delta\nng \simeq 0.05$ and $\Delta\nng\simeq 0.35$ for the local and equilateral model respectively, quite close to those expected from the ideal CMB mission.

We remark that the constraint on the extra running parameter come, in principle, at {\it no cost} with respect to the constraints on the amplitude. In fact, given a specific observable and a specific model it is always possible to choose the optimal pivot point that removes any degeneracy between the two parameters.  However, it is useful to determine an optimal pivot point that can be used by different probes. We notice that, assuming our definition for the scale-dependent non-Gaussian parameter $\fnl(k)$ given in Eq.~(\ref{eq:runningfnl}), for an experiment like the proposed CMBPol mission, quite close to the ideal CMB experiment, the optimal pivot point is about $k_p\simeq 0.3$ Mpc$^{-1}$ for the local model and $k_p\simeq 0.4$ Mpc$^{-1}$ for the equilateral model, indicating the different distribution of the signal across the range of observable scales of the two models.

We also studied the constraints achievable by future large-scale structure surveys, finding results comparable to those from CMB measurements both for local and equilateral non-Gaussianity. In particular, for the local model we focused on the effect of non-Gaussian initial conditions on the bias of halos and galaxies which received considerable attention in the recent literature, \citep{DalalEtal2008,MatarreseVerde2008,SlosarEtal2008,McDonald2008,AfshordiTolley2008,TaruyaKoyamaMatsubara2008} and which lead to constraints on the local $\fnll$ from current observations already comparable to those from the CMB, \citep{SlosarEtal2008,AfshordiTolley2008}. We considered an expression for the correction to the galaxy linear bias parameter given in terms of the initial matter bispectrum, as those proposed  in \citep{MatarreseVerde2008,TaruyaKoyamaMatsubara2008}, which naturally allows for a scale-dependent $\fnl(k_1,k_2,k_3)$, and derived the expected constraints on the non-Gaussian parameters $\fnl$ and $\nng$ from measurements of the galaxy power spectrum in future spectroscopic and photometric surveys. We find that large-volume surveys such as LSST and Euclid should be able to provide constraints smaller than those expected from Planck and somehow able to improve those expected from the proposed CMBPol mission both on $\fnl$ as well as on the running parameter $\nng$, possibly achieving a $1$-$\sigma$ uncertainty below $\Delta\nng\simeq 0.05$ in a joint analysis. Moreover, we noticed that the CMB and LSS constraints show a marked complementarity in the specific case of local non-Gaussianity. 

The effects of equilateral non-Gaussianity on the galaxy power spectrum, on the other hand, are expected on theoretical grounds, to be negligible, \citep{TaruyaKoyamaMatsubara2008}. In this case, we turned our attention to the galaxy bispectrum. Recent studies of the effects of non-Gaussian initial conditions on the bispectrum of biased populations have shown that large-scale corrections, similar to the one affecting the power spectrum in the local case, are present {\it both} for local {\it and} equilateral non-Gaussianity \citep{JeongKomatsu2009,Sefusatti2009}. However, these preliminary results still need to be properly tested against numerical simulations before providing reliable predictions for the galaxy bispectrum. We therefore considered the simple tree-level expression for the galaxy bispectrum in perturbation theory where the only non-Gaussian correction is represented by the primordial component to the matter bispectrum. This results in a quite conservative Fisher matrix analysis since the neglected (1-loop) corrections due to non-linear bias are expected to significantly increase the non-Gaussian signature on the galaxy bispectrum, in principle for any non-Gaussian model. Nevertheless, our analysis, which constitutes an extension to the case of running non-Gaussianity of the analysis of \citep{SefusattiKomatsu2007}, shows that future large spectroscopic surveys like Euclid could provide constraints on the $\nng$ parameter of the order of $\Delta\nng\simeq 0.1$, comparable or even better than those from CMBPol, while presenting a sensible complementarity.

It is to be stressed that these simple and preliminary results do not constitutes a proper analysis of the ability of large-scale structure observations to constrain non-Gaussianity. They show, however, that the large-scale galaxy distribution can be an as sensitive, although less direct, probe of the early Universe and that a possible detection of a significant level of non-Gaussianity in the CMB should find a confirmation in the large-scale structure. Moreover, they indicate that extra information, as a possible scale-dependence, on such non-Gaussian component can be extracted by a joint analysis of CMB and LSS data. Clearly a complete analysis of large-scale galaxy correlators should take into account the power spectrum as well as the bispectrum and their covariance, which we neglected here. Still, we expect our results to motive further studies in the same direction.

Finally we considered a specific, string-motivated, inflationary model---DBI inflation---as a case study for running non-Gaussianity. This model naturally predicts a large non-Gaussian signature with a significant scale-dependence. Interestingly, for two common DBI inflaton Lagrangians, a smaller non-Gaussian amplitude corresponds to a stronger running. In this respect, constraints on this specific model should consistently take into account the peculiar scale-dependence of the predicted curvature bispectrum. We showed, in fact,
that the determination of the running parameter $\nng$ can place additional constraints on the parameters of the specific model of inflation breaking degeneracies otherwise present from a simple determination of the overall amplitude of the primordial bispectrum. Although an explicit realization of DBI inflation in string theory has not been found, under a purely phenomenological point of view, our analysis shows how the extra information provided by the running parameter $\nng$ directly translates into extra information on the parameters of the model.

\acknowledgements

We are particularly grateful to Matias Zaldarriaga for comments on an early draft of the paper. We thank Xingang Chen, Marilena LoVerde, Liam McAllister, Federico Piazza, Sarah Shandera, Henry Tye, Licia Verde, Jiajun Xu, for useful discussions. M.~G.~J. and E.~S. are supported in part by the U.S. Department of Energy at Fermilab. E.~S. acknowledges support by the French Agence National de la Recherche under grant BLAN07-1-212615. The research of E.~P. is supported by U.S. N.S.F. grant PHY-0355005. E.~S. expresses its gratitude to the Department of Applied Mathematics and Theoretical Physics of the University of Cambridge, where this work was started, for their kind hospitality.

\bibliography{Bibliography}

\appendix

%%%%%%%%%%%%%%%%%%%%%%%%%%%%%%%%%%%%%%%%%%%%%%%%%%%%%%%%%%%%%%%%%%%%%%%%%%%%%%%%
%%%%%%%%%%%%%%%%%%%%%%%%%%%%%%%%%%%%%%%%%%%%%%%%%%%%%%%%%%%%%%%%%%%%%%%%%%%%%%%%
\section{Scale-dependence and triangular configurations} 
\label{app:triangles}

\begin{figure}[t]
\begin{center}
{\includegraphics[width=0.48\textwidth]{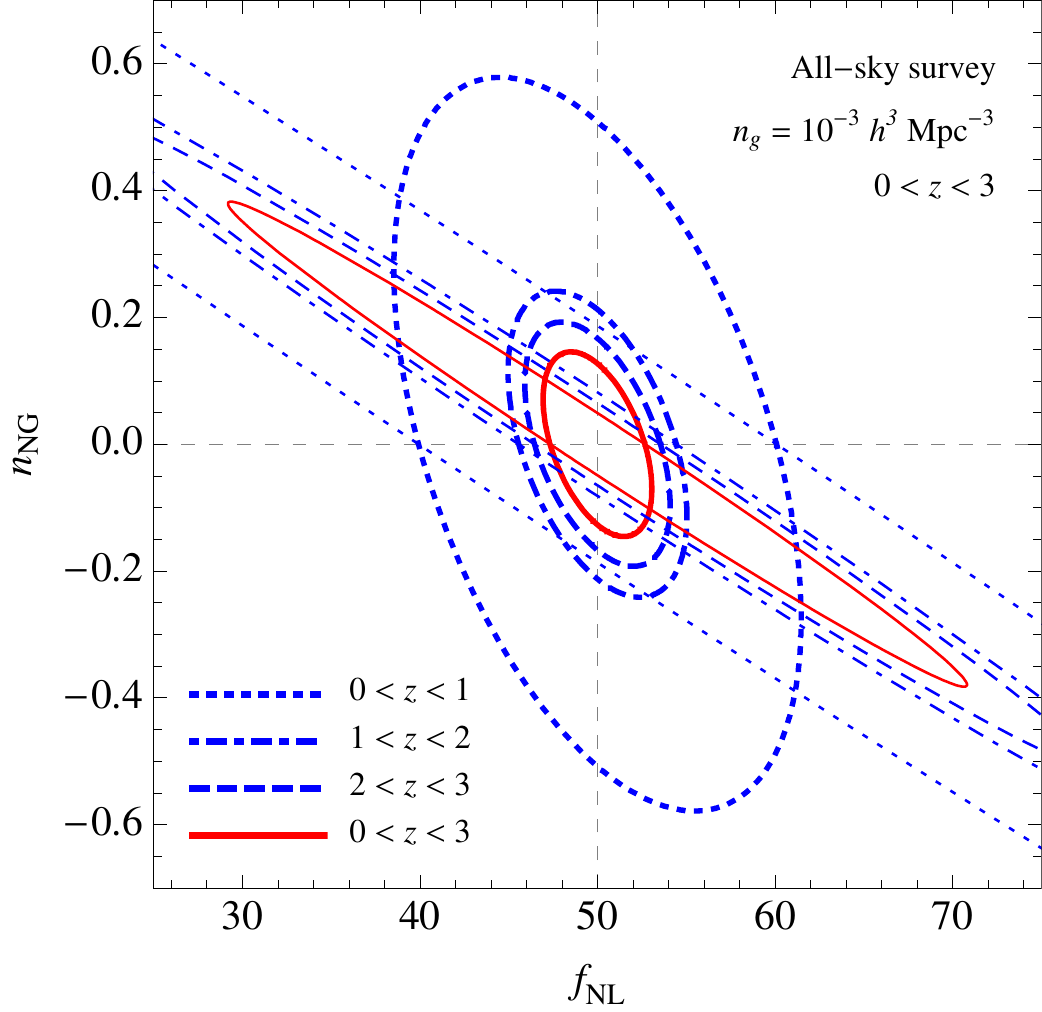}}
{\includegraphics[width=0.48\textwidth]{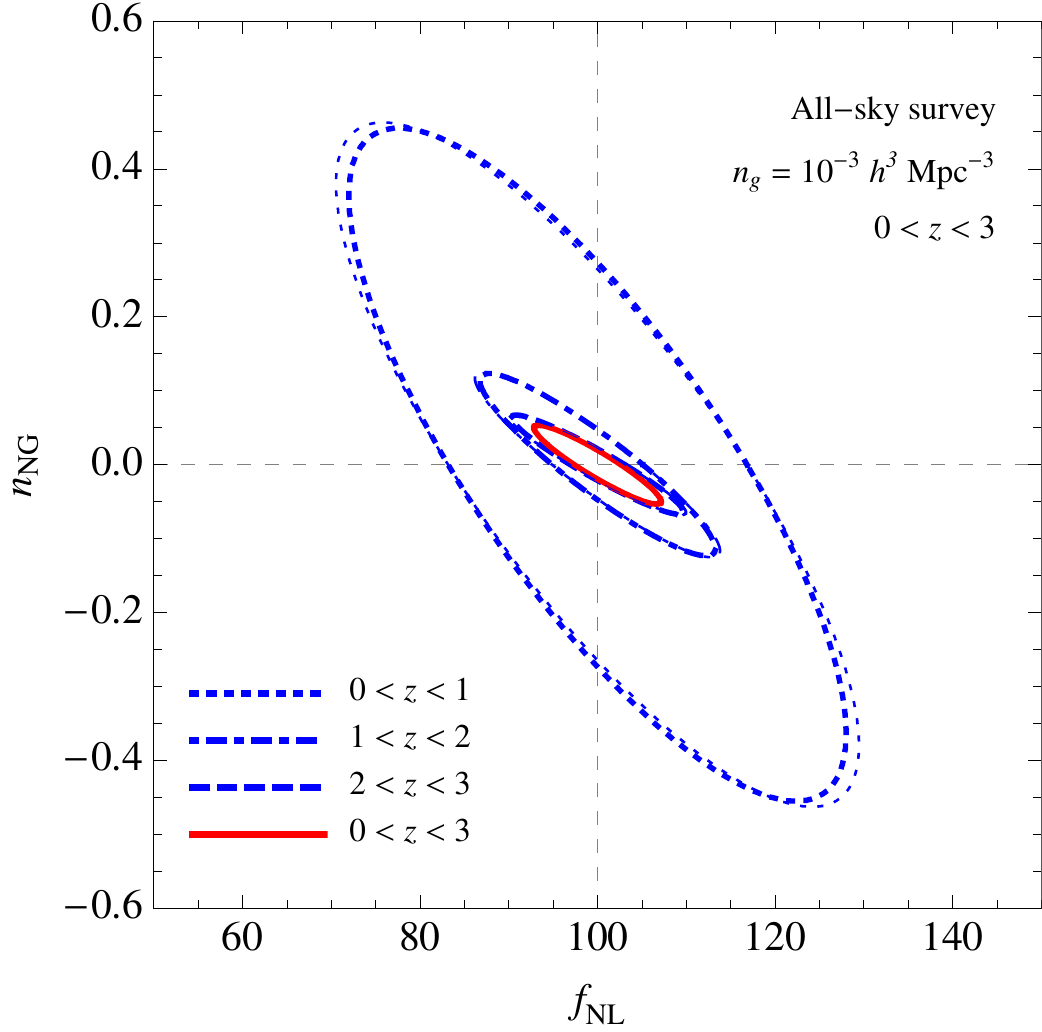}}
\caption{Correlation coefficient $c_{12}$ between the two non-Gaussian parameters $\fnl$ and $\nng$ as a function of $l_{max}$ assuming several values of the pivot point $k_p$ in the $T+E$ case for CMBPol. {\it Left panel}: local model. {\it Right panel}: equilateral model.}
\label{fig:app:kG_vs_kA}
\end{center}
\end{figure}

In this Appendix we compare the results from the all-sky Fisher matrix analyses of the galaxy power spectrum (for local non-Gaussianity) and of the galaxy bispectrum (for equilateral non-Gaussianity) obtained assuming the geometric mean, adopted in this work to define the scale-dependent non-Gaussian parameter $\fnl(k_1,k_2,k_3)\equiv\fnl(K)$, given by
\be\label{eq:geom}
K=(k_1k_2k_3)^{1/3},
\ee
with the same results obtained instead assuming the arithmetic mean
\be\label{eq:arit}
K=(k_1+k_2+k_3)/3,
\ee
assumed by other works and derived as the ``correct" form of the curvature bispectrum predicted by DBI inflation. In Fig.~\ref{fig:app:kG_vs_kA} the thick lines correspond to the geometric choice of Eq.~(\ref{eq:geom}) while the thin lines to the arithmetic one of Eq.~(\ref{eq:arit}). On the left panel, we show the analysis for the galaxy power spectrum and local non-Gaussianity, while on the right panel we show the analysis for the galaxy bispectrum and equilateral non-Gaussianity. In the case of equilateral non-Gaussianity, the difference is minimal (and, as expected, almost unnoticeable) since most of the signal is coming from equilateral configurations. The two choices show instead a significant difference in the case of local non-Gaussianity, both in terms of degeneracy and overall marginalized errors. While the factorizable, geometric mean of Eq.~(\ref{eq:geom}) is motivated by the easier implementation of the CMB estimators, we can assume this choice to have a purely phenomenological value. On the other hand, in the equilateral model, we can expect the CMB results to hold even assuming the arithmetic mean, better motivated under the theoretical point-of-view. A proper implementation of an estimator for the CMB bispectrum described by a running non-Gaussian parameter $\fnl(K)$ with $K$ given by Eq.~(\ref{eq:arit}) will be considered in a future work.  

%%%%%%%%%%%%%%%%%%%%%%%%%%%%%%%%%%%%%%%%%%%%%%%%%%%%%%%%%%%%%%%%%%%%%%%%%%%%%%%%
%%%%%%%%%%%%%%%%%%%%%%%%%%%%%%%%%%%%%%%%%%%%%%%%%%%%%%%%%%%%%%%%%%%%%%%%%%%%%%%%
\section{The choice of the pivot point} 
\label{app:pivot}

\begin{figure}[t]
\begin{center}
{\includegraphics[width=0.48\textwidth]{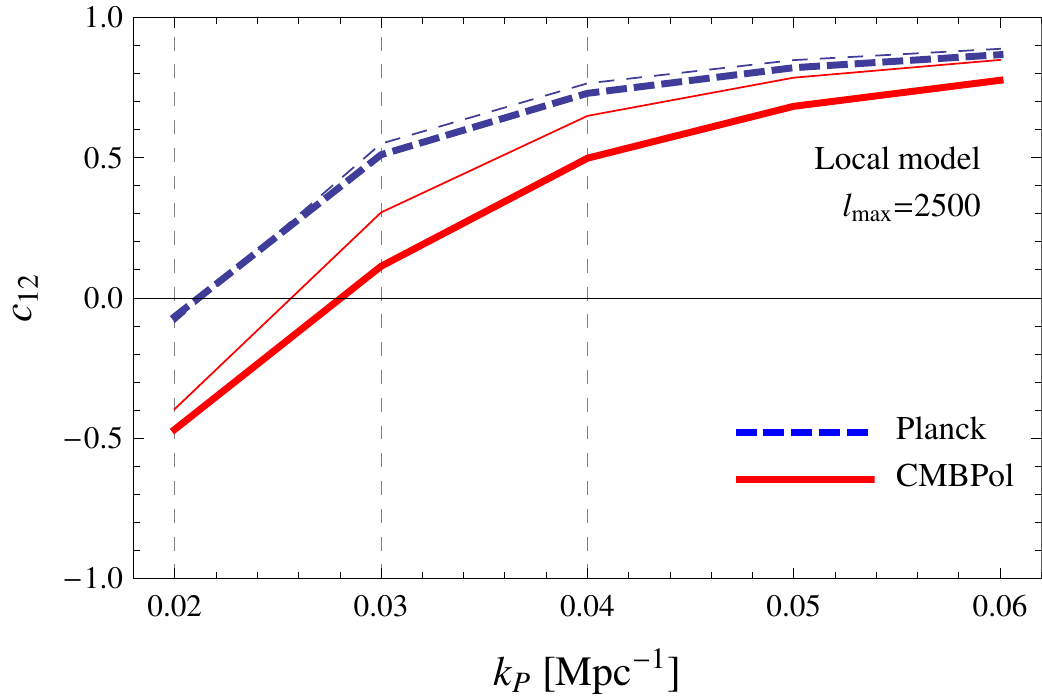}}
{\includegraphics[width=0.48\textwidth]{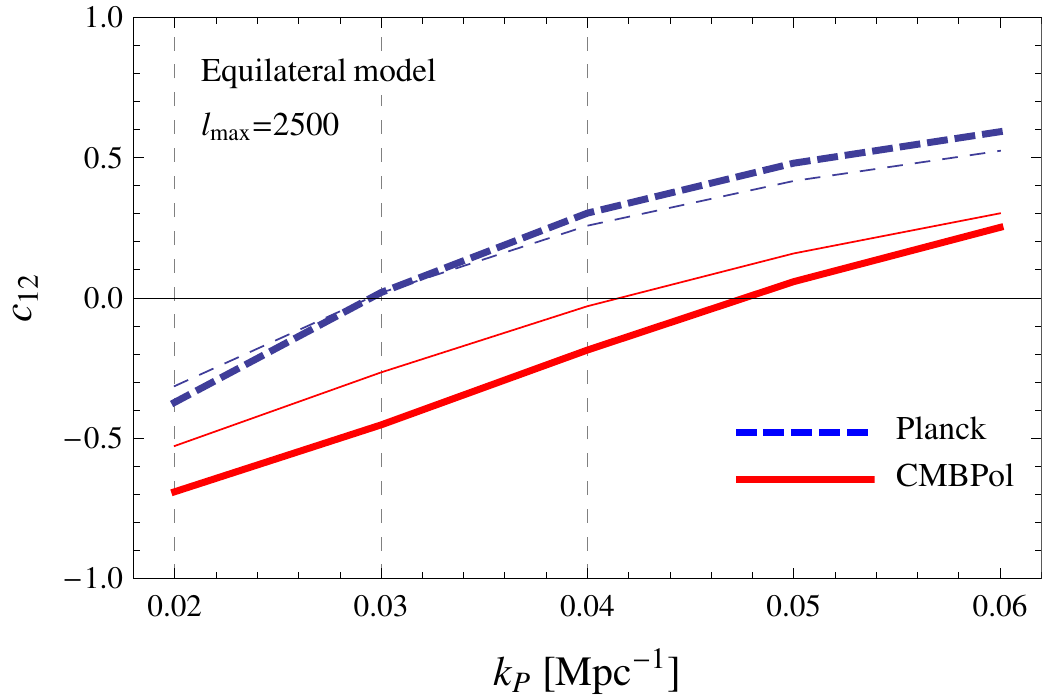}}
\caption{Correlation coefficient $c_{12}$ between the two non-Gaussian parameters $\fnl$ and $\nng$ (see text) as a function of the pivot point $k_p$ for the temperature ({\it thin lines}) and temperature and polarization analysis ({\it thick lines}) assuming the Planck ({\it dashed lines}) and CMBPol ({\it continuous lines}) experiments with $l_{max}=2500$. {\it Left panel}: local model. {\it Right panel}: equilateral model.}
\label{fig:cc}
\end{center}
\end{figure}

The degree of degeneracy between the non-Gaussian parameters $\fnl$ and $\nng$ strongly depends on the specific choice for the pivot point the enters the definition of Eq.~(\ref{eq:runningfnl}). Since such a choice is arbitrary, it is convenient to choose a value of $k_p$ that minimizes the correlation between the parameters. However, the best value obtained in this way depends in turn on the observable considered and on its characteristic signal-to-noise over the range of scales probed, which, in addition, is specific to the non-Gaussian model under consideration. Since the CMB bispectrum is currently providing the best constraints on primordial non-Gaussianity and also constitutes its best-studied probe to date, in this work we determine the value of pivot point $k_p$ to be the value that minimizes the degeneracy between $\fnl$ and $\nng$ as derived from the CMB analysis. 

Given the CMB Fisher matrix $F_{ij}$, the correlation coefficient for the parameters is given by
\be
c_{12}=\frac{\left(F^{-1}\right)_{12}}{\sqrt{\left(F^{-1}\right)_{11}\left(F^{-1}\right)_{22}}},
\ee
where the indices 1 and 2 label the parameters $\fnl$ and $\nng$, respectively. Notice that since $(F^{-1})_{12}\sim\fnl^{-1}$, $(F^{-1})_{22}\sim\fnl^{-2}$ and $(F^{-1})_{11}$ does not depend on $\fnl$, the correlation coefficient is independent on the fiducial value of $\fnl$ chosen.

In Fig.~\ref{fig:cc} we plot the correlation coefficient $c_{12}$ between the two non-Gaussian parameters $\fnl$ and $\nng$ as a function of the pivot point $k_p$ for the Planck experiment ({\it dashed, blue lines}) and the CMBPol experiment ({\it continuous, red lines}) assuming in both cases $l_{max}=2500$.  The thin lines correspond to the temperature analysis while thick lines correspond to the combined temperature and polarization information. Finally, on the left panel we assume local non-Gaussianity while on the right panel equilateral non-Gaussianity. As can be expected, the local model prefers a lower value for $k_p$ of about $0.03$ Mpc$^{-1}$ where $c_{12}\simeq 0.1$ for CMBPol, due to the higher signal-to-noise in configurations correlating small and large angular scales. For the equilateral model we find instead a best value around $k_p=0.05$ Mpc$^{-1}$, where $c_{12}\simeq 0.06$. The corresponding values for Planck are naturally lower in both cases.

A high value of $k_p$ can be expected to be a better choice for an analysis of large-scale structure information, since the galaxy distribution and even more cluster abundance probe smaller scales. For simplicity we could choose to determine $k_p$ as the optimal value for the CMB bispectrum analysis, so that the expected errors on $\fnl$ from future CMB experiments, {\it marginalized over the running parameter} $\nng$, are not significantly different from the values already determined in the literature under the assumption of a constant $\fnl$. In this respect the choice of $k_p\simeq 0.04$ Mpc$^{-1}$, proposed already by \citet{LoVerdeEtal2008}, seems to be a particularly reasonable choice as a compromise between the two models of non-Gaussianity and in the perspective of future CMB missions beyond Planck.

%%%%%%%%%%%%%%%%%%%%%%%%%%%%%%%%%%%%%%%%%%%%%%%%%%%%%%%%%%%%%%%%%%%%%%%%%%%%%%%%
%%%%%%%%%%%%%%%%%%%%%%%%%%%%%%%%%%%%%%%%%%%%%%%%%%%%%%%%%%%%%%%%%%%%%%%%%%%%%%%%
\section{Running of non-Gaussianity in angular DBI inflation}
\label{a:mia}

In this Appendix we obtain an analytical estimate for the running of non-Gaussianity in the simplest model of angular DBI inflation, {\it i.e.}~with constant $f$ and $V=m^2 \phi^2/2$. Our starting point is
\be \label{nng}
n_{NG}\simeq -2 s \equiv -2 \frac{\dot c_s}{c_s H}=2 \frac{\dot \gamma}{\gamma H}\simeq2 \frac{H''}{H'H \sqrt{f}}\,,
\ee
where in the last step we used the fact that we are in the ultra-relativistic regime
\be \label{gamma}
\gamma=\sqrt{1+4 f (H')^2}\simeq 2 H' \sqrt{f} \gg1\,.
\ee
Using the Hamilton-Jacobi formalism it is straightforward to solve for $H$ at linear order in $\phi$
\be 
H(\phi)\simeq \frac{m \phi}{\sqrt{6}}\equiv\frac{m}{\sqrt{6}}(\pin - \dep)\,,
\ee
where we have introduced the distance $\dep$ traveled by the inflaton since the CMB perturbations have exited the horizon at $\phi=\pin$. To obtain the running $n_{NG}$ we need to go beyond the linear approximation. We therefore make the ansatz
\be
H(\phi)= \left(\frac{m \pin}{\sqrt{6}}+ h_0\right)+ \left(\frac{m}{\sqrt{6}}+h_1 \right) \dep+h_2\dep^2\,.
\ee
We now expand 
\be 
3 H^2=V + \frac{\gamma-1}{f}
\ee
at second order in $\dep$, using the exact expression for $\gamma$ given in Eq.~\ref{gamma}. Then we linearize in $h_0$, $h_1$ and $h_2$ and solve for them. The result is
\be 
\pin^2 h_0=\pin h_1=h_2=\frac{1}{3 \sqrt{f}\pin^3}\,.
\ee
Plugging this solution into Eq.~\ref{nng} we get the estimate for the running of non-Gaussianity produced close to $\phi=\pin$
\be 
n_{NG}\simeq - \frac{8}{f \pin^4 m^2}\,.
\ee
We have numerically checked that this is a good estimate for $\gamma\gg1$ and $\dep\ll 1$.

\end{document}